\documentclass[12pt]{article}

\usepackage{axodraw}
\usepackage{epsfig}

\makeatletter

\def\paragraph{\@startsection{paragraph}{4}{\z@}{+2.00ex plus
 +1ex minus +.2ex}{1.5ex plus .2ex}{\it\normalsize}}

\def\section{\@startsection {section}{1}{\z@}{+3.0ex plus +1ex minus
  +.2ex}{2.3ex plus .2ex}{\normalsize\bf\boldmath}}
\def\subsection{\@startsection{subsection}{2}{\z@}{+2.5ex plus +1ex
minus +.2ex}{1.5ex plus .2ex}{\normalsize\bf\boldmath}}
\def\subsubsection{\@startsection{subsubsection}{3}{\z@}{+3.25ex plus
 +1ex minus +.2ex}{1.5ex plus .2ex}{\normalsize\it}}

\expandafter\ifx\csname mathrm\endcsname\relax\def\mathrm#1{{\rm #1}}\fi

\newcounter{saveeqn}

\@addtoreset{equation}{section}

\newcount\@tempcntc
\def\@citex[#1]#2{\if@filesw\immediate\write\@auxout{\string\citation{#2}}\fi
  \@tempcnta\z@\@tempcntb\m@ne\def\@citea{}\@cite{\@for\@citeb:=#2\do
    {\@ifundefined
       {b@\@citeb}{\@citeo\@tempcntb\m@ne\@citea
        \def\@citea{,\penalty\@m\ }{\bf ?}\@warning
       {Citation `\@citeb' on page \thepage \space undefined}}%
    {\setbox\z@\hbox{\global\@tempcntc0\csname
b@\@citeb\endcsname\relax}%
     \ifnum\@tempcntc=\z@ \@citeo\@tempcntb\m@ne
       \@citea\def\@citea{,\penalty\@m}
       \hbox{\csname b@\@citeb\endcsname}%
     \else
      \advance\@tempcntb\@ne
      \ifnum\@tempcntb=\@tempcntc
      \else\advance\@tempcntb\m@ne\@citeo
      \@tempcnta\@tempcntc\@tempcntb\@tempcntc\fi\fi}}\@citeo}{#1}}

\def\@citeo{\ifnum\@tempcnta>\@tempcntb\else\@citea
  \def\@citea{,\penalty\@m}%
  \ifnum\@tempcnta=\@tempcntb\the\@tempcnta\else
   {\advance\@tempcnta\@ne\ifnum\@tempcnta=\@tempcntb \else
\def\@citea{--}\fi
    \advance\@tempcnta\m@ne\the\@tempcnta\@citea\the\@tempcntb}\fi\fi}

\def\nl{\nonumber\\}

\def\nln{\\*[-1ex]}

\newcommand{\lsim}
{\mathrel{\raisebox{-.3em}{$\stackrel{\displaystyle <}{\sim}$}}}

\def\asymp#1%
{\mathrel{\raisebox{-.4em}{$\widetilde{\scriptstyle #1}$}}}

\def\Nequal#1%
{\mathrel{\raisebox{-.5em}{$\stackrel{=}{\scriptstyle\rm#1}$}}}
\newcommand{\dsl}[1]{\not \hspace{-0.7mm}#1}
\def\dsl{\mathpalette\make@slash}
\def\make@slash#1#2{\setbox\z@\hbox{$#1#2$}%
  \hbox to 0pt{\hss$#1/$\hss\kern-\wd0}\box0}

\def\beq{\begin{equation}}
\def\eeq{\end{equation}}
\def\beqar{\begin{eqnarray}}
\def\eeqar{\end{eqnarray}}
\def\barr#1{\begin{array}{#1}}
\def\earr{\end{array}}
\def\bfi{\begin{figure}}
\def\efi{\end{figure}}
\def\btab{\begin{table}}
\def\etab{\end{table}}
\def\bce{\begin{center}}
\def\ece{\end{center}}
\def\nn{\nonumber}
\def\disp{\displaystyle}
\def\text{\textstyle}

\def\al{\alpha}
\def\be{\beta}

\def\ga{\gamma}
\def\de{\delta}

\def\eps{\epsilon}
\def\veps{\varepsilon}
\def\la{\lambda}

\def\si{\sigma}

\def\refeq#1{\mbox{(\ref{#1})}}

\def\reffi#1{\mbox{Figure~\ref{#1}}}
\def\reffis#1{\mbox{Figures~\ref{#1}}}
\def\refta#1{\mbox{Table~\ref{#1}}}
\def\reftas#1{\mbox{Tables~\ref{#1}}}
\def\refse#1{\mbox{Section~\ref{#1}}}

\def\refapp#1{\mbox{App.~\ref{#1}}}
\def\citere#1{\mbox{Ref.~\cite{#1}}}
\def\citeres#1{\mbox{Refs.~\cite{#1}}}

\newcommand{\TeV}{\unskip\,\mathrm{TeV}}
\newcommand{\GeV}{\unskip\,\mathrm{GeV}}
\newcommand{\MeV}{\unskip\,\mathrm{MeV}}
\newcommand{\keV}{\unskip\,\mathrm{keV}}
\newcommand{\pba}{\unskip\,\mathrm{pb}}
\newcommand{\fba}{\unskip\,\mathrm{fb}}

\newcommand{\ri}{{\mathrm{i}}}
\newcommand{\rd}{{\mathrm{d}}}

\newcommand{\rE}{{\mathrm{E}}}

\newcommand{\Oa}{\mathswitch{{\cal{O}}(\alpha)}}
\newcommand{\Oaa}{\mathswitch{{\cal{O}}(\alpha^2)}}
\renewcommand{\L}{{\cal{L}}}
\newcommand{\M}{{\cal{M}}}

\def\mathswitchr#1{\relax\ifmmode{\mathrm{#1}}\else$\mathrm{#1}$\fi}

\newcommand{\PW}{\mathswitchr W}
\newcommand{\Pw}{\mathswitchr w}
\newcommand{\PZ}{\mathswitchr Z}

\newcommand{\Pg}{\mathswitchr g}
\newcommand{\PH}{\mathswitchr H}
\newcommand{\Pe}{\mathswitchr e}

\newcommand{\Pd}{\mathswitchr d}
\newcommand{\Pdbar}{\bar{\mathswitchr d}}
\newcommand{\Pu}{\mathswitchr u}

\newcommand{\Ps}{\mathswitchr s}

\newcommand{\Pc}{\mathswitchr c}

\newcommand{\Pt}{\mathswitchr t}

\newcommand{\Pep}{\mathswitchr {e^+}}
\newcommand{\Pem}{\mathswitchr {e^-}}
\newcommand{\PWp}{\mathswitchr {W^+}}
\newcommand{\PWm}{\mathswitchr {W^-}}
\newcommand{\PWpm}{\mathswitchr {W^\pm}}

\def\mathswitch#1{\relax\ifmmode#1\else$#1$\fi}

\newcommand{\MW}{\mathswitch {M_\PW}}

\newcommand{\MZ}{\mathswitch {M_\PZ}}
\newcommand{\MH}{\mathswitch {M_\PH}}
\newcommand{\Me}{\mathswitch {m_\Pe}}

\newcommand{\Mt}{\mathswitch {m_\Pt}}
\newcommand{\GW}{\Gamma_{\PW}}
\newcommand{\GZ}{\Gamma_{\PZ}}

\newcommand{\sw}{\mathswitch {s_\Pw}}
\newcommand{\cw}{\mathswitch {c_\Pw}}

\newcommand{\GF}{\mathswitch {G_\mu}}

\def\solid{\raise.9mm\hbox{\protect\rule{1.1cm}{.2mm}}}
\def\dash{\raise.9mm\hbox{\protect\rule{2mm}{.2mm}}\hspace*{1mm}}

\def\ie{i.e.\ }
\def\eg{e.g.\ }
\def\cf{cf.\ }

\newcommand{\sing}{{\mathrm{sing}}}
\newcommand{\finite}{{\mathrm{finite}}}
\newcommand{\YFS}{{\mathrm{YFS}}}

\newcommand{\DPA}{{\mathrm{DPA}}}

\newcommand{\Born}{{\mathrm{Born}}}
\newcommand{\born}{{\mathrm{Born}}}

\newcommand{\virt}{{\mathrm{virt}}}
\newcommand{\real}{{\mathrm{real}}}
\newcommand{\soft}{{\mathrm{soft}}}
\newcommand{\coll}{{\mathrm{coll}}}
\newcommand{\sub}{{\mathrm{sub}}}

\newcommand{\fact}{{\mathrm{fact}}}
\newcommand{\nonfact}{{\mathrm{nfact}}}

\newcommand{\ffp}{\mathswitch{\mathrm{f\/f}'}}
\newcommand{\mfp}{\mathswitch{\mathrm{mf}'}}
\newcommand{\mmp}{\mathswitch{\mathrm{mm}'}}

\newcommand{\recomb}{{\mathrm{rec}}}

\newcommand{\U}{\mathrm{U}}

\def\Li{\mathop{\mathrm{Li}_2}\nolimits}

\def\Re{\mathop{\mathrm{Re}}\nolimits}

\def\sgn{\mathop{\mathrm{sgn}}\nolimits}

\newcommand{\eeWW}{{\Pe^+ \Pe^-\to \PW^+ \PW^-}}
\newcommand{\Wpff}{{\PW^+ \to f_1\bar f_2}}
\newcommand{\Wmff}{{\PW^- \to f_3\bar f_4}}
\newcommand{\eeWWffff}{\Pep\Pem\to\PW\PW\to 4f}
\newcommand{\eeffff}{\Pep\Pem\to 4f}
\newcommand{\eeffffg}{\eeffff\ga}
\renewcommand{\O}{{\cal O}}

\newcommand{\sfo}{\tilde{s}}

\newcommand{\qfo}{\tilde{q}}

\newcommand{\Phifo}{\tilde{\Phi}}
\newcommand{\kon}{\hat{k}}
\newcommand{\bkon}{{\bf \kon}}
\newcommand{\ton}{\hat{t}}

\newcommand{\SPppka}{\langle p_+ \kon_1 \rangle}
\newcommand{\SPppkb}{\langle p_+ \kon_2 \rangle}
\newcommand{\SPppkc}{\langle p_+ \kon_3 \rangle}
\newcommand{\SPppkd}{\langle p_+ \kon_4 \rangle}
\newcommand{\SPpmka}{\langle p_- \kon_1 \rangle}
\newcommand{\SPpmkb}{\langle p_- \kon_2 \rangle}
\newcommand{\SPpmkc}{\langle p_- \kon_3 \rangle}
\newcommand{\SPpmkd}{\langle p_- \kon_4 \rangle}
\newcommand{\SPkakb}{\langle \kon_1 \kon_2 \rangle}
\newcommand{\SPkakc}{\langle \kon_1 \kon_3 \rangle}
\newcommand{\SPkakd}{\langle \kon_1 \kon_4 \rangle}
\newcommand{\SPkbkc}{\langle \kon_2 \kon_3 \rangle}
\newcommand{\SPkbkd}{\langle \kon_2 \kon_4 \rangle}

\newcommand{\CSPppka}{\SPppka^*}
\newcommand{\CSPppkb}{\SPppkb^*}

\newcommand{\CSPppkd}{\SPppkd^*}
\newcommand{\CSPpmka}{\SPpmka^*}
\newcommand{\CSPpmkb}{\SPpmkb^*}

\newcommand{\CSPpmkd}{\SPpmkd^*}
\newcommand{\CSPkakb}{\SPkakb^*}

\newcommand{\CSPkakd}{\SPkakd^*}
\newcommand{\CSPkbkc}{\SPkbkc^*}
\newcommand{\CSPkbkd}{\SPkbkd^*}

\newcommand{\bp}{{\bf p}}
\newcommand{\bk}{{\bf k}}
\newcommand{\bare}{{\mathswitchr{bare}}}
\newcommand{\calo}{{\mathswitchr{calo}}}

\def\draftdate{\relax}
\def\mda{\relax}
\def\mua{\relax}
\def\mla{\relax}
\def\draft{
\def\thtystars{******************************}
\def\sixtystars{\thtystars\thtystars}
\typeout{}
\typeout{\sixtystars**}
\typeout{* Draft mode!
         For final version remove \protect\draft\space in source file *}
\typeout{\sixtystars**}
\typeout{}
\def\draftdate{\today}
\def\mua{\marginpar[\boldmath\hfil$\uparrow$]%
                   {\boldmath$\uparrow$\hfil}%
                    \typeout{marginpar: $\uparrow$}\ignorespaces}
\def\mda{\marginpar[\boldmath\hfil$\downarrow$]%
                   {\boldmath$\downarrow$\hfil}%
                    \typeout{marginpar: $\downarrow$}\ignorespaces}
\def\mla{\marginpar[\boldmath\hfil$\rightarrow$]%
                   {\boldmath$\leftarrow $\hfil}%
                    \typeout{marginpar: $\leftrightarrow$}\ignorespaces}
\def\Mua{\marginpar[\boldmath\hfil$\Uparrow$]%
                   {\boldmath$\Uparrow$\hfil}%
                    \typeout{marginpar: $\uparrow$}\ignorespaces}
\def\Mda{\marginpar[\boldmath\hfil$\Downarrow$]%
                   {\boldmath$\Downarrow$\hfil}%
                    \typeout{marginpar: $\downarrow$}\ignorespaces}
\def\Mla{\marginpar[\boldmath\hfil$\Rightarrow$]%
                   {\boldmath$\Leftarrow $\hfil}%
                    \typeout{marginpar: $\leftrightarrow$}\ignorespaces}
\overfullrule 5pt
\oddsidemargin -15mm
\marginparwidth 29mm
}

\def\stars{\strut\leaders\hbox{*}\hfill\strut}
\def\starline{\hfil\strut\hfil\hbox to \textwidth {\stars}\hfil}

\begin{document}
\thispagestyle{empty}
\def\thefootnote{\fnsymbol{footnote}}
\setcounter{footnote}{1}
\null
\draftdate\hfill BI-TP 2000/06 \\
\strut\hfill  LU-ITP 2000/003\\
\strut\hfill PSI-PR-00-11\\
\strut\hfill UR-1604 \\
\strut\hfill hep-ph/0006307
\vfill
\begin{center}
{\large \bf\boldmath
Electroweak radiative corrections to $\Pep\Pem\to\PW\PW\to4\mbox{ fermions}$ 
\\[.5em]
in double-pole approximation --- the {\sc RacoonWW} approach
\par} \vskip 2em
\vspace{1cm}
{\large
{\sc A.\ Denner$^1$, S.\ Dittmaier$^2$, M. Roth$^{3}$ and
D.\ Wackeroth$^4$} } 
\\[.5cm]
$^1$ {\it Paul-Scherrer-Institut\\
CH-5232 Villigen PSI, Switzerland} 
\\[0.3cm]
$^2$ {\it Theoretische Physik, Universit\"at Bielefeld \\
D-33615 Bielefeld, Germany}
\\[0.3cm]
$^3$ {\it Institut f\"ur Theoretische Physik, Universit\"at Leipzig\\
D-04109 Leipzig, Germany}
\\[0.3cm]
$^4$ {\it Department of Physics and Astronomy, University of Rochester\\
Rochester, NY 14627-0171, USA}
\par 
\end{center}\par
\vskip .5cm {\bf Abstract:} \par We calculate the complete $\O(\al)$
electroweak radiative corrections to $\eeWWffff$ in the electroweak
Standard Model
in the double-pole approximation. We give analytical results for the
non-factorizable virtual corrections and express the factorizable
virtual corrections in terms of the known corrections to on-shell
W-pair production and W decay. The calculation of the bremsstrahlung
corrections, \ie the processes $\eeffffg$ in lowest order, is based on
the full matrix elements. 
The matching of soft and collinear singularities between virtual and real
corrections is done
alternatively in two different ways, namely by using a subtraction
method and by applying phase-space slicing.  The $\O(\al)$ corrections
as well as higher-order initial-state photon radiation are implemented
in the Monte Carlo generator {\sc RacoonWW}.  Numerical results of
this program are presented
for the W-pair-production cross section, angular and W-invariant-mass
distributions at LEP2.  We also discuss the intrinsic theoretical
uncertainty of our approach.
\par
\vskip .5cm
\noindent
June 2000   
\null
\setcounter{page}{0}
\clearpage
\def\thefootnote{\arabic{footnote}}
\setcounter{footnote}{0}

\section{Introduction}
\label{se:intro}

At present, the focus of electroweak
Standard Model
tests lies on
\PW-boson-pair production at LEP2, $\eeWWffff$
\cite{wwrev,lep2rep,lep2mcws}. While the
invariant-mass distributions of the final-state fermion pairs allow
for precise measurements of the \PW-boson mass, the 
total cross section and the
angular
distributions can be used to obtain information on the non-abelian
couplings of the \PW~bosons. LEP2 provides us with $\O(10^4)$
\PW-boson pairs, thus leading to a typical experimental accuracy of
1\% \cite{lep2mcws}.
At a future $\Pep\Pem$ linear collider, 
the W-boson yield could
be increased by two orders of magnitude \cite{bu99}, and the
experimental accuracy will reach the level of some 0.1\%.

The Monte Carlo generators used in the past \cite{lep2repWevgen} 
for \PW-pair production are based on the lowest-order matrix elements
and typically include only the universal electroweak corrections.
These comprise the running of the electromagnetic coupling,
corrections associated with the $\rho$~parameter, the Coulomb
singularity close to threshold, and photonic corrections in
leading-logarithmic approximation.  For the total cross section these
generators have a precision at the level of 1--2\% at LEP2 energies
\cite{bo92,lep2repWcs,di97}; for distributions the accuracy in general
is worse.
While this was sufficient for the analysis of LEP2 data in the past,
it becomes already insufficient for the present analysis and will be
definitely inadequate for the final LEP2 analysis.
Since the
non-universal corrections increase with energy, the
accuracy of generators including only universal corrections is at the
level of 5--20\% \cite{di97} 
in the total cross section and thus not adequate at 
linear collider energies.

In this paper we aim at a theoretical accuracy of $\lsim0.5\%$ 
at not too large energies%
\footnote{Above 0.5--$1\TeV$ at least the leading electroweak 
logarithms at the two-loop level should be taken into account.}, 
which is sufficient for
LEP2 and reasonable for a $500\GeV$ linear collider.  This requires,
on the one hand, to include the complete set of lowest-order diagrams
for $\eeffff$, which is already implemented in many existing
generators.  Moreover, the complete $\Oa$ corrections for $\eeWWffff$
have to be taken into account, and care must be taken that also the
leading higher-order corrections are included.

The full treatment of the processes $\Pep\Pem\to 4f$ at the one-loop
level is of enormous complexity. 
Nevertheless, 
there is ongoing work in this direction \cite{Vi98}.
Moreover, the requirement to include
the finite width of the W bosons poses severe theoretical problems
with gauge invariance. 

An economic approach for the calculation of the $\Oa$ corrections to
$\eeWWffff$ consists in using the double-pole approximation (DPA).  In
DPA, only those terms are kept that are enhanced by two resonant
\PW~propagators and thus by 
a factor $\MW/\GW$ with respect
to all other contributions. With respect to the leading contributions,
the corrections to the non-doubly-resonant contributions are typically
of the order $\al/\pi \times \GW/\MW$. Even when taking into account a
conservative safety factor,
this approximation is sufficient for our precision tag of
$0.5\%$ as long as the doubly-resonant contributions dominate the
cross section. This is the case at LEP2 energies 
sufficiently above threshold if forward-scattered
electrons or positrons in the final state (single-\PW~production%
\footnote{A discussion of state-of-the-art calculations for single-W
production can be found in \citere{lep2mcws}.})
and, for processes involving the neutral current,
fermion--antifermion pairs with low invariant masses 
are excluded.
At higher energies, where diagrams without two resonant
\PW~bosons become more sizeable, it may be necessary to suppress those
by 
appropriate cuts, \eg 
invariant-mass cuts, 
on the final-state fermions. These types of
cuts must be applied anyhow in order to extract the physically
interesting $\PW\PW$ signal.

The DPA has several advantages as far as the virtual corrections are
concerned. Besides reducing the number of diagrams drastically, it
naturally provides a gauge-invariant answer. In contrast to the
corrections to the full process $\eeffff$, the virtual corrections in DPA 
depend on the fermion flavours of the final state in a universal way,
i.e.\ there is only one generic set of diagrams for all final states.
Moreover, the invariant functions that build up the
corrections do not depend on the decay angles but only on the
production angle of the \PW~bosons, thus allowing to speed up their
calculation in the Monte Carlo generators considerably. Finally, the
virtual factorizable 
corrections in DPA can be directly obtained from the
existing results for on-shell W-pair production
\cite{rcwprod1,rcwprod2} and \PW-boson decay
\cite{rcwdecay2,rcwdecay1}. 
The virtual non-factorizable corrections,
\ie those virtual corrections that cannot be associated with either
\PW-pair production or \PW-boson decay,
have a simple structure \cite{me96,nfc2,nfc1a,ro99}.

On the other hand, the definition of the DPA is problematic for the
bremsstrahlung corrections. Photon radiation from the initial and from
the final state leads to \PW~propagators that become resonant at
different locations in phase space.  The situation depends on the
regime of the photon energy. If the radiated photons are hard
($E_\ga\gg\GW$), these locations are well separated, and the photon
radiation can be unambiguously assigned to the \PW-pair production
subprocess (at least theoretically) or to one of the \PW-decay
subprocesses. For soft photons ($E_\ga\ll\GW$), the resonances
coincide, and the DPA is identical to the one without photon. However,
for semi-soft photons ($E_\ga\sim\GW$), the resonant propagators
overlap, and it is not obvious how the DPA can be applied without
omitting or double-counting doubly-resonant contributions.  For these
reasons, we have preferred to employ the exact $\eeffffg$ matrix
element for the real bremsstrahlung corrections and the exact
$\eeffff$ matrix element for the
soft and collinear singular virtual corrections, 
and to apply the DPA only to the non-leading
part of the virtual
corrections.
Furthermore, all observables for real photons are based on a full
lowest-order calculation for $\eeffffg$ \footnote{A survey of
  state-of-the-art calculations for $\eeffffg$ is also contained in
  \citere{lep2mcws}.}.
 
A possible strategy for a DPA for the $\Oa$ corrections to pair
production of unstable particles was already proposed in \citere{Ae94}.
Recently different versions of DPAs have been used in the literature.

A first complete calculation of the $\Oa$ corrections for off-shell
\PW-pair production, including a numerical study of leptonic final
states, was presented in \citere{Be98}. In that work the DPA is
applied both to the virtual and real corrections using a
semi-analytical approach. Moreover, also the phase space is treated in
DPA. The proposed DPA for the
bremsstrahlung process 
is based on the distinction between hard,
semi-soft, and soft photons, without a common treatment of
photons of these domains for all observables. 
A realistic photon recombination for collinear photon emission is not
considered there. For instance, the invariant masses of the W~bosons are
reconstructed from the momenta of the corresponding decay fermions alone,
without recombining soft and collinear photons.
The resulting large shifts in the peak position of the W~line shape,
namely $-20\MeV$, $-39\MeV$, and $-77\MeV$ for $\tau^+\nu_\tau$,
$\mu^+\nu_\mu$, and $\Pep\nu_\Pe$ final states at a centre-of-mass
(CM) energy of $184\GeV$, respectively, result from mass-singular
logarithms of the form $\alpha\ln(m_l/\MW)$ and are due to the absence of
collinear photons in the definition of the invariant masses.

$\Oa$ corrections for \PW-pair production in DPA have also been
implemented in the Monte Carlo generator {\sc YFSWW3}
\cite{ja97,ja99}. In
\citere{ja97} the $\Oa$ corrections to on-shell W-pair production have
been combined with the exponentiation of photonic corrections from the
initial and intermediate (\PW\PW) states. In \citere{ja99}, the
final-state corrections have been added in the leading-logarithmic
approximation. 
The normalization of the W decays is fixed using the corrected
branching ratios.
The non-factorizable corrections are not fully included but only as an
approximation in terms of the screened Coulomb ansatz \cite{Ch99}.
The spin correlations between the two W~decays  have been neglected 
in the contributions of the non-leading corrections.
In \citere{ja99} the results of \citere{Be98} have been confirmed
qualitatively. Moreover, it has been found that the shifts of the
peaks 
in the W-invariant-mass distributions
are reduced by taking into account collinear photons in the
definition of the invariant masses, which effectively replaces the
mass-singular logarithms by logarithms of a minimum opening angle for
collinear photon emission.

Another approximate version of the DPA was presented in \citere{ku99}
and discussed for a purely hadronic final state. 
Both virtual and real corrections were treated within DPA, but
non-factorizable corrections were neglected. The main emphasis was put
on linear-collider energies, in particular on the comparison to a
high-energy approximation \cite{be93} for the one-loop correction.

In this paper we present the first complete calculation of the $\Oa$
corrections for off-shell \PW-pair production in the DPA that has been
implemented in a Monte Carlo generator.  Numerical results of this
generator, which is called {\sc RacoonWW}, have already been presented
in \citeres{de99a,lep2mcws} and \cite{de99b} for LEP2 and
linear-collider energies, respectively.  The lowest-order cross
section is calculated including the complete diagrams for any
four-fermion final state or in the CC03 approximation, \ie including
only the doubly-resonant W-pair production diagrams.  The complete
virtual corrections to \PW-pair production and \PW~decay and the
virtual non-factorizable corrections are included in the DPA. As far
as the real corrections are concerned, the matrix elements for the
minimal gauge-invariant subset of $\eeffffg$ including the
doubly-resonant contributions corresponding to W-pair production (the
CC11 subset) are included. The real corrections and the virtual
corrections are matched in such a way that soft and collinear
singularities cancel as far as they should.  The treatment of these
singularities has been implemented in two different ways, one of which
uses the subtraction method described in \citeres{di99,ro99}, the
other one uses phase-space slicing.  We use the exact four-fermion
phase space throughout.  Finally, we note that all parts of the
calculations have been performed in two independent ways.

The paper is outlined as follows: in \refse{se:strategy} we describe
the general strategy of our calculation. In \refse{se:virt} details on
the virtual factorizable and non-factorizable corrections in DPA are
given.  In \refse{se:sing} we explain the matching of soft and
collinear singularities between virtual and real corrections, both for
the subtraction method and for phase-space slicing. The inclusion of
higher-order initial-state radiation is described in \refse{se:isr},
and \refse{se:qcd} contains a discussion of QCD corrections.
Numerical results are presented in \refse{se:numres}, including
predictions for the W-pair production cross section, angular and W
invariant-mass distributions at LEP2, 
a detailed discussion of the intrinsic ambiguities of the
proposed DPA, and comparisons to results of other
authors. Our conclusions are presented in \refse{se:concl}, and
the appendices provide some useful explicit results.

\section{Strategy of the calculation}
\label{se:strategy}

We consider the process
\beqar\label{eq:ee4f}
\Pep(p_+,\si_+)+\Pem(p_-,\si_-) &\;\to\;&
\PWp(k_+,\la_+)+\PWm(k_-,\la_-)
\nn\\
&\;\to\;&
f_1(k_1,\si_1)+\bar f_2(k_2,\si_2)+f_3(k_3,\si_3)+\bar f_4(k_4,\si_4).
\eeqar
The arguments label the momenta $p_\pm$, $k_i$ and helicities
$\si_i=\pm1/2$, $\la_j=0,\pm1$ of the corresponding particles. We
often use only the signs to denote the helicities. The fermion masses
are neglected whenever possible.  As a consequence, the helicities of
the incoming electrons and positrons are related as $\si=\si_-=-\si_+$
(in the absence of collinear photon emission). For the W-pair-mediated
diagrams, and more generally for the graphs of the CC11 class, the
helicities of the outgoing fermions are fixed, $\si_{1,3} = -\si_{2,4}
= -1/2$, owing to the left-handed coupling of the \PW~bosons. In
general this does not hold for other background diagrams.

The lowest-order cross section is calculated using the complete
lowest-order matrix elements $\M^{\eeffff}_\born$ of \citere{ee4fa}.

The radiative corrections to \refeq{eq:ee4f} consist of virtual
corrections, resulting from loop diagrams, as well as of real
corrections, originating from the process
\beqar\label{eq:ee4fg}
\Pep(p_+,\si_+)+\Pem(p_-,\si_-) &\;\to\;& \PWp(k_+,\la_+)+\PWm(k_-,\la_-)
\; ({}+\ga)
\nn\\
&\;\to\;& f_1(k_1,\si_1)+\bar f_2(k_2,\si_2)+f_3(k_3,\si_3)+\bar
f_4(k_4,\si_4) +\ga(k,\la_\gamma).
\nn\\
\eeqar
Both have to be combined properly in order to ensure necessary
cancellations of soft and collinear singularities.

\subsection{Virtual corrections}
\label{se:virt1}

We treat the non-leading
virtual corrections in double-pole approximation
(DPA), \ie we take only those terms into account that are enhanced by
two resonant \PW~propagators (doubly-resonant corrections). 

\subsubsection{Virtual factorizable corrections}
\label{se:virtfac}

In the DPA, there are two types of contributions, the {\em
  factorizable} and the {\em non-factorizable} ones.  The former are
the ones that can be associated to one of the production or decay
subprocesses, the latter are the ones that connect these subprocesses.
This splitting cannot be completely done on the basis of Feynman
diagrams; otherwise the splitting would not be gauge-invariant. To
define this splitting, we start by inspecting only those graphs that
retain two resonant \PW~propagators outside the loops, \ie all
diagrams of the generic structure shown in \reffi{fig:fRCsdiag}.
\begin{figure}
\centerline{
\begin{picture}(200,105)(0,0)
\ArrowLine(30,50)( 5, 95)
\ArrowLine( 5, 5)(30, 50)
\Photon(30,50)(150,80){2}{11}
\Photon(30,50)(150,20){2}{11}
\ArrowLine(150,80)(190, 95)
\ArrowLine(190,65)(150,80)
\ArrowLine(190, 5)(150,20)
\ArrowLine(150,20)(190,35)
\GCirc(30,50){10}{.5}
\GCirc(90,65){10}{1}
\GCirc(90,35){10}{1}
\GCirc(150,80){10}{.5}
\GCirc(150,20){10}{.5}
\DashLine( 70,0)( 70,100){2}
\DashLine(110,0)(110,100){2}
\put(50,26){W}
\put(50,68){W}
\put(115,13){W}
\put(115,82){W}
\put(-12, 0){$\Pem$}
\put(-12,95){$\Pep$}
\put(195, 1){$\bar f_4$}
\put(195,34){$f_3$}
\put(195,60){$\bar f_2$}
\put(195,95){$f_1$}
\put(-25,-15){\footnotesize On-shell production}
\put(120,-15){\footnotesize On-shell decays}
\end{picture}
} 
\vspace*{1em}
\caption{Diagrammatic structure of virtual factorizable corrections to
$\eeWWffff$}
\label{fig:fRCsdiag}
\efi
Since the loop corrections in such graphs can be associated either to
the production of the \PW-boson pair or to the decay of one of the
\PW~bosons, this subset of graphs can be associated with the virtual
factorizable corrections.  The corresponding amplitude is of the form
\beq
\label{eq:Mstruc}
{\cal M} =
\frac{R(k_+^2,k_-^2,\theta)}{(k_+^2-\MW^2+\ri\MW\GW)(k_-^2-\MW^2+\ri\MW\GW)},
\eeq
where $k_\pm^2$ are the invariant masses of the virtual
\PW~bosons, $\theta$ represents all other kinematical variables,
$\MW$ is the renormalized \PW-boson mass, for which we take the
on-shell mass, and $\GW$ is the width of the \PW~boson.

The diagrams of \reffi{fig:fRCsdiag} do not form a
gauge-invariant subset. However, in DPA a 
gauge-invariant contribution can be extracted by replacing the
numerator with the gauge-invariant residue \cite{polescheme},
\beq\label{eq:defDPA}
{\cal M}_{\DPA}^{\eeWWffff} =
\frac{R(\MW^2,\MW^2,\theta)}{(k_+^2-\MW^2+\ri\MW\GW)(k_-^2-\MW^2+\ri\MW\GW)}.
\eeq
In principle, the amplitude should be expanded about the complex poles
in $k_\pm^2$.  Replacing the complex pole position by
$\MW^2-\ri\MW\GW$ with on-shell mass and width introduces an error of
$\Oaa$ (see App.~D 
of \citere{bhf2}). The neglect of the width in
the numerator introduces an error of order $\GW/\MW$. Since we apply
the DPA only to the corrections, this leads to an error of
$\O(\al\GW/\MW)$ which is the order of the uncertainty of our
calculation.  Note that additional 
infrared (IR) singularities appear in the
factorizable corrections when taking the on-shell limit in the
numerator.
 
In lowest order, the matrix element in DPA factorizes into the one for
the production of the two on-shell \PW~bosons, $\M^\eeWW_\born$, the
(transverse parts of the) propagators of these bosons, and the matrix
elements for the decays of these on-shell bosons, $\M^\Wpff_\born$ and
$\M^\Wmff_\born$:
\beq\label{eq:mbornDPA}
{\cal M}^{\eeWWffff}_{\born,\DPA} =
\sum_{\la_+,\la_-}\frac{\M^\eeWW_\born \M^\Wpff_\born \M^\Wmff_\born}{K_+ K_-},
\eeq
where we introduced the abbreviations
\beq
K_\pm = k_\pm^2-M^2, \qquad M^2 = \MW^2-\ri\MW\GW
\eeq
for the off-shellness of the W~bosons. Equation \refeq{eq:mbornDPA}
contains the coherent sum over the physical polarizations $\la_\pm$ of
the \PWpm~bosons. Note that the polarizations have to be defined in
the same way for the production and the decay matrix elements.

\begin{sloppypar}
The definition of the DPA for factorizable diagrams \refeq{eq:defDPA}
implies
that the squared momenta of the W-boson legs of the
production and decay vertex functions, which are hidden in the shaded
blobs of \reffi{fig:fRCsdiag}, have to be set to their on-shell values
$\MW^2$.  The W~self-energies, marked by open blobs, have to be
expanded about $\MW^2$,
and the corresponding residues have to be
distributed equally to the production and decay parts.  
Finally, we can express the factorizable
doubly-resonant corrections by the product of
gauge-independent on-shell matrix elements for W-pair production and
\PW~decays,
and the (transverse parts of the) \PW~propagators,
\beqar\label{eq:mvirtDPA}
\de\M^{\eeWWffff}_{\virt,\fact,\DPA} &=&
\sum_{\la_+,\la_-} \frac{1}{K_+ K_-}
\Bigl( \de\M^\eeWW \M^\Wpff_\born \M^\Wmff_\born
\nl && {}\quad
+ \M^\eeWW_\born \de\M^\Wpff \M^\Wmff_\born
\nl && {}\quad
+ \M^\eeWW_\born \M^\Wpff_\born \de\M^\Wmff \Bigr),
\eeqar
where 
$\delta\M$ 
denote one-loop contributions.
Details on the actual calculation of
$\M^{\eeWWffff}_{\virt,\fact,\DPA}$ can be found in \refse{se:fRC}.
\end{sloppypar}

\subsubsection{Virtual non-factorizable corrections}
\label{se:virtnonfac}

All loop diagrams that are not of the generic form of
\reffi{fig:fRCsdiag} belong to the 
virtual non-factorizable corrections.
Simple power counting \cite{nfc1a} reveals that only diagrams with
photon exchange in the loop can give rise to doubly-resonant (virtual)
non-factorizable corrections. The considered set of diagrams is not
gauge-invariant even in the DPA. Therefore, we have defined the
non-factorizable doubly-resonant corrections by subtracting the
factorizable doubly-resonant corrections from the complete
doubly-resonant corrections \cite{nfc1a}.  The so-defined
non-factorizable corrections do not only receive contributions from
diagrams in which the photon links the production and decay
subprocesses, i.e.\ manifestly non-factorizable diagrams [graphs like
(a), (b), (c) of \reffi{fig:nfRCsdiags}], but also from factorizable
diagrams [graphs like (d), (e), (f), (g) of \reffi{fig:nfRCsdiags}].
The latter contributions arise by subtracting the factorizable
contributions of those graphs, which contain the artificial IR
singularities mentioned above, 
from the complete graphs in DPA,
in which the IR singularities are replaced by
logarithms of the form $\ln K_\pm$.
A representative set of diagrams contributing to the non-factorizable
corrections to $\eeWWffff$ is shown in \reffi{fig:nfRCsdiags}.
\bfi
\begin{center}
\begin{picture}(360,510)(0,10)
\Text(0,495)[lb]{(a) type (\mfp)}
\put(20,390){
\begin{picture}(150,100)(0,0)
\ArrowLine(30,50)( 5, 95)
\ArrowLine( 5, 5)(30, 50)
\Photon(30,50)(90,20){2}{6}
\Photon(30,50)(90,80){-2}{6}
\Vertex(60,65){2.0}
\GCirc(30,50){10}{.5}
\Vertex(90,80){2.0}
\Vertex(90,20){2.0}
\ArrowLine(90,80)(120, 95)
\ArrowLine(120,65)(90,80)
\ArrowLine(120, 5)( 90,20)
\ArrowLine( 90,20)(105,27.5)
\ArrowLine(105,27.5)(120,35)
\Vertex(105,27.5){2.0}
\Photon(60,65)(105,27.5){-2}{5}
\put(86,50){$\gamma$}
\put(63,78){$W$}
\put(40,65){$W$}
\put(52,18){$W$}
\put(10, 5){$\mathrm{e}^-(p_-)$}
\put(10,90){$\mathrm{e}^+(p_+)$}
\put(125,90){$f_1(k_1)$}
\put(125,65){$\bar f_2(k_2)$}
\put(125,30){$f_3(k_3)$}
\put(125, 5){$\bar f_4(k_4)$}
\end{picture}
}
\Text(210,495)[lb]{(b) type (\ffp)}
\put(230,390){
\begin{picture}(120,100)(0,0)
\ArrowLine(30,50)( 5, 95)
\ArrowLine( 5, 5)(30, 50)
\Photon(30,50)(90,80){-2}{6}
\Photon(30,50)(90,20){2}{6}
\GCirc(30,50){10}{.5}
\Vertex(90,80){2.0}
\Vertex(90,20){2.0}
\ArrowLine(90,80)(120, 95)
\ArrowLine(120,65)(105,72.5)
\ArrowLine(105,72.5)(90,80)
\Vertex(105,72.5){2.0}
\ArrowLine(120, 5)( 90,20)
\ArrowLine( 90,20)(105,27.5)
\ArrowLine(105,27.5)(120,35)
\Vertex(105,27.5){2.0}
\Photon(105,27.5)(105,72.5){2}{4.5}
\put(93,47){$\gamma$}
\put(55,73){$W$}
\put(55,16){$W$}
\end{picture}
}
\Text(0,365)[lb]{(c) type (if)}
\put(20,260){
\begin{picture}(120,100)(0,0)
\ArrowLine(27,55)(15, 75)
\Vertex(15,75){2.0}
\ArrowLine(15,75)( 3, 95)
\ArrowLine( 3, 5)(30, 50)
\Photon(30,50)(90,80){-2}{6}
\Photon(30,50)(90,20){2}{6}
\GCirc(30,50){10}{.5}
\Vertex(90,80){2.0}
\Vertex(90,20){2.0}
\ArrowLine(90,80)(105,87.5)
\ArrowLine(105,87.5)(120, 95)
\ArrowLine(120,65)(90,80)
\ArrowLine(120, 5)( 90,20)
\ArrowLine( 90,20)(120,35)
\Vertex(105,87.5){2.0}
\PhotonArc(66.25,36.25)(64.25,52.9,142.9){2}{8}
\put(55,90){$\gamma$}
\put(68,55){$W$}
\put(55,16){$W$}
\end{picture}
}
\Text(210,365)[lb]{(d) type (\mmp)}
\put(230,260){
\begin{picture}(120,100)(0,0)
\ArrowLine(30,50)( 5, 95)
\ArrowLine( 5, 5)(30, 50)
\Photon(30,50)(90,80){-2}{6}
\Photon(30,50)(90,20){2}{6}
\Photon(70,30)(70,70){2}{3.5}
\Vertex(70,30){2.0}
\Vertex(70,70){2.0}
\GCirc(30,50){10}{.5}
\Vertex(90,80){2.0}
\Vertex(90,20){2.0}
\ArrowLine(90,80)(120, 95)
\ArrowLine(120,65)(90,80)
\ArrowLine(120, 5)( 90,20)
\ArrowLine( 90,20)(120,35)
\put(76,47){$\gamma$}
\put(45,68){$W$}
\put(45,22){$W$}
\put(72,83){$W$}
\put(72,11){$W$}
\end{picture}
}
\Text(0,235)[lb]{(e) type (im)}
\put(20,130){
\begin{picture}(150,100)(0,0)
\ArrowLine(27,55)(15, 75)
\Vertex(15,75){2.0}
\ArrowLine(15,75)( 3, 95)
\ArrowLine( 3, 5)(30, 50)
\Photon(30,50)(90,20){2}{6}
\Photon(30,50)(90,80){-2}{6}
\Vertex(60,65){2.0}
\GCirc(30,50){10}{.5}
\Vertex(90,80){2.0}
\Vertex(90,20){2.0}
\ArrowLine(90,80)(120, 95)
\ArrowLine(120,65)(90,80)
\ArrowLine(120, 5)( 90,20)
\ArrowLine( 90,20)(120,35)
\PhotonArc(32.5,47.5)(32.596,32.47,122.47){2}{4.5}
\put(36,92){$\gamma$}
\put(75,61){$W$}
\put(51,48){$W$}
\put(52,18){$W$}
\end{picture}
}
\Text(210,235)[lb]{(f) type (mf)}
\put(230,130){
\begin{picture}(120,100)(0,0)
\ArrowLine(30,50)( 3, 95)
\ArrowLine( 3, 5)(30, 50)
\Photon(30,50)(90,80){-2}{6}
\Photon(30,50)(90,20){2}{6}
\Vertex(70,70){2.0}
\GCirc(30,50){10}{.5}
\Vertex(90,80){2.0}
\Vertex(90,20){2.0}
\ArrowLine(90,80)(105,87.5)
\Vertex(105,87.5){2.0}
\ArrowLine(105,87.5)(120, 95)
\ArrowLine(120,65)(90,80)
\ArrowLine(120, 5)(90,20)
\ArrowLine(90,20)(120,35)
\PhotonArc(87.5,78.75)(19.566,26.565,206.565){2}{6}
\put(57,86){$\gamma$}
\put(77,62){$W$}
\put(50,48){$W$}
\put(55,16){$W$}
\end{picture}
}
\Text(0,105)[lb]{(g) type (mm)}
\put(20,0){
\begin{picture}(120,100)(0,0)
\ArrowLine(30,50)( 5, 95)
\ArrowLine( 5, 5)(30, 50)
\Photon(30,50)(90,80){-2}{6}
\Photon(30,50)(90,20){2}{6}
\Vertex(75,72.5){2.0}
\Vertex(50,60){2.0}
\GCirc(30,50){10}{.5}
\Vertex(90,80){2.0}
\Vertex(90,20){2.0}
\ArrowLine(90,80)(120, 95)
\ArrowLine(120,65)(90,80)
\ArrowLine(120, 5)( 90,20)
\ArrowLine( 90,20)(120,35)
\PhotonArc(62.5,66.25)(13.975,26.565,206.565){-2}{3.5}
\put(55,90){$\gamma$}
\put(44,45){$W$}
\put(62,54){$W$}
\put(82,64){$W$}
\put(55,16){$W$}
\end{picture}
}
\end{picture}
\end{center}
\caption{A representative set of diagrams contributing to the 
virtual non-factorizable corrections}
\label{fig:nfRCsdiags}
\efi

Note that the non-factorizable corrections do not contain a product of
two independent \PW~propagators and that these corrections involve 
logarithms of the form $\ln(K_\pm)$, which have to be kept exactly. 
Therefore, it is not possible to define a residue similar to 
$R(\MW^2,\MW^2,\theta)$ for the double resonance as done for the 
factorizable corrections. The resonance structure of the non-factorizable 
corrections has the form of a homogeneous polynomial of order two in 
$K_+$ and $K_-$ in the denominator.

The non-factorizable corrections yield a simple correction factor
$\de^{\virt}_{\nonfact,\DPA}$ to the lowest-order cross section.
Its explicit form is given in \refse{se:nfRC}.

\subsubsection{On-shell projection}
\label{se:onshellprojection}

The gauge-invariant definition of the factorizable corrections
requires the introduction of an associated phase-space point with
on-shell W~bosons for each phase-space point with general off-shell
W~bosons. In our formulas, we leave the form of the on-shell
projection open and mark all momenta of the associated on-shell point
by carets. In particular, the on-shell projections of the momenta
$k_+=k_1+k_2$ and $k_-=k_3+k_4$ of the W~bosons obey
\beq
\kon_+ = \kon_1+\kon_2, \qquad
\kon_- = \kon_3+\kon_4, \qquad
\kon_\pm^2 = \MW^2.
\eeq
The actual form of this on-shell projection is not uniquely
determined, but involves some freedom. However, this intrinsic
ambiguity leads to differences of the order of $\alpha\GW/(\pi\MW)$
for different versions of the projection, i.e.\ the ambiguity remains
below the desired level of accuracy for the DPA. The explicit
on-shell projection used in the numerics is given in
\refapp{app:onshell}.  For the on-shell kinematics we define the
Mandelstam variables $s$ and $\ton$ as usual,
\beqar
s &=& (p_+ + p_-)^2 = (\kon_+ + \kon_-)^2,
\nn\\
\ton &=& (p_+-\kon_+)^2 = (p_- -\kon_-)^2 = 
\MW^2-\frac{s}{2}(1-\beta\cos\theta),
\eeqar
where $\beta=\sqrt{1-4\MW^2/s}$ and $\theta$ are the velocity and the
scattering angle of the outgoing W~bosons, respectively.

\subsection{Real photonic corrections}
\label{se:real1}

A substantial part of the $\O(\alpha)$ corrections to four-fermion
production is due to real photon emission. Since detectors are not
able to detect photons that are soft or collinear to fermions (other
than muons), we have events with {\it visible} and {\it invisible}
photons.  The former are events in which a photon of finite energy is
separated from all other particles in the detector, the latter
correspond to events in which no separate photon is seen. Events with
visible photons can be taken into account via the processes
$\eeffffg$, for invisible photons a careful treatment of soft and
collinear singularities is required in addition.

\subsubsection{The bremsstrahlung process $\eeffffg$}

For the bremsstrahlung process $\eeffffg$ the complete lowest-order
matrix elements $\M^{\eeffffg}$ are taken into account.  More
precisely, we employ the matrix elements for the $4f+\gamma$ final
states of the CC11 class, which is the minimal gauge-invariant subset
of graphs containing all doubly-resonant diagrams corresponding to
W-pair production. We use the explicit results presented in
\citere{ee4fa}.

\subsubsection{Invisible photons}

The emission of invisible photons contributes to the $\O(\alpha)$
corrections to $\eeffff$.
They include soft and collinear singularities, which are regularized
by an infinitesimal photon mass $\la$ and small fermion masses $m_f$,
respectively.  Since the singularities of real and virtual corrections
are related to each other, these corrections have to be combined
carefully, in order to avoid a mismatch in the singularity structure.
We use two different procedures of treating soft and collinear photon
emission: one is based on a subtraction method, the other on
phase-space slicing. The precise implementation of these procedures is
discussed in \refse{se:sing}.

\subsection{The master formula}
\label{se:master}

The total cross section is composed of the following contributions
\begin{equation}\label{eq:crosssection0}
\int\rd \sigma =
\int \rd \sigma_{\Born}^{\eeffff}+
\int \rd \sigma_{\virt}^{\eeffff}
+\int \rd\sigma^{\eeffffg}.
\end{equation}
Here $\rd\sigma_{\Born}^{\eeffff}$ is the full lowest-order 
cross section to $\eeffff$,
\begin{equation}
\rd\sigma_{\Born}^{\eeffff}=
\frac{1}{2 s} \rd\Phi_{4f} |\M^{\eeffff}_\born|^2 
\end{equation}
with the corresponding matrix element $\M^{\eeffff}_\born$ and the
four-particle phase-space element $\rd \Phi_{4f}$.  Similarly,
$\rd\sigma^{\eeffffg}$, which describes the real corrections,
is the full lowest-order cross section to $\eeffffg$,
and $\rd \sigma_{\virt}^{\eeffff}$ denotes the virtual one-loop
corrections.

Both the virtual and the real corrections involve soft and collinear
singularities. These singularities are extracted by separating the
cross sections into finite and singular parts:
\beqar\label{eq:finsing}
\rd\sigma_{\virt}^{\eeffff} &=&
\rd\sigma_{\virt,\finite}^{\eeffff} + 
\rd\sigma_{\virt,\sing}^{\eeffff},\nl
\rd\sigma^{\eeffffg} &=&
\rd\sigma_{\finite}^{\eeffffg} + 
\rd\sigma_{\sing}^{\eeffffg}.
\eeqar
The singular part of the virtual corrections factorizes into the
lowest-order cross section and a simple correction factor.  Moreover,
the singularities in the real part, $\rd\sigma_{\sing}^{\eeffffg}$,
can be split off, and the five-particle phase-space element
$\rd\Phi_{4f\ga}$ can be decomposed into $\rd\Phi_{4f}$ and the
phase-space element of the photon $\rd\Phi_{\ga}$.  The
integration over $\rd\Phi_{\ga}$ 
can be partially performed, and the result can be written as a
convolution of a structure function with the lowest-order cross
section $\rd\sigma_{\Born}^{\eeffff}$.  When adding both
contributions, all soft singularities and all collinear singularities
associated with the final state cancel, and the resulting ``singular''
cross section,
\beq
\rd\sigma_{\virt+\real,\sing}^{\eeffff}
=\rd\sigma_{\virt,\sing}^{\eeffff}+\rd\sigma_{\sing}^{\eeffffg},
\eeq
contains, apart from finite terms, only collinear singularities
associated with the initial state, \ie
leading logarithms of the form $\ln(s/\Me^2)$,
at least for inclusive enough observables.

In {\sc RacoonWW}, the DPA is only applied to the finite part of 
the virtual corrections,
\begin{eqnarray}
\label{eq:virtfinite}
\rd\sigma_{\virt,\finite}^{\eeffff} &\to& 
\rd\sigma_{\virt,\finite,\DPA}^{\eeWWffff}=
\rd\sigma_{\virt,\DPA}^{\eeWWffff}
-\rd\sigma_{\virt,\sing,\DPA}^{\eeWWffff},
\end{eqnarray}
with $\rd\sigma_{\virt,\sing,\DPA}^{\eeWWffff}$ defined in \refse{se:YFS}.
The doubly-resonant virtual corrections are composed as follows,
\begin{eqnarray}
\rd\sigma_{\virt,\DPA}^{\eeWWffff}&=&
\frac{1}{2 s} \rd\Phi_{4f} \left[2\Re \Bigl((\M^{\eeWWffff}_{\born,\DPA})^*
\de\M^{\eeWWffff}_{\virt,\fact,\DPA}\Bigr) \right.
\nonumber\\
&& {}
+\left. |\M^{\eeWWffff}_{\born,\DPA}|^2 \de^{\virt}_{\nonfact,\DPA} \right],
\end{eqnarray}
where $\M^{\eeWWffff}_{\virt,\fact,\DPA}$ denotes the matrix element for
the factorizable virtual corrections (\refse{se:fRC}) and
$\de^{\virt}_{\nonfact,\DPA}$ is the factor describing the
non-factorizable virtual corrections (\refse{se:nfRC}).

Finally, we arrive at the master formula for the cross section:
\begin{equation}\label{eq:crosssection}
\int\rd \sigma =
\int \rd \sigma_{\Born}^{\eeffff}+
\int \rd \sigma_{\virt,\finite,\DPA}^{\eeWWffff}
+\int \rd\sigma_{\virt+\real,\sing}^{\eeffff}
+\int \rd\sigma_{\finite}^{\eeffffg}.
\end{equation}
Since the contribution $\rd\sigma_{\virt+\real,\sing}^{\eeffff}$ 
is not treated in DPA, the leading-logarithmic photonic corrections
resulting from initial-state radiation (ISR) are treated exactly in our
approach. 

The separation of the singularities can be done in different ways.  In
{\sc RacoonWW}, we have implemented two possibilities  to extract the
singularities from the real corrections, one is based on a
subtraction method and discussed in \refse{se:singsub}, the other is
based on the phase-space-slicing method and described in
\refse{se:singsli}. Both procedures are equivalent.
For the extraction of the singularities from the
virtual corrections, also two 
choices
have been implemented. 
One is
motivated by the subtraction method, the other is inspired by the
definition of a 
U(1)-gauge-invariant virtual photon part following
\citere{Yennie:1961ad} (\cf\refse{se:YFS}). 
Since the definition of
a finite part of the virtual corrections differs by finite
terms between both cases, 
and since the singular parts are treated
exactly but the finite parts in DPA, both approaches differ by terms
of the order of the intrinsic 
uncertainty
of the DPA. These differences are
due to finite terms that are either taken into account in DPA or
exactly.

\section{Virtual corrections in double-pole approximation}
\label{se:virt}

\subsection{Factorizable corrections}
\label{se:fRC}

We have calculated the contribution of the virtual corrections to the
transition matrix elements in DPA,
$\de\M^{\eeWWffff}_{\virt,\fact,\DPA}$, in two different ways, the
results of which are in 
perfect numerical agreement.  In a first approach
we made use of the results of \citeres{rcwprod1} and \cite{rcwdecay1}
for the pair production and decay of on-shell W~bosons, respectively.
The virtual corrections in DPA are
given by \refeq{eq:mvirtDPA} in this approach.
When using the existing on-shell results,
particular care is needed concerning the polarization of the
intermediate W~bosons. In \citeres{rcwprod1,rcwdecay1} the polarization 
vectors for the W~bosons 
were introduced
in a way convenient for each process.
In order to keep the spin correlation
between production and decay subprocesses, it is, however, necessary to
take one and the same polarization vectors for the production and the
decay of a W~boson and to perform the summation over the W~polarizations
coherently, i.e.\ at the level of the unsquared matrix element.
Hence, the results for on-shell W~bosons cannot be taken as black boxes,
but have to be carefully combined.

In a second approach we have performed a completely new and independent
calculation of all relevant one-loop diagrams for the pair production
and the decay of on-shell W~bosons. 
The Feynman graphs have been generated with {\sl FeynArts} \cite{FA}.
The actual calculation has been done twice
using our own {\sl Mathematica} \cite{math} routines,
once in the `t~Hooft--Feynman gauge using the Feynman rules and
the renormalization scheme of \citere{de93} and once in the
background-field formalism using the results of \citere{de95}. The
evaluation of the one-loop diagrams follows the methods described in
\citere{de93} in both calculations, and the two results are in perfect
numerical agreement. In the following we briefly sketch the strategy of
these calculations which exploits the factorization property of the
amplitudes.

\subsubsection{Structure of the matrix elements}

The DPA matrix elements are decomposed
into a set of so-called standard matrix elements (SMEs) $\M^\sigma_n$, 
which contain the spin structure of the external particles, and invariant
coefficients $F^\sigma_n$. Since the 
relevant
diagrams factorize into
W-pair production and W~decay, the SMEs for W-pair production
can be chosen as ($\si=\pm$)
\beqar
\M^\sigma_1 &=& \bar v(p_+)\dsl\veps^*_+(\dsl{\kon}_+ -\dsl p_+)
\dsl\veps^*_- \omega_\sigma u(p_-),
\nn\\
\M^\sigma_2 &=& \bar v(p_+)\text\frac{1}{2}(\dsl{\kon}_+ -\dsl{\kon}_-)
\omega_\sigma u(p_-) (\veps^*_+\veps^*_-),
\nn\\
\M^\sigma_3 &=& \bar v(p_+)\dsl\veps^*_+\omega_\sigma u(p_-)
(\veps^*_- \kon_+)
-\bar v(p_+)\dsl\veps^*_-\omega_\sigma u(p_-)
(\veps^*_+ \kon_-),
\nn\\
\M^\sigma_4 &=& \bar v(p_+)\dsl\veps^*_+\omega_\sigma u(p_-)
(\veps^*_- p_-)
-\bar v(p_+)\dsl\veps^*_-\omega_\sigma u(p_-)
(\veps^*_+ p_+),
\nn\\
\M^\sigma_5 &=& \bar v(p_+)\text\frac{1}{2}(\dsl{\kon}_+ -\dsl{\kon}_-)
\omega_\sigma u(p_-) (\veps^*_+ \kon_-) (\veps^*_- \kon_+),
\nn\\
\M^\sigma_6 &=& \bar v(p_+)\text\frac{1}{2}(\dsl{\kon}_+ -\dsl{\kon}_-)
\omega_\sigma u(p_-) (\veps^*_+ p_+) (\veps^*_- p_-),
\nn\\
\M^\sigma_7 &=& \bar v(p_+)\text\frac{1}{2}(\dsl{\kon}_+ -\dsl{\kon}_-)
\omega_\sigma u(p_-) 
\left[ (\veps^*_+ \kon_-) (\veps^*_- p_-) + 
(\veps^*_+ p_+) (\veps^*_- \kon_+) \right]
\label{eq:smes}
\eeqar
with ``effective W-polarization vectors''
\beqar
\veps^{*,\mu}_+ &=& \frac{e}{\sqrt{2}\sw} \,
\frac{1}{K_+} \,
\bar u(\kon_1)\gamma^\mu\omega_- v(\kon_2),
\nn\\
\veps^{*,\mu}_- &=& \frac{e}{\sqrt{2}\sw} \,
\frac{1}{K_-} \,
\bar u(\kon_3)\gamma^\mu\omega_- v(\kon_4).
\label{eq:effpols}
\eeqar
The 14
SMEs given in \refeq{eq:smes} are exactly those defined in 
\citere{de93},
where 
$\bar v(p_+)$ and $u(p_-)$ are the Dirac spinors of the massless
$\Pep\Pem$ initial state. The chirality projectors are given by
$\omega_\pm = \frac{1}{2}(1\pm\gamma_5)$. The effective W-polarization
vectors of \refeq{eq:effpols} are a formal shorthand for the
W~propagators and the tree-level decay matrix elements. Of course, the
off-shellness of the W~bosons has to be kept in these W~propagators, where
$k_+=k_1+k_2$ and $k_-=k_3+k_4$ are inserted, but the W~decays are 
calculated with the on-shell kinematics, i.e.\ the Dirac spinors 
$\bar u(\kon_i)$ and $v(\kon_i)$ depend on the on-shell-projected
momenta. Owing to the consistent neglect of fermion masses, the
polarization vectors are transverse,
\beq
\kon_\pm \veps^*_\pm = 0.
\eeq
This transversality and $\kon_\pm^2=\MW^2$ ensure that the algebraic
reduction of the $\O(\alpha)$ corrections to the W-pair production 
subprocess to SMEs and invariant coefficients can be performed in the
same way as for pure on-shell pair production, i.e.\ without
knowing any details of the W~decays. Explicitly, the decomposition of DPA
matrix elements reads
\beq
\M^{\eeWWffff,\si}_{\virt,\fact,\DPA}(p_+,p_-,\kon_+,\kon_-,k_+^2,k_-^2)=
\sum_{n=1}^{7} F^\sigma_n(s,\ton) 
\M^\sigma_n(p_+,p_-,\kon_+,\kon_-,k_+^2,k_-^2).
\eeq
It is important to realize that the invariant coefficients
$F^\sigma_n(s,\ton)$ depend only on the scalar products of the 
momenta of the on-shell-projected
W~bosons, but not on the momenta of their decay products. 
The explicit expressions of the SMEs are listed in \refapp{app:smes}.

A few remarks on the choice of SMEs are appropriate. The choice
\refeq{eq:smes} with \refeq{eq:effpols} is obtained by taking all
products of SMEs for the subprocesses $\Pep\Pem\to\PWp\PWm$, 
$\PWp\to f_1\bar f_2$, and $\PWm\to f_3\bar f_4$ for massless fermions
(cf.\ (5.11) and (5.14) of \citere{de93}), and by subsequently replacing 
$\veps^*_{\pm,\mu}\veps_{\pm,\nu}$ by the W~propagators. 
Owing to the purely left-handed coupling of the \PW~boson and the use
of massless fermions, there is only one SME for each decay subprocess.
Moreover, we
have included the lowest-order coupling factor for the decays so that
the invariant functions $F^\sigma_n$ for the lowest-order amplitudes and
the loop corrections for the production process are normalized in the
same way as for on-shell W-pair production. In particular, the 
non-vanishing coefficients $F^\sigma_{n,\Born}(s,\ton)$ of the lowest-order 
diagrams (see \reffi{fig:sigdiags}) in DPA are given by
\bfi
\centerline{
\begin{picture}(135,80)(0,0)
\ArrowLine(30,40)(10,70)
\ArrowLine(10,10)(30,40)
\Vertex(30,40){1.2}
\Photon(30,40)(70,40){2}{4}
\Vertex(70,40){1.2}
\Photon(70,40)(100,60){2}{4}
\Photon(70,40)(100,20){2}{4}
\Vertex(100,60){1.2}
\Vertex(100,20){1.2}
\ArrowLine(100,60)(120,70)
\ArrowLine(120,50)(100,60)
\ArrowLine(100,20)(120,30)
\ArrowLine(120,10)(100,20)
\put(-5,58){$\Pep$}
\put(-5,12){$\Pem$}
\put(40,27){$\gamma,Z$}
\put(75,55){$W$}
\put(75,15){$W$}
\put(125,70){$f_1$}
\put(125,45){$\bar f_2$}
\put(125,30){$f_3$}
\put(125, 5){$\bar f_4$}
\end{picture}
\hspace{3em}
\begin{picture}(105,80)(0,0)
\ArrowLine(40,60)(10,70)
\ArrowLine(40,20)(40,60)
\ArrowLine(10,10)(40,20)
\Vertex(40,60){1.2}
\Vertex(40,20){1.2}
\Photon(40,60)(70,60){2}{3.5}
\Photon(40,20)(70,20){2}{3.5}
\Vertex(70,60){1.2}
\Vertex(70,20){1.2}
\ArrowLine( 70,60)( 90,70)
\ArrowLine( 90,50)( 70,60)
\ArrowLine( 70,20)( 90,30)
\ArrowLine( 90,10)( 70,20)
\put(-5,58){$\Pep$}
\put(-5,12){$\Pem$}
\put(45,35){$\nu_{\Pe}$}
\put(50,68){$W$}
\put(50, 5){$W$}
\put( 95,70){$f_1$}
\put( 95,45){$\bar f_2$}
\put( 95,30){$f_3$}
\put( 95, 5){$\bar f_4$}
\end{picture} }
\caption{Doubly-resonant lowest-order diagrams for 
$\eeWWffff$}
\label{fig:sigdiags}
\efi
\beqar\label{eq:Fborn}
F^-_{1,\Born}(s,\ton) &=& \frac{e^2}{2\sw^2 \ton}, 
\nn\\
F^\sigma_{3,\Born}(s,\ton) &=& 
-F^\sigma_{2,\Born}(s,\ton) =
\frac{2e^2}{s}
-\frac{2e^2}{s-\MZ^2}\biggl(1-\frac{\delta_{\sigma-}}{2\sw^2}\biggr).
\eeqar
The chosen set of 14 SMEs is not minimal; more precisely, only 12 SMEs are
independent. The reduction to independent SMEs can be performed by
making use of two relations between several SMEs that follow
from the four-dimensionality of space time (cf.\ (11.8) of
\citere{de93}).

The contributions of the virtual corrections to the \PW-pair
production subprocess to the invariant coefficients $F^\si_n$ can be
directly read of from the results of \citere{rcwprod1}. The
contributions of the $\PW$ decays are obtained by multiplying the
corrections to the decay matrix elements, which can be extracted for
instance from \citere{rcwdecay1}, with the lowest-order
coefficients $F^\sigma_{n,\Born}(s,\ton)$ of \refeq{eq:Fborn}.
Note that care has to be taken that the relevant phases are taken into
account properly.

\subsubsection{Evaluation of coefficient functions}

Finally, we add some remarks on the actual numerical evaluation of the
invariant coefficients $F^\sigma_n(s,\ton)$. The expressions for the
one-loop contributions to these coefficients are rather involved (see
\citeres{rcwprod1,rcwprod2}) so that the corresponding computer codes
are lengthy and slow. Moreover, the employed
Passarino--Veltman reduction \cite{pa79} of tensor integrals to scalar
integrals breaks down at the boundary of phase space ($\cos\theta=\pm
1$), necessarily leading to numerical instabilities in the very
forward and backward directions of the W-production angle $\theta$. In
order to solve these problems of CPU time and numerical instabilities,
we have made use of the fact that all invariant coefficients
$F^\sigma_n(s,\ton)$ 
depend only on $\theta$ for a fixed
scattering energy and fixed input parameters, such as masses and
couplings. Before starting the multi-dimensional Monte Carlo
integration over the four-particle phase space, we calculate a set of
generalized Fourier coefficients
\beq
c^\sigma_{n,l}(s) = \frac{2l+1}{2}\int_{-1}^{+1}\rd\cos\theta\;
\ton \, F^\sigma_n(s,\ton) P_l(\cos\theta)
\label{eq:cint}
\eeq
by performing the integrals with the Legendre polynomials 
\beq
P_l(x) = \frac{1}{2^l l!}\frac{\rd^l}{\rd x^l}\left[(x^2-1)^l\right],
\qquad l=0,1,\ldots,
\eeq
numerically. Using, for instance, Gaussian integration for these
integrations yields sufficiently precise results without the need to
enter the region of numerical instabilities in the coefficients
$F^\sigma_n(s,\ton)$. During the Monte Carlo integration, the coefficients
are numerically reconstructed by making use of the generalized Fourier series
\beq
F^\sigma_n(s,\ton) = 
\sum_{l=0}^\infty \, \frac{1}{\ton} \, c^\sigma_{n,l}(s) P_l(\cos\theta),
\label{eq:Fsuml}
\eeq
which involves only trivial algebra. Thus, its evaluation is extremely
fast.  In practice, relatively few generalized Fourier coefficients
$c^\sigma_{n,l}(s)$ are needed; for example, taking $l=0,\ldots,20$
reproduces the full calculation of $F^\sigma_n(s,\ton)$ within roughly
six digits for moderate scattering angles and LEP2 energies.  Note
also that the generalized Fourier series remains stable in the forward
and backward directions where the original evaluation of
$F^\sigma_n(s,\ton)$ breaks down.  The explicit factor $\ton$ in
\refeq{eq:cint} and \refeq{eq:Fsuml} was included in order to account
for a $t$-channel pole in some of the $F^\sigma_n(s,\ton)$; without
this factor the expansion is less efficient and requires much more
terms in the sum over $l$ in \refeq{eq:Fsuml}.

\subsection{Non-factorizable corrections}
\label{se:nfRC}

In \citeres{me96,nfc2,nfc1b,nfc1a} the non-factorizable corrections
have been discussed in detail. In these references, however, the
virtual non-factorizable corrections have been combined with their
counterparts involving real photon emission, in order to obtain an
IR-safe correction that can be discussed separately. In our
present approach we do not separate the non-factorizable parts from
the full real corrections so that we need the virtual non-factorizable
corrections separately.
Here also those graphs must be taken into account that are cancelled
by their real counterparts [graphs (c), (e), (f), and (g) of
\reffi{fig:nfRCsdiags}] in the approach of
\citeres{me96,nfc2,nfc1b,nfc1a}.  The complete expression for the
virtual non-factorizable corrections is given below%
\footnote{The complete expression for the virtual non-factorizable
corrections can also be found in \citere{ro99}.}%
, closely following the notation and conventions of \citere{nfc1a}.

The virtual non-factorizable corrections can be written in terms of a
correction factor $\delta_{\nonfact,\DPA}^\virt$ to the lowest-order
cross section. Analogously to \citere{nfc1a},
$\delta_{\nonfact,\DPA}^\virt$ is decomposed into contributions that
are associated with a pair of final-state fermions,
\beq
\delta_{\nonfact,\DPA}^\virt = \sum_{a=1,2} \, \sum_{b=3,4} \, (-1)^{a+b+1} \, Q_a Q_b \,
\frac{\alpha}{\pi} \, \Re\{\Delta^\virt(p_+,p_-;k_+,k_a;k_-,k_b)\},
\label{eq:nffac}
\eeq
and we only give $\Delta^\virt=\Delta^\virt(p_+,p_-;k_+,k_2;k_-,k_3)$,
since the other terms follow by obvious substitutions. The function
$\Delta^\virt$ receives contributions from diagrams that have been
classified into seven different types in \reffi{fig:nfRCsdiags},
\beq
\Delta^\virt = \Delta^\virt_{\mfp} 
+ \Delta^\virt_{\ffp}        + \Delta^\virt_{\mathrm{if}} 
+ \Delta^\virt_{\mmp}        + \Delta^\virt_{\mathrm{im}}
+ \Delta^\virt_{\mathrm{mf}} + \Delta^\virt_{\mathrm{mm}}.
\label{eq:Delta}
\eeq
We note that in contrast to the sum of the virtual and real
contributions given in \citeres{me96,nfc2,nfc1b,nfc1a} the pure
virtual non-factorizable corrections depend both on the final and
initial state of the reaction. The contributions of the different
types of diagrams are given by
\beqar\label{eq:Delta1}
\Delta^\virt_{\mfp} &\sim& {}
- (s_{23}+s_{24})K_+ D_0(-k_-,k_+,k_2,0,M,M,m_2)
\nn\\ && {}
- (s_{13}+s_{23})K_- D_0(-k_3,-k_-,k_+,0,m_3,M,M),
\\[.5em]
\Delta^\virt_{\ffp} &\sim& {}
-s_{23}K_+K_- E_0(-k_3,-k_-,k_+,k_2,\lambda,m_3,M,M,m_2),
\\[.5em]
\Delta^\virt_{\mathrm{if}} &\sim& {}
-t_{+2}K_+ D_0(p_+,k_+,k_2,\lambda,\Me,M,m_2)
+u_{-2}K_+ D_0(p_-,k_+,k_2,\lambda,\Me,M,m_2)
\nn\\ && {}
-t_{-3}K_- D_0(p_-,k_-,k_3,\lambda,\Me,M,m_3)
+u_{+3}K_- D_0(p_+,k_-,k_3,\lambda,\Me,M,m_3),
\nn\\ && {}
\\[.5em]
\Delta^\virt_{\mmp} &\sim& {} (2\MW^2-s)\biggl\{ C_0(k_+,-k_-,0,M,M)
- \Bigl[C_0(k_+,-k_-,\la,\MW,\MW)\Bigr]_{k_\pm^2=\MW^2} \biggr\},
\hspace{2em}
\\[.5em]
\Delta^\virt_{\mathrm{im}} &\sim& {}
-(t-\MW^2)\biggl\{ \phantom{{}+{}} C_0(p_+,k_+,0,\Me,M)
-\Bigl[C_0(p_+,k_+,\lambda,\Me,\MW)\Bigr]_{k_+^2=\MW^2} 
\nn\\ && \phantom{ {}-(t-\MW^2)\biggl\{ }
+ C_0(p_-,k_-,0,\Me,M)
-\Bigl[C_0(p_-,k_-,\lambda,\Me,\MW)\Bigr]_{k_-^2=\MW^2} \biggr\}
\nn\\ && {}
+(u-\MW^2)\biggl\{ \phantom{{}+{}} C_0(p_+,k_-,0,\Me,M)
-\Bigl[C_0(p_+,k_-,\lambda,\Me,\MW)\Bigr]_{k_-^2=\MW^2} 
\nn\\ && \phantom{ {}-(u-\MW^2)\biggl\{ }
+ C_0(p_-,k_+,0,\Me,M)
-\Bigl[C_0(p_-,k_+,\lambda,\Me,\MW)\Bigr]_{k_+^2=\MW^2} \biggr\},
\\[.5em]
\Delta^\virt_{\mathrm{mf}} &\sim& {}
\MW^2\biggl\{ \phantom{{}+{}} C_0(k_+,k_2,0,M,m_2)
-\Bigl[C_0(k_+,k_2,\lambda,\MW,m_2)\Bigr]_{k_+^2=\MW^2} 
\nn\\ && \phantom{ {}\MW^2\biggl\{ }
+C_0(k_-,k_3,0,M,m_3)
-\Bigl[C_0(k_-,k_3,\lambda,\MW,m_3)\Bigr]_{k_-^2=\MW^2} \biggr\},
\\[.5em]
\Delta^\virt_{\mathrm{mm}} &\sim& {}
2\MW^2\Biggl\{ \frac{B_0(k_+^2,0,M)-B_0(M^2,0,M)}{k_+^2-M^2}
+ \frac{B_0(k_-^2,0,M)-B_0(M^2,0,M)}{k_-^2-M^2}
\nn\\ && \phantom{ {}2\MW^2\Biggl\{ }
-2B'_0(\MW^2,\lambda,\MW) \Biggr\},
\label{eq:Delta2}
\eeqar
with the kinematical invariants
\beq
\begin{array}[b]{rlrlrl}
s_{ij} &= (k_i + k_j)^2, \qquad &&& i,j &=1,2,3,4,
\nn\\
t_{+i} &= (p_+ - k_i)^2, \qquad &
u_{-i} &= (p_- - k_i)^2, \qquad & i &=1,2,
\nn\\
t_{-i} &= (p_- - k_i)^2, \qquad &
u_{+i} &= (p_+ - k_i)^2, \qquad & i &=3,4,
\nn\\
t &= (p_+ - k_+)^2, \qquad &
u &= (p_+ - k_-)^2. &&
\earr
\eeq
The sign ``$\sim$'' in \refeq{eq:Delta1}--\refeq{eq:Delta2} 
indicates that the limits $k_\pm^2\to\MW^2$ and 
$\GW\to 0$ are implicitly understood whenever 
possible. The definition of the scalar integrals 
$B_0$, $C_0$, $D_0$, $E_0$ and of their arguments can be found in
\citeres{de93,nfc1a}. The explicit expressions of these functions have
been given in \citere{nfc1a} for the $\mfp$, $\ffp$, and $\mmp$ parts;
the ones of the remaining scalar integrals are listed in
\refapp{app:scalints}.

In order to facilitate the evaluation of $\Delta^\virt$ as much as
possible, we insert the explicit expressions for the scalar integrals
into the different contributions.
For $\Delta^\virt_{\mmp}$ we can simply take over the result of 
\citere{nfc1a}; specifically, the combination of $C_0$
functions in $\Delta^\virt_{\mmp}$ is given there in (C.1) and (C.2).
For $\Delta^\virt_{\mathrm{mm}}$ we obtain
\beq
\Delta^\virt_{\mathrm{mm}} \sim
2\ln\biggl(\frac{\lambda\MW}{-K_+}\biggr)
+2\ln\biggl(\frac{\lambda\MW}{-K_-}\biggr)+4
\eeq
using \refeq{eq:B0} of the appendix below. 
For the remaining parts it is convenient to add up
all contributions, resulting in
\beqar
\lefteqn{ \Delta^\virt_{\mfp} + \Delta^\virt_{\ffp} 
+ \Delta^\virt_{\mathrm{if}} + \Delta^\virt_{\mathrm{im}}
+ \Delta^\virt_{\mathrm{mf}} } \hspace*{2em}
\nn\\
&\sim& 
-\frac{K_+K_-s_{23}\det(Y_0)}{\det(Y)}D_0(-k_4,k_+ +k_3,k_2+k_3,0,M,M,0)
\nn\\ && {}
- \frac{K_+ \det(Y_3)}{\det( Y)} F_3
- \frac{K_- \det(Y_2)}{\det( Y)} F_2
+\ln\biggl(\frac{\lambda^2}{\MW^2}\biggr)
\ln\biggl(-\frac{s_{23}}{\MW^2}-\ri\epsilon\biggr)
\nn\\ && {}
+2\ln\biggl(\frac{-K_+}{\lambda\MW}\biggr)
\ln\biggl[\frac{u_{-2}(\MW^2-t)}{t_{+2}(\MW^2-u)}\biggr]
+2\ln\biggl(\frac{-K_-}{\lambda\MW}\biggr)
\ln\biggl[\frac{u_{+3}(\MW^2-t)}{t_{-3}(\MW^2-u)}\biggr]
\nn\\ && {}
+\Li\biggl(1+\frac{\MW^2-t}{t_{+2}}\biggr)
+\Li\biggl(1+\frac{\MW^2-t}{t_{-3}}\biggr)
\nn\\ && {}
-\Li\biggl(1+\frac{\MW^2-u}{u_{-2}}\biggr)
-\Li\biggl(1+\frac{\MW^2-u}{u_{+3}}\biggr),
\eeqar
where the determinants $\det(Y)$, $\det(Y_i)$ and the functions
$D_0(\ldots)$, $F_i$ are given in \citere{nfc1a} [see (3.36), (4.10),
and (C.3) there]. Finally, we note that in
$\delta^\virt_{\nonfact,\DPA}$ all 
fermion-mass
singularities of the  initial- and final-state fermions drop out,
although some non-factorizable graphs contain such 
fermion-mass singularities.

\section{Treatment of soft and collinear photon emission}
\label{se:sing}

In the following we describe two procedures of treating soft and
collinear photon emission: one is based on a subtraction method, the 
other on phase-space slicing. In both cases soft and collinear
singularities are regularized by an infinitesimal
photon mass and small fermion masses, respectively.

For convenience, we introduce a second, generic notation 
for the momenta and helicities of the external particles:
\beqar
\Pep(q_1, \kappa_1)+\Pem(q_2, \kappa_2 ) &\;\to\;& 
f_1(q_3, \kappa_3)+\bar f_2(q_4, \kappa_4)+f_3(q_5, \kappa_5)
+\bar f_4(q_6, \kappa_6) \, ({}+\ga(k, \la_\ga))
\nn\\
\eeqar
in addition to \refeq{eq:ee4f} and \refeq{eq:ee4fg}.
Moreover, the masses of the external fermions are denoted by $m_i$
($q_i^2=m_i^2\to0$), and we use the invariants 
$s_{ij}=2q_i q_j$ in the following.

\subsection{The dipole subtraction approach}
\label{se:singsub}

The idea of so-called subtraction methods 
is to subtract a simple auxiliary function from the singular integrand
of the 
bremsstrahlung integral
and to add this contribution back again after partial analytic
integration.
This auxiliary function has to be chosen in such a way that it cancels all
singularities of the original integrand so that the phase-space
integration of the difference can be performed numerically, even over the
singular regions of the original integrand, which is
$|\M^{\eeffffg}|^2$ in our case. In this difference $\M^{\eeffffg}$ can be
evaluated without regulators for  soft or collinear singularities, i.e.\ 
we can make use of the results of \citere{ee4fa} for
$\M^{\eeffffg}$ with massless fermions.
The auxiliary function has to be simple enough so that it can
be integrated over the singular regions analytically, when the
subtracted contribution is added again. This part contains the singular
contributions and requires regulators, i.e.\ photon and fermion masses
have to be reintroduced there.
In {\sc RacoonWW} we have applied the {\it dipole subtraction formalism},
which is a process-independent approach that was first proposed 
\cite{ca96} within QCD for massless unpolarized partons and
subsequently generalized to photon radiation of massive polarized
fermions \cite{di99}. 
We only need the limit of small fermion
masses \cite{di99,ro99} in which the application of the method is 
relatively simple.
In order to keep the description of the method transparent, we describe 
only the basic structure of the individual terms in \refeq{eq:crosssection} 
explicitly and refer to \citeres{di99,ro99}%
\footnote{The subtraction functions of \citere{di99} and \citere{ro99} 
are slightly different. As default, the subtraction function
of \citere{di99} are implemented in {\sc RacoonWW}.}
for the details.

In the dipole subtraction formalism the subtraction function is
constructed from contributions that are labelled by ordered pairs $ij$ of
charged fermions, so-called ``dipoles''. The fermions $i$ and $j$ are called
{\it emitter} and {\it spectator}, respectively, since 
by construction
only the kinematics of the emitter $i$ leads to collinear singularities.
The auxiliary 
functions $|\M_{\sub,ij}|^2$ that are subtracted from the
original integrand $|\M^{\eeffffg}|^2$ are given by
\beqar
|\M_{\sub,ij}(\Phi_{4f\gamma})|^2 &=& - (-1)^{i+j} Q_i Q_j e^2 
\sum_{\tau=\pm} g_{ij,\tau}^{(\sub)}(q_i,q_j,k) \, 
|\M^{\eeffff}_{\born}(\Phifo_{4f,ij},\tau \kappa_i)|^2,
\eeqar
where $\tau \kappa_i$ denotes the helicity of the emitter%
\footnote{We suppress the polarization arguments as far as possible.}.
Here and in the following the indices $i,j$ run over $1,\ldots,6$ if not
stated otherwise.
Note that $|\M_{\sub,ij}|^2$ is a function on the entire $4f\gamma$ phase
space $\Phi_{4f\gamma}$, while the lowest-order matrix element 
$\M^{\eeffff}_{\born}$ on the r.h.s.\ requires momenta of the 
non-radiative $4f$ phase space $\Phi_{4f}$, i.e.\ $\Phi_{4f}$ has to be
embedded in $\Phi_{4f\gamma}$ by an appropriate mapping. This mapping is
chosen differently for different $ij$ pairs, as indicated 
by $\Phifo_{4f,ij}$ in the argument of $\M^{\eeffff}_{\born}$.
The mappings have to ensure that the kinematics in $\Phi_{4f\gamma}$ and
$\Phifo_{4f,ij}$ asymptotically approach each other in the {\it soft limit},
\begin{eqnarray}
\label{eq:subsoft}
q_i & \asymp{k\to 0} & \qfo_i, \qquad 
q_j  \asymp{k\to 0}  \qfo_j, 
\end{eqnarray}
and in the
{\it collinear limits},
\begin{eqnarray}
\label{eq:subcoll}
q_i - k \asymp{kq_i\to 0} & \qfo_i, & \qquad q_j \asymp{kq_i\to 0} \qfo_j, 
\qquad i=1,2, 
\\
q_i + k \asymp{kq_i\to 0} & \qfo_i, & \qquad q_j \asymp{kq_i\to 0} \qfo_j,
\qquad i=3,\ldots,6, 
\end{eqnarray}
as required by the 
factorization theorems for mass singularities.
The momenta $\qfo_i,\qfo_j$ 
have to respect momentum conservation and 
mass-shell conditions everywhere. 
The process-independent radiator functions $g^{(\sub)}_{ij,\pm}$ 
behave in the soft limit as
\beqar
\label{eq:softlimit}
g^{(\sub)}_{ij,+}(q_i,q_j,k) & \asymp{k\to 0} &
\frac{2(q_i q_j)}{(q_i k)(q_i k+q_j k)}
-\frac{m_i^2}{(q_i k)^2}, \qquad
g^{(\sub)}_{ij,-}(q_i,q_j,k) \asymp{k\to 0} \O(1)
\eeqar
and in the collinear limits as
\beqar
g^{(\sub)}_{ij,+}(q_i,q_j,k) & \asymp{q_i k\to 0} & 
\frac{1}{q_i k} 
\left[\frac{1}{x_i} P_{ff}(x_i)
-\frac{1+x_i^2}{x_i}\frac{m_i^2}{2 q_i k}\right], 
\nn\\
g^{(\sub)}_{ij,-}(q_i,q_j,k) & \asymp{q_i k\to 0} & 
\frac{(1-x_i)^2}{x_i} \frac{m_i^2}{2(q_i k)^2}, 
\qquad x_i = 1-\frac{k^0}{q_i^0}, \qquad i=1,2,
\\[.5em]
g^{(\sub)}_{ij,+}(q_i,q_j,k) & \asymp{q_i k\to 0} & 
\frac{1}{q_i k} 
\left[P_{ff}(z_i)-\frac{1+z_i^2}{z_i}\frac{m_i^2}{2 q_i k}\right], 
\nn\\ 
g^{(\sub)}_{ij,-}(q_i,q_j,k) & \asymp{q_i k\to 0} & 
\frac{(1-z_i)^2}{z_i}\frac{m_i^2}{2(q_i k)^2}, 
\qquad z_i = \frac{q_i^0}{q_i^0+k^0}, \qquad i=3,\ldots ,6,
\eeqar
where $P_{ff}(y)$ is the usual splitting function,
\beq\label{eq:splitting_function}
P_{ff}(y) = \frac{1+y^2}{1-y}.
\eeq 
Taking into account charge conservation, it is
easy to show that the sum of the subtraction functions $|\M_{\sub,ij}|^2$ 
and $\sum_{\lambda_\gamma}|\M^{\eeffffg}|^2$ 
are asymptotically equal in the singular limits,
\beq
\sum_{\lambda_\gamma} 
|\M^{\eeffffg}(\Phi_{4f\gamma})|^2
 \;\sim\; 
\sum_{i,j=1 \atop i \ne j}^6 
|\M_{\sub,ij}(\Phi_{4f\gamma})|^2
\qquad \mbox{for} \quad k\to 0 
\quad \mbox{or} \quad q_i k\to 0,
\eeq
so that the integral of their difference becomes integrable in those
regions and the photon and fermion masses can be neglected.

Now we are able to define the finite part of the real corrections,
\begin{eqnarray}
\label{eq:subreal}
\int \rd \sigma_{\finite}^{\eeffffg}&=&
\frac{1}{2s} \int \rd \Phi_{4f\gamma} \Bigg[
\sum_{\lambda_\gamma}|\M^{\eeffffg}|^2 \Theta(\Phi_{4f\gamma})
-\sum_{i,j=1 \atop i \ne j}^6
|\M_{\sub,ij}|^2  \Theta(\Phifo_{4f,ij})
\Bigg],
\nn\\
\end{eqnarray}
where the photon recombination procedure and separation cuts 
are included in the observable defined by $\Theta$.
We define $\Theta(\Phi_{4f(\gamma)})=1$ if the event passes the separation 
cuts after 
eventual photon recombination and $\Theta(\Phi_{4f(\gamma)})=0$ 
otherwise. 
In contrast to the first term, the observable in the second term depends 
on the $4f$ phase space, $\Phifo_{4f,ij}$, and is independent of the 
photon recombination procedure. The reason for the different arguments 
is that the final-state mass singularities included in the subtraction 
functions have to match the 
corresponding singularities of the virtual corrections 
exactly. On the other hand, 
the different arguments of the terms
require that all photons have to be 
combined with the nearest charged fermion in the collinear limits 
in order to obtain $\Theta(\Phi_{4f\gamma})=\Theta(\Phifo_{4f,ij})$
in the soft and collinear limits 
[see \refeq{eq:subsoft} and \refeq{eq:subcoll}].

The calculation of distributions is
similar to the application of cuts, since a histogram of a distribution
is nothing but a series of cuts. Hence, the histogram routine that
generates the desired distribution during the Monte Carlo integration
has to handle each column of the histogram in the same way as a cutted
contribution to the integrated cross section. Note that 
in this procedure
the original differential cross section and the subtraction
functions 
may contribute to different columns of the histogram for one
and the same event. The final result for each column is nevertheless
finite, because such events are in general far away from the singular
regions%
\footnote{At the edges of the histogram columns this can also occur for
``singular events''. The finiteness of such contributions is guaranteed
by the suppression of phase space for those events.}.

In addition to the finite part of the real corrections, which is given
in \refeq{eq:subreal}, we have to determine the singular part of the real 
corrections, $\rd\sigma_{\sing}^{\eeffffg}$. Therefore, we have to 
evaluate the phase-space integral of all contributions $|\M_{\sub,ij}|^2$ 
at least over the singular regions for finite mass regulators $m_i$ and 
$\lambda$. To this end, 
we split the five-particle phase space into 
the four-particle phase space and the remaining photonic parts:
\beqar
\int \rd \Phi_{4f\ga}&=&\int_0^1 \rd x 
\int \rd \Phifo_{4f,ij}(x) \int \rd \Phi_{\ga,ij}.
\eeqar

If $i$ and/or $j$ are initial-state particles, the  momentum of one of
these is reduced  by a factor $x$, 
\ie $\qfo_1=x q_1$ or $\qfo_2=x q_2$. This is indicated by the
argument $x$ in $\Phifo_{4f,ij}(x)$.
(For final-state emitter and spectator  $\rd \Phi_{\ga,ij}$           
includes the function $\delta (1-x)$ which fixes $x=1$.)
Then, the singular part of the real corrections can be written as
\beqar
\label{eq:sub}
\lefteqn{
\int \rd \si_{\sing}^{\eeffffg}
= -\frac{\alpha}{2\pi} \sum_{i,j=1 \atop i \ne j}^6 
\sum_{\tau=\pm} (-1)^{i+j} Q_i Q_j}
\\
&& {} \times 
\int_0^1\rd x\, 
\int\rd\Phifo_{4f,ij}(x)\, 
{\cal G}^{(\sub)}_{ij,\tau}(\sfo_{ij},x)\, 
\frac{1}{2xs} \, \left|\M^{\eeffff}_{\born}
\Big(\Phifo_{4f,ij}(x),\tau\kappa_i\Big)\right|^2 
\Theta\Big(\Phifo_{4f,ij}(x)\Big),
\nn
\eeqar
where the functions ${\cal G}^{(\sub)}_{ij,\tau}$ originate from the
photonic phase-space integral over the radiator functions
$g^{(\sub)}_{ij,\tau}$, 
\beqar
{\cal G}^{(\sub)}_{ij,\tau}(\sfo_{ij},x)&=&
8 \pi^2 \int \rd \Phi_{\ga,ij}\,x
\, g^{(\sub)}_{ij,\tau}(q_i,q_j,k),
\eeqar
and $\sfo_{ij}=2\qfo_i\qfo_j$. The momenta $\qfo_i,\qfo_j$ result from
mapping the momenta $q_i,q_j,k$ to $\Phifo_{4f,ij}$.

The functions ${\cal G}^{(\sub)}_{ij,\tau}$
involve mass singularities from the initial state,
\beq
{\cal G}^{(\sub)}_{ij,+}(\sfo_{ij},x)\Big|_{\mathrm{sing}} =
P_{ff}(x)\ln\biggl(\frac{\sfo_{ij}}{\Me^2}\biggr),\qquad i=1,2,
\eeq
which are associated to the splitting function. These singular terms are
exactly the well-known universal logarithms of the structure-function
approach for leading-logarithmic ISR.

For both emitter and
spectator from the final state ($i,j=3,\ldots,6$) the convolution
over $x$ in \refeq{eq:sub} is absent, 
so that the integrand depends on the 
$4f$ kinematics of the original CM system in this case. For the other
cases, where at least the emitter or the 
spectator are from the initial state, the soft
singularities appear at the endpoint $x\to 1$ in the convolution
over $x$. This IR-sensitive 
endpoint contributions can be separated from the convolution
as follows,
\beqar
\label{eq:sub2}
\lefteqn{
\int \rd \si_{\sing}^{\eeffffg}
= -\frac{\alpha}{2\pi} \sum_{i,j=1 \atop i \ne j}^6 
\sum_{\tau=\pm} (-1)^{i+j} Q_i Q_j}
\\
&& {} \times 
\int_0^1\rd x\, 
\Biggl[
\int\rd\Phifo_{4f,ij}(x)\, 
{\cal G}^{(\sub)}_{ij,\tau}(\sfo_{ij},x)\, 
\frac{1}{2xs} \, \left|\M^{\eeffff}_{\born}
\Big(\Phifo_{4f,ij}(x),\tau\kappa_i\Big)\right|^2 
\Theta\Big(\Phifo_{4f,ij}(x)\Big)
\nn\\
&& \phantom{ {}\times \Biggl\{ \int_0^1\rd x\, \Biggl[ } {}
-\int\rd\Phifo_{4f,ij}(1)\, 
{\cal G}^{(\sub)}_{ij,\tau}(\sfo_{ij},x)\, \frac{1}{2s} \,
\left|\M^{\eeffff}_{\born}
\Big(\Phifo_{4f,ij}(1),\tau\kappa_i\Big)\right|^2 
\Theta\Big(\Phifo_{4f,ij}(1)\Big) 
\Biggr]\nl
&& {}
-\frac{\alpha}{2\pi} 
\sum_{i,j=1 \atop i\ne j}^6 \sum_{\tau=\pm}
(-1)^{i+j} Q_i Q_j
\frac{1}{2 s} \int\rd\Phi_{4f} G^{(\sub)}_{ij,\tau}(s_{ij}) 
\left|\M^{\eeffff}_{\born}
\Big(\Phi_{4f},\tau \kappa_i \Big)\right|^2 \Theta(\Phi_{4f}).\nn
\eeqar
If both $i>2$ and $j>2$, ${\cal G}^{(\sub)}_{ij,\tau}$ is proportional
to $\de(1-x)$ and thus does not contribute to the second and third
line 
of \refeq{eq:sub2}.
The second term within square brackets in \refeq{eq:sub2}
represents the endpoint contribution (with opposite sign) at $x\to 1$ 
that is split off.
All the endpoint contributions,
which are contained in the last term of \refeq{eq:sub2},
obey the $4f$ kinematics of the original CM system,
since $\Phi_{4f}=\Phifo_{4f,ij}(1)$.
The functions $G^{(\sub)}_{ij,\tau}$ originate from the integral 
of the functions ${\cal G}^{(\sub)}_{ij,\tau}$ over $x$,
\beqar
\label{eq:subG}
G^{(\sub)}_{ij,\tau}(\sfo_{ij})&=&
\int_0^1 \rd x\,{\cal G}^{(\sub)}_{ij,\tau}(\sfo_{ij},x).
\eeqar 
As mentioned above, the photon mass $\lambda$ and the fermion masses
$m_i$ have to be kept finite, in order to regularize 
soft and collinear
singularities in \refeq{eq:subG}.
Among these integrals of course only the non-spin-flip parts receive
singular contributions,
\beqar\label{eq:Gend}
G^{(\sub)}_{ij,+}(\sfo_{ij}) &=&
\L(\sfo_{ij},m_i^2)+C_{ij}-\frac{1}{2},
\nn\\
G^{(\sub)}_{ij,-}(\sfo_{ij}) &=& \frac{1}{2},
\eeqar
where
\beq\label{eq:defL}
\L(\sfo_{ij},m_i^2)=
\ln\biggl(\frac{m_i^2}{\sfo_{ij}}\biggr)
\ln\biggl(\frac{\la^2}{\sfo_{ij}}\biggr)
+ \ln\biggl(\frac{\la^2}{\sfo_{ij}}\biggr)
- \frac{1}{2}\ln^2\biggl(\frac{m_i^2}{\sfo_{ij}}\biggr)
+ \frac{1}{2}\ln\biggl(\frac{m_i^2}{\sfo_{ij}}\biggr).
\eeq
These are exactly the soft and collinear singularities of the virtual
${\cal O}(\alpha)$ corrections with opposite sign. The constants
$C_{ij}$, which are specific for the subtraction terms of  
\citere{di99}, read
\beqar
\label{eq:subcij}
C_{ab}&=&-\frac{\pi^2}{3}+2, \quad
C_{ak}=\frac{\pi^2}{6}-1, \quad
C_{ka}=-\frac{\pi^2}{2}+\frac{3}{2}, \quad
C_{kl}=-\frac{\pi^2}{3}+\frac{3}{2}
\eeqar
with $a,b=1,2$ and $k,l=3,\ldots,6$.

Owing to their kinematical structure, the sum of all endpoint
contributions, \ie the last line of \refeq{eq:sub2},
 can be directly combined with the virtual 
corrections before integrating over the $4f$ phase space. 

\subsection{The phase-space-slicing approach}
\label{se:singsli}

\newcommand{\delsoft}{\de_{\mathrm{s}}}
\newcommand{\delcoll}{\de_{\mathrm{c}}}

The idea of the phase-space-slicing method is
to divide the $4f\gamma$ phase space into singular and non-singular regions, 
then to evaluate the singular regions analytically
and to perform an explicit cancellation of the arising 
soft
and collinear singularities against their counterparts in the virtual
corrections. The finite remainder can 
be evaluated by using the usual Monte Carlo techniques. 
For the actual implementation of this well-known procedure 
(see e.g.\ \citere{be82}) we closely
follow the approaches of \citeres{bo93,ba99}.
We divide the five-particle phase space into soft and collinear regions
by introducing the cut-off parameters 
$\delsoft$ and $\delcoll$, respectively. We decompose the 
real corrections as
\begin{equation}\label{eq:si_4fg}
\rd\sigma^{\eeffffg}= \frac{1}{2s} \rd\Phi_{4f\gamma} |\M^{\eeffffg}|^2 
= \rd\sigma_{\soft}+ \rd\sigma_{\coll}+\rd\sigma_{\finite}^{\eeffffg}.
\end{equation}
Here $\rd\sigma_{\soft}$ describes the contribution of
the soft photons, 
\ie of photons with energies 
$E_{\gamma} < \delsoft \sqrt{s}/2= \Delta E$ in the CM frame, and 
$\rd\sigma_{\coll}$ describes real photon radiation outside
the soft-photon region ($E_{\gamma}>\Delta E$) 
but collinear to a charged fermion.
We define the collinear region by 
$1>\cos\theta_{\ga f} > 1-\delcoll$,
where $\theta_{\ga f}$ 
is the angle between the charged fermion and the emitted
photon in the CM frame. The remaining part, which is free of
singularities, is denoted by $\rd\sigma_{\finite}^{\eeffffg}$. 

In the soft and collinear regions, the squared matrix element 
$|\M^{\eeffffg}|^2$ factorizes into
the leading-order squared matrix element $|\M^{\eeffff}_{\Born}|^2$ 
and a soft or collinear factor. 
Also the five-particle phase space factorizes 
into a four-particle and a soft or collinear part, 
so that the integration over the photon phase space
can be performed analytically. 

In the soft-photon region, we apply the soft-photon approximation to
$|\M^{\eeffffg}|^2$, \ie 
the photon four-momentum $k$ is omitted 
everywhere but in the IR-singular 
propagators. Since we neglect $k$
also in the resonant gauge-boson propagators we have to assume
$E_{\gamma}< \Delta E \ll \Gamma_W$.
In this region
$\rd\sigma^{\eeffffg}$ can be written as 
\cite{Yennie:1961ad,de93}
\begin{eqnarray}
\rd\sigma_{\soft} &=& \rd\sigma_{\Born}^{\eeffff}
\frac{\alpha}{4\pi^2} \sum_{i=1}^6 \sum_{j=i+1}^6 
(-1)^{i+j} Q_i Q_j
\int_{E_{\gamma}< \Delta E \atop |{\bf k}|^2=E_{\gamma}^2-\lambda^2} 
\frac{\rd^3 \bk}{E_{\gamma}}
\left(\frac{q_i^{\mu}}{kq_i}-\frac{q_j^{\mu}}{kq_j}\right)^2
\nonumber\\
&=& \rd\sigma_{\Born}^{\eeffff}
\frac{\alpha}{2 \pi} \sum_{i=1}^6 \sum_{j=i+1}^6 
(-1)^{i+j} Q_i Q_j
\left[ I_{ii}+I_{jj}-2I_{ij} \right]
\end{eqnarray}
with the basic integrals
\begin{equation}
I_{ij}=\frac{1}{2\pi} 
\int_{E_{\gamma}< \Delta E \atop |{\bf k}|^2=E_{\gamma}^2-\lambda^2} 
\frac{\rd^3 {\bf k}}{E_{\gamma}} \frac{q_j q_j}{(kq_i) (kq_j)}.
\end{equation}
The explicit expression for the integrals $I_{ij}$ 
can be found in \citeres{'tHooft:1979xw,de93}.
For our purpose it is sufficient to keep the fermion masses only as
regulators for the collinear singularities ($E_i\gg m_i$). In this limit
we obtain
\begin{eqnarray}\label{eq:si_soft}
\rd\sigma_{\soft}&=&\rd\sigma_{\born}^{\eeffff} \frac{\alpha}{2 \pi} 
\sum_{i=1}^6 \sum_{j=i+1}^6 (-1)^{i+j} Q_i Q_j
\nonumber\\
&& \times  \left\{ 2 \ln\left(\frac{2\Delta E}{\lambda}\right) 
\, \left[2-\ln\left(\frac{s_{ij}^2}{m_i^2 m_j^2}\right)\right]
-2 \ln\left(\frac{4 E_i E_j}{m_i m_j}\right)\right.
\nonumber\\
&&{}+ \left. 
\frac{1}{2}\ln^2\left(\frac{4 E_i^2}{m_i^2}\right)+
\frac{1}{2} \ln^2\left(\frac{4 E_j^2}{m_j^2}\right)+
\frac{2\pi^2}{3}+2 \Li\left(1-\frac{4 E_i E_j}{s_{ij}}\right)
\right\}.         
\end{eqnarray}

Now we turn to the collinear singularities.
In the collinear region, we consider
an initial-state (final-state) fermion with momentum $q_i$  
being split into a collinear photon and a fermion with 
the resulting momentum $\qfo_i$ 
after (before) photon radiation,
\ie for {\em initial-state radiation}
\[\Pe(q_i) \to 
\gamma\Big(k=(1-x_i) q_i\Big)+\Pe\Big(\qfo_i=x_i q_i\Big),\qquad i=1,2,\] 
and for
{\em final-state radiation} 
\[f\Big(\qfo_i=q_i/z_i\Big) \to  \gamma\Big(k=q_i (1-z_i)/z_i\Big)+f(q_i), 
\qquad i=3,\ldots,6. \]
In the asymptotic limit, $|\M^{\eeffffg}|^2$
factorizes into the leading-order squared matrix element $|\M^{\eeffff}|^2$
and a collinear factor 
describing collinear initial-state and final-state radiation, respectively,
as long as $\delcoll$ is sufficiently small.
In the collinear region also the five-particle phase space factorizes into
a four-particle phase space and a collinear part,
so that the cross section for hard photon radiation ($E_{\gamma}>\Delta E$)
in the collinear region $1>\cos\theta_{\ga f}>1-\delcoll$ 
reads
\begin{equation}\label{eq:si_collin}
\rd\sigma_{\coll} = \rd\sigma_{\coll}^{\mathrm{initial}}+\rd\sigma_{\coll}^{\mathrm{final}}
\end{equation}
with $E_i\gg m_i$, $\theta_{\ga f}={\cal O}(m_i/E_i)$ 
and
\begin{eqnarray}\label{collini}
\rd\sigma_{\coll}^{\mathrm{initial}} = & \disp
\sum_{i=1,2} \frac{\alpha}{2\pi} \int_0^{1-\delsoft} \rd x_i \,
\Biggl\{ & \phantom{{}+{}}
\rd\sigma^{\eeffff}_{\Born}(x_i q_i, +\kappa_i )
P_{ff}(x_i) \left[ \ln\left(\frac{\hat s}{\Me^2} 
\frac{\delcoll}{2}\frac{1}{x_i}\right)-1 \right]
\nn\\
&&
{}+\rd\sigma^{\eeffff}_{\Born}(x_i q_i, -\kappa_i ) \, (1-x_i) \Biggr\}
\end{eqnarray}
and
\begin{eqnarray}\label{eq:si_collfin}
\rd\sigma_{\coll}^{\mathrm{final}} = & \disp
\sum_{i=3}^6  \frac{\alpha}{2\pi} Q_i^2 
\Biggl\{ & \phantom{{}+{}}
\rd\sigma^{\eeffff}_{\born}(q_i,+\kappa_i) 
\int_0^{1-\Delta E/E_i} \rd z_i \,
P_{ff}(z_i) \left[ \ln\left(\frac{4 E_i^2}{m_i^2}
\frac{\delcoll}{2} z_i^2 \right)- 1\right]
\nn\\
&& {}
+ \rd\sigma^{\eeffff}_{\born}(q_i,-\kappa_i)
\int_0^1 \rd z_i \, (1-z_i)
\Biggr\}
\nonumber\\
=& \disp
\sum_{i=3}^6 \frac{\alpha}{2 \pi} Q_i^2  
\Biggl\{ & \phantom{{}+{}}
\rd\sigma^{\eeffff}_{\born}(q_i,+\kappa_i) 
\nn\\
&& \phantom{{}+{}} {} \times
\left(
\left[\frac{3}{2}+2\ln\left(\frac{\Delta E}{E_i}\right)\right] 
\left[1-\ln\left(\frac{4 E_i^2}{m_i^2} \frac{\delcoll}{2}\right)\right]
+\frac{5}{2}-\frac{2\pi^2}{3} \right)
\nn\\
&& {}
+ \rd\sigma^{\eeffff}_{\born}(q_i,-\kappa_i)\frac{1}{2}
\Biggr\} 
\end{eqnarray}
with the splitting function $P_{ff}$ of
\refeq{eq:splitting_function}.
In the case of initial-state radiation, the
four-particle phase space is generated 
in the CM system of the hard scattering process
after the emission of the collinear photon, \ie in the system with
CM energy $\tilde s=x_{1,2} s$,
 and the four-momenta are then Lorentz-boosted 
back to the laboratory frame. In the case of final-state radiation the 
CM energy before and after collinear photon radiation
are the same, so that $\rd \sigma_{\born}^{\eeffff}$ does not depend on $z_i$.
Thus, the integration over $z_i$ can be performed analytically, as done in the
last equation of \refeq{eq:si_collfin}. Note that this procedure
implicitly assumes that photons within 
small cones collinear to charged
final-state fermions will never be separated from those collinear fermions.

Subtracting the soft and collinear cross sections \refeq{eq:si_soft}
and
\refeq{eq:si_collin} from the cross section 
of the process $\eeffffg$ \refeq{eq:si_4fg} yields the finite cross section
$\rd\sigma_{\finite}^{\eeffffg}$. 
As usual in the phase-space-slicing approach,
this subtraction is done in practice
by imposing cuts on the $4f\gamma$
phase space, \ie a photon-energy
cut, $E_{\gamma}>\Delta E=\delsoft\sqrt{s}/2$, 
and a cut on the angles between the photon and charged fermions,
$-1<\cos\theta_{\ga f}<1-\delcoll$,
but by using the exact matrix elements 
for
$\eeffffg$ of \citere{ee4fa}.
The collinear and soft cross sections are added,
\begin{equation}
\rd\sigma_{\sing}^{\eeffffg}=\rd\sigma_{\soft}+\rd\sigma_{\coll},
\end{equation}
and combined with the singular part of the virtual cross section 
$\rd\sigma_{\virt,\sing}^{\eeffff}$ of \refse{se:YFS},
as described in \refse{se:master}. 

The full radiative corrections are split into a $4 f$ part 
with no or 
an invisible photon,
$\rd \sigma_{\virt,\finite,\DPA}^{\eeWWffff}
+\rd \sigma_{\virt+\real,\sing}^{\eeffff}$, 
and a $4f\ga$ part 
with a visible photon, $\rd\sigma_{\finite}^{\eeffffg}$.
Both contributions
depend on the cut-off parameters $\delsoft,\delcoll$.
The dependence on these technical cuts cancels in the sum
when the cut-off parameters are chosen to be small enough so that
the soft-photon and leading-pole approximations apply.
In \reffi{fi:cutoff} we display the cut-off dependence of 
the total cross sections to 
both parts separately and also illustrate
the cancellation of the cut-off dependence in their sum.
\begin{figure}
{\centerline{
\setlength{\unitlength}{1cm}
\begin{picture}(14,7.8)
\put(-4.2,-13.5){\includegraphics{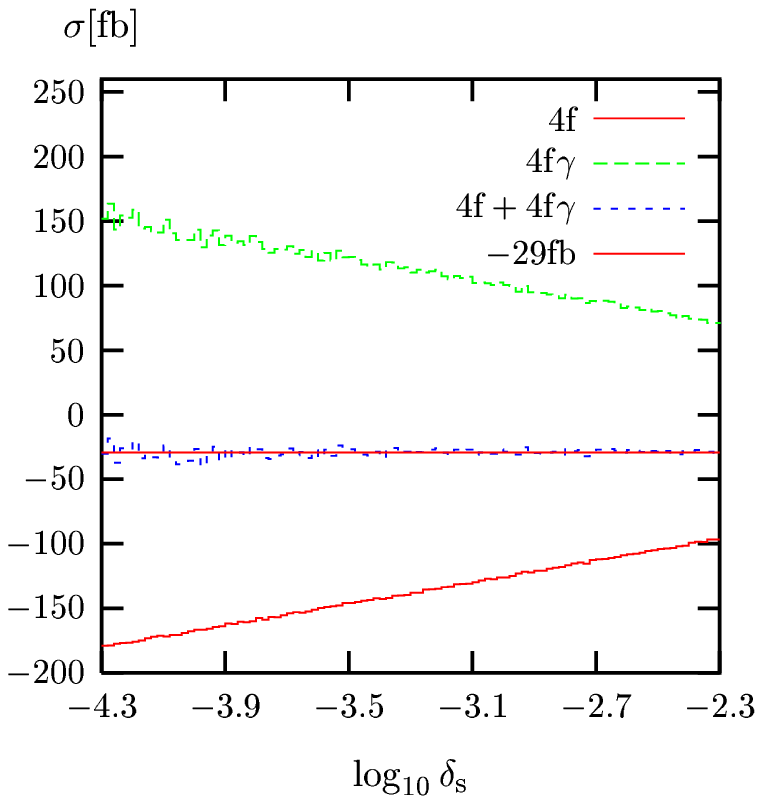}}
\put( 2.8,-13.5){\includegraphics{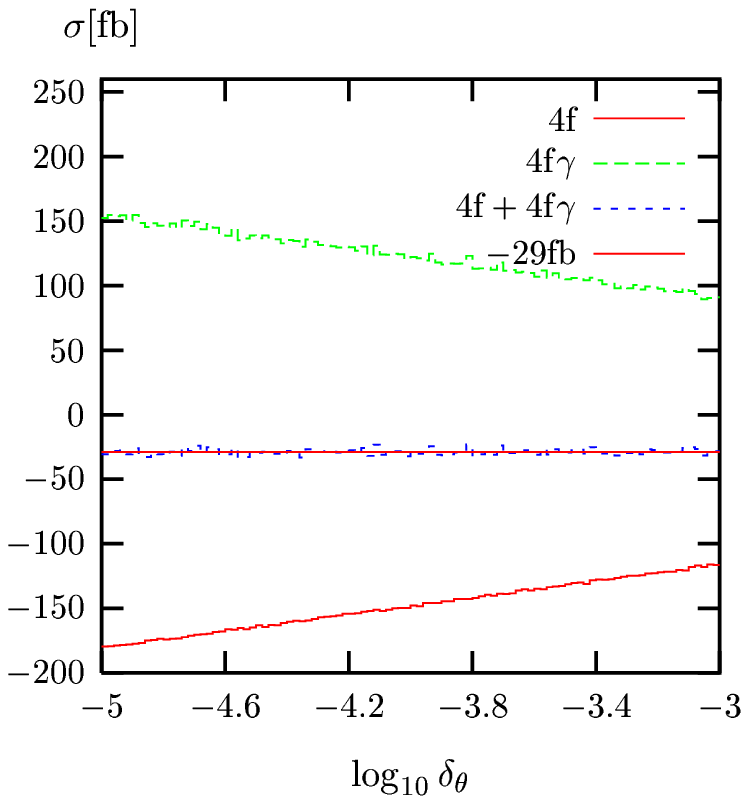}}
\end{picture} }}
\caption[]{Cut-off dependence of the total cross section in the
  phase-space-slicing approach for $\Pep\Pem\to \Pu \Pdbar \mu^-
  \bar\nu_{\mu}$ at $\protect\sqrt{s}=200\,\protect\GeV$.  The $4f$
  part (without Born), the $4f\ga$ part, and their sum are shown as a
  function of $\log_{10} \delsoft$ (l.h.s.) (with $\delcoll=0.0005$)
  and $\log_{10} \delcoll$ (r.h.s.) (with $\delsoft=0.001$).}
\label{fi:cutoff}
\efi

\subsection{Definition of finite virtual corrections}
\label{se:YFS}

Next, we turn to the extraction of the singular contributions from the
virtual corrections, \ie the definition of 
$\rd \sigma_{\virt,\sing}^{\eeffff}$ in \refeq{eq:finsing}.
Here we can make use of the well-known cancellation for soft and
collinear singularities between real and virtual corrections (KLN
theorem \cite{Ki62}).  For inclusive photons, the only uncancelled
singularities are the collinear singularities originating from the
continuum part of initial-state radiation ($x_i\ne 1$). In other
words, the singularities of the virtual corrections are exactly given
by the endpoint parts defined in \refse{se:singsub} within the
subtraction approach, but with opposite sign. Therefore, we can define
the singular part of the virtual corrections by
\begin{eqnarray}\label{eq:virt_sing_sub}
\rd\sigma_{\virt,\sing,\sub}^{\eeffff} &=& 
\rd\sigma_{\Born}^{\eeffff}
\frac{\alpha}{2\pi} \sum_{i=1}^6 \sum_{j=i+1}^6 
(-1)^{i+j} Q_i Q_j
\left(\L(s_{ij},m_i^2)+\L(s_{ij},m_j^2)+C_{ij}+C_{ji} \right)\nn
\nln
\end{eqnarray}
with
$\L(s_{ij},m_i^2)$ and $C_{ij}$ defined in \refeq{eq:defL} and
\refeq{eq:subcij} of~\refse{se:singsub}, respectively.  Note that we
do not include the polarization-dependent
contributions of the endpoint parts [the terms $\pm1/2$ in
\refeq{eq:Gend}] in this definition. 
Since the polarization-dependent terms are non-singular,
the simple formulae of this section hold also for the polarized case.
Subtracting \refeq{eq:virt_sing_sub} from the virtual corrections
yields the finite virtual cross section in the
subtraction-method-inspired approach.  We note that it is
gauge-invariant by construction, since only the Born cross section
enters without modification.

The definition of the virtual singular cross section is a matter of
convention as far as finite contributions are concerned, since finite
terms can be redistributed between singular and finite parts.  Another
possibility is to define a finite virtual photon contribution by
following the approach of \citere{Yennie:1961ad}. In this YFS-inspired
approach, the IR-singular contributions are extracted with the help of
an explicitly $\U(1)$-gauge-invariant current as follows,
\begin{eqnarray} \label{eq:qedsoft}
\rd\sigma_{\soft,\YFS}^{\eeffff}&=& 
\rd\sigma_{\Born}^{\eeffff} \Re \left\{
(-\ri 4\pi \alpha) \sum_{i=1}^6 \sum_{j=i+1}^6 
(-1)^{i+j} Q_i Q_j \right.
\nonumber\\
&& \left. {} \times\int \frac{\rd^4 k}{(2 \pi)^4}
\left[\frac{(k-2\theta_iq_i)^{\mu}}{k^2-2\theta_iq_ik+\ri\eps}
-\frac{(k+2\theta_jq_j)^{\mu}}{k^2+2\theta_jq_jk+\ri\eps}\right]^2
\frac{1}{k^2-\lambda^2+\ri\eps} \right\}
\nonumber\\ 
&=&
\rd\sigma_{\Born}^{\eeffff}
\frac{\alpha}{2\pi} \sum_{i=1}^6 \sum_{j=i+1}^6 
(-1)^{i+j} Q_i Q_j
\\
&& {}
\times
\left[\L(s_{ij},m_i^2)+\L(s_{ij},m_j^2)+2\ln\left(\frac{s_{ij}}{m_i m_j}\right)
-\frac{\pi^2}{3}- \theta(\theta_i \theta_j) \, \pi^2+2 \right],\nn
\end{eqnarray}
where $\theta_i=+1(-1)$ if $i$ is a final-state (initial-state)
fermion.  Again we keep the fermion masses only as regulators for the
collinear singularities.  As can easily be seen from
\refeq{eq:qedsoft}, the $\U(1)$ gauge invariance is guaranteed by the
existence of a conserved current.  This YFS factor takes care only of
the IR-singular logarithms and the double-logarithmic mass
singularities. Since the single-logarithmic mass-singular terms build
a separately gauge-invariant subset we can simply add them to the YFS
factor as follows \cite{Wackeroth:1997hz},
\begin{eqnarray}\label{eq:virt_sing_yfs} 
\rd\sigma_{\virt,\sing,\YFS}^{\eeffff}&=&
\rd \sigma_{\soft,\YFS}^{\eeffff}
-\rd\sigma_{\Born}^{\eeffff}
\frac{\alpha}{\pi} \sum_{i=1}^6 \sum_{j=i+1}^6 
(-1)^{i+j} Q_i Q_j
\ln\left(\frac{s_{ij}}{m_i m_j}\right) .
\end{eqnarray}
This defines the singular virtual cross section in the YFS-inspired
approach.

While the requirement of the cancellation of the
soft and collinear singularities
in \refeq{eq:virtfinite}
unambiguously fixes the logarithmic terms, the two definitions of the
virtual singular contribution differ by finite, non-logarithmic terms
\begin{eqnarray}
\lefteqn{\rd\sigma_{\virt,\sing,\sub}^{\eeffff}-
\rd\sigma_{\virt,\sing,\YFS}^{\eeffff}=}
\quad\\
&=&\rd\sigma_{\Born}^{\eeffff} 
\frac{\alpha}{2\pi} \sum_{i=1}^6 \sum_{j=i+1}^6 
(-1)^{i+j} Q_i Q_j
\left[\frac{\pi^2}{3}+\theta(\theta_i \theta_j) \, \pi^2 -2
+C_{ij}+C_{ji}\right],\nn
\end{eqnarray}
as can been seen by comparing \refeq{eq:virt_sing_sub} and
\refeq{eq:virt_sing_yfs}.
Since the singular term 
$\rd\sigma_{\virt,\sing,(\sub,\YFS)}^{\eeffff}$ is treated once in
DPA, when subtracted from the doubly-resonant virtual corrections
$\rd\sigma_{\virt,\DPA}^{\eeWWffff}$ [see \refeq{eq:virtfinite}], and
once exactly, when included in
$\rd\sigma_{\virt+\real,\sing}^{\eeffff}$, the cross sections
calculated with the subtraction-method-inspired and in the
YFS-inspired approach differ by non-doubly-resonant terms
\begin{eqnarray}\label{eq:diff_sl_sub}
\Delta 
&=&\left(\rd\sigma_{\Born}^{\eeffff}-\rd\sigma_{\Born,\DPA}^{\eeWWffff}\right)
\nonumber\\
&& {} \times 
\frac{\alpha}{2\pi} \sum_{i=1}^6 \sum_{j=i+1}^6 
(-1)^{i+j} Q_i Q_j
\left[\frac{\pi^2}{3}+\theta(\theta_i \theta_j) \, \pi^2 -2
+C_{ij}+C_{ji}\right].
\end{eqnarray}
As can be seen from \refeq{eq:diff_sl_sub}, the ambiguity induced by
these finite terms is of the order of the 
uncertainty of the DPA. A
numerical discussion of this ambiguity for observables can be found
in~\refse{se:amb}. 

While the two different definitions of the virtual singular
corrections are inspired by the subtraction method and the YFS
treatment of IR singularities, their use is independent of the method
employed for the treatment of singularities in the bremsstrahlung
contribution. 
This means that 
for the splitting between singular and finite virtual corrections
both definitions $\rd\sigma_{\virt,\sing,(\sub,\YFS)}^{\eeffff}$ 
can be used in the subtraction and phase-space-slicing approaches. 

\section{Higher-order initial-state radiation}
\label{se:isr}

The emission of photons collinear to the incoming electrons or
positrons leads to corrections that are enhanced by large logarithms.
In order to achieve an accuracy at the few $0.1\%$ level, the
corresponding higher-order contributions, i.e.\ contributions beyond
$\Oa$, must be taken into account. This can be done in the
structure-function method \cite{sf,lep2repWcs}.  According to the
mass-factorization theorem, the leading-logarithmic (LL) initial-state
QED corrections can be written as a convolution of the lowest-order
cross section with structure functions, and the corresponding
differential cross section reads
\newcommand{\LL}{\mathrm{LL}}
\beq\label{sigmaLL}
  \int \rd\sigma^{\LL} =
  \int^1_0 \rd x_1 \int^1_0 \rd x_2 \,
  \Gamma_{\Pe\Pe}^{\LL}(x_1,Q^2)\Gamma_{\Pe\Pe}^{\LL}(x_2,Q^2)
  \int \rd\sigma_\born^{\eeffff}(x_1 p_+,x_2 p_-).
\eeq
Here $x_1$ and $x_2$ denote the fractions of the longitudinal momentum
carried by the incoming electron and positron momenta just before the
hard scattering process occurs.  This means that the incoming momenta
$p_\pm$ before emission of the collinear photon are rescaled by
$x_{1,2}$, and the CM frame of the hard scattering process with the
lowest-order cross section $\rd\sigma_\born^{\eeffff}(x_1 p_+,x_2
p_-)$ is boosted along the beam axis.  The LL structure function
including $\O(\al^3)$ terms is given by \cite{lep2repWcs}
\newcommand{\zetal}{\be_\Pe}
\beqar\label{SFexp}
  \Gamma_{\mathrm{ee}}^{\LL}(x,Q^2) &=&    
    \frac{\exp\left(-\frac{1}{2}\zetal\gamma_{\rE} +
        \frac{3}{8}\zetal\right)}
{\Gamma\left(1+\frac{1}{2}\zetal\right)}
    \frac{\zetal}{2} (1-x)^{\frac{\zetal}{2}-1} - \frac{\zetal}{4}(1+x) 
\nn\\
&&  {} - \frac{\zetal^2}{32} \biggl\{ \frac{1+3x^2}{1-x}\ln(x)
    + 4(1+x)\ln(1-x) + 5 + x \biggr\}
\nn\\
&&  {} - \frac{\zetal^3}{384}\biggl\{
      (1+x)\left[6\Li(x)+12\ln^2(1-x)-3\pi^2\right] 
\nn\\
&& \quad\quad {}
+\frac{1}{1-x}\biggl[ \frac{3}{2}(1+8x+3x^2)\ln(x) 
+6(x+5)(1-x)\ln(1-x)
\nn\\
&& \quad\quad\quad {}
+12(1+x^2)\ln(x)\ln(1-x)-\frac{1}{2}(1+7x^2)\ln^2(x)
\nn\\
&& \quad\quad\quad  {}
+\frac{1}{4}(39-24x-15x^2)\biggr] \biggr\}
\eeqar
with
\beq
  \zetal = \frac{2\alpha}{\pi} \left(L-1\right),
\eeq
and the leading logarithm
\beq
L = \ln\frac{Q^2}{\Me^2}.
\eeq
Note that the scale $Q^2$ is not fixed within LL approximation, but
has to be set to a typical scale of the underlying process; for the
numerics we use $Q^2=s$.  In \refeq{SFexp} $\gamma_E$ is the Euler
constant and $\Gamma(y)$ the gamma function, which should not be
confused with the structure functions.  Note that some non-leading
terms are incorporated, taking into account the fact that the residue
of the soft-photon pole is proportional to $L-1$ rather than $L$ for
the initial-state photon radiation.

We add the cross section \refeq{sigmaLL} to the one-loop result and
subtract the lowest-order and one-loop contributions 
$\rd\sigma^{\LL,1}$ already contained 
within this formula, 
\beqar\label{sigmaLL1}
  \int \rd\sigma^{\LL,1} &=&
  \int^1_0 \rd x_1\rd x_2 
  \Bigl[\de(1-x_1)\de(1-x_2)
  +\Gamma_{\Pe\Pe}^{\LL,1}(x_1,Q^2)\de(1-x_2)
\nl
  &&\qquad
  {}+\de(1-x_1)\Gamma_{\Pe\Pe}^{\LL,1}(x_2,Q^2)\Bigr]
  \int \rd\sigma_\born^{\eeffff}(x_1p_+,x_2p_-),
\eeqar
in order to avoid double counting.
The one-loop contribution to the structure function reads
\beqar
  \Gamma_{\mathrm{ee}}^{\LL,1}(x,Q^2) &=&
  \frac{\zetal}{4} \left(\frac{1+x^2}{1-x}\right)_+ 
\nl
  &=& \frac{\zetal}{4} \lim_{\eps\to 0} 
  \left[\delta(1-x)\left(\frac{3}{2}+2\ln\eps\right) 
  + \theta(1-x-\eps)\frac{1+x^2}{1-x}\right].
\eeqar
Note that the uncertainty that is connected with the choice of $Q^2$
enters now in ${\cal O}(\alpha^2)$, since all $\Oa$ corrections,
including constant terms, are taken into account.

\section{QCD corrections}
\label{se:qcd}

QCD corrections enter the processes $\eeWWffff$ in two different
places.  On the one hand, the hadronic W~width receives a QCD
correction, namely a factor $(1+\alpha_{\mathrm{s}}/\pi)$
\cite{rcwdecay2,rcwdecay1} in ${\cal O}(\alpha_{\mathrm{s}})$.  This
affects the total W~width $\GW$ in the resonant W~propagators.
On the other hand, each hadronically decaying W~boson receives a QCD
correction to the $Wq\bar q'$ vertex.  If the full phase space for
gluon emission is integrated over, this correction reduces to a
multiplicative correction factor $(1+\alpha_{\mathrm{s}}/\pi)$ for
each hadronically decaying W~boson in ${\cal O}(\alpha_{\mathrm{s}})$.
The application of this inclusive factor to distributions is usually
called ``naive QCD correction''.
In the total cross section these naive QCD factors cancel against the
corresponding factors in the W~width in the resonant W~propagators.
Note that one-gluon exchange%
\footnote{Multiple gluon exchange, which is intrinsically connected to 
colour reconnection, is not considered here. It should be included in 
the hadronization simulation.}
between quarks from different W~bosons vanishes exactly for pure 
charged-current (CC) reactions, because of the colour structure.

If the gluon phase space is not integrated over, QCD corrections have to
be calculated from the virtual vertex corrections and from real gluon
emission with the matrix elements $\eeffff+\Pg$. This option is also
supported by {\sc RacoonWW} for four-fermion final states of the CC11
class. The calculation is technically similar to the one
for the photonic corrections \cite{ee4fa}.

\section{Numerical results}
\label{se:numres}

For the numerical results, with the exception of the comparison to the
results of other groups, we used (as in
\citeres{de99a,de99b,lep2mcws}) the following parameters:
\beq\label{eq:pars}
\begin{array}[b]{rlrl}
\GF =& 1.16637\times 10^{-5} \GeV^{-2}, \qquad&\alpha=&1/137.0359895, \\
\MW =& 80.35\GeV,& \GW =& 2.08699\ldots\GeV, \\
\MZ =& 91.1867\GeV,& \GZ =& 2.49471\GeV, \\
\Mt =& 174.17 \GeV,&\MH=& 150\GeV, \\
\Me =& 510.99907 \keV.
\end{array}
\eeq
We work in the fixed-width scheme and fix
the weak mixing angle by $\cw=\MW/\MZ$, $\sw^2=1-\cw^2$.
The parameter set \refeq{eq:pars} is over-complete but
self-consistent. Instead of $\alpha$ we use $\GF$ to parame\-trize the
lowest-order matrix element, \ie we use the effective coupling
\beq
\alpha_{\GF} = \frac{\sqrt{2}\GF\MW^2\sw^2}{\pi}
\eeq
in the lowest-order matrix element. This parametrization has the
advantage that all higher-order contributions associated with the
running of the electromagnetic coupling from
zero to $\MW^2$ and the leading universal two-loop $\Mt$-dependent
corrections are
already absorbed in the Born cross section.  In the relative $\Oa$
corrections, on the other hand, we use $\al=\al(0)$, since in the real
corrections and in the mass-singular virtual corrections, which yield
the bulk of the remaining corrections, the scale of the real or
virtual photon is zero.  The W-boson width given above is calculated
including the electroweak and QCD one-loop corrections with
$\alpha_{\mathrm s}=0.119$.

As default, we use the following set of separation and recombination cuts:
\begin{enumerate}
\item All photons within a cone of 5 degrees around the beams are
  treated as invisible, \ie their momenta are disregarded when
  calculating angles, energies, and invariant masses.
\item Next, the invariant masses $M_{f\gamma}$ of the photon with each
  of the charged final-state fermions are calculated. If the smallest
  $M_{f\gamma}$ is smaller than a certain cutoff $M_\recomb$ or if the
  energy of the photon is smaller than $1\GeV$, the photon is combined
  with the
charged final-state fermion that leads to the smallest $M_{f\gamma}$,
\ie the momenta of the photon and the fermion are added and associated
with the momentum of the fermion,
and the photon is discarded%
\footnote{Except for the $1\GeV$ cut, the described cut and recombination
procedure coincides with the one used in \citeres{de99a,de99b}.}.
\item Finally, all events are discarded in which one of the charged
  final-state fermions is within a cone of 10 degrees around the
  beams.  No other cuts are applied.
\end{enumerate}

In \citeres{de99a,de99b} we have already presented a short survey of
numerical results for ${\cal O}(\alpha)$ corrections to $\eeWWffff$
obtained with {\sc RacoonWW} for LEP2 and linear-collider energies.
In the meantime we extended {\sc RacoonWW} by taking into account
higher-order ISR and by considering QCD one-loop corrections as
described in \refse{se:isr} and \refse{se:qcd}, respectively.  Here we
concentrate on LEP2 energies and provide results for the total cross
sections and distributions obtained with this extended version of {\sc
  RacoonWW}.  Additional numerical results can also be found in
\citere{lep2mcws}.  Unless stated otherwise, for the lowest-order
contributions only the CC03 diagrams are taken into account. Our
``best'' results comprise the CC03 Born cross sections and the
corrected cross sections including all the radiative corrections
described in this paper, i.e.\ electroweak one-loop corrections in
DPA, exact ${\cal O}(\alpha)$ photon radiation based on the full
matrix element for $\eeffffg$, higher-order ISR up to ${\cal
  O}(\alpha^3)$ and ``naive'' QCD ${\cal O}(\alpha_{\mathrm{s}})$
corrections (see \refse{se:qcd}).

As mentioned before, {\sc RacoonWW} involves two branches for the
treatment of soft and collinear singularities, one following the
subtraction (see \refse{se:singsub}) and one the phase-space-slicing
method (see \refse{se:singsli}).  While these two branches use the
same matrix elements, the Monte Carlo integration is performed
completely independently, thus providing us with a powerful numerical
check of {\sc RacoonWW}. In the following we present numerical results
of both branches of {\sc RacoonWW}, starting with the total cross
sections to $\eeWWffff$ at LEP2 CM energies.  If not stated otherwise,
the finite virtual corrections that are treated in DPA are defined by
the subtraction-method-inspired approach \refeq{eq:virt_sing_sub}.

\subsection{The total W-pair production cross section at LEP2}
\label{se:totcs}

In \reftas{tab:lep2cs2} and \ref{tab:lep2cs3} we list the 
predictions of {\sc RacoonWW}
for the different W-pair production channels as well as for the 
total CC03 W-pair production cross section for LEP2 CM energies.
\begin{table}
\centerline{
\begin{tabular}{|l||l|l|l||l|}
\hline
$\sqrt{s}/\GeV$ & $\sigma_{\mathrm{leptonic}}/\fba$ & 
$\sigma_{\mathrm{semileptonic}}/\fba$ & 
$\sigma_{\mathrm{hadronic}}/\fba$ & $\sigma^{\PW\PW}/\pba$ \\
\hline\hline
172.086 & 142.088(71)  &  442.50(36) &  1376.14(67)  &  12.0934(76) \\
\hline
176.000 & 160.076(78)  &  498.03(25) &  1550.04(75)  &  13.6171(67)\\
\hline
182.655 & 180.697(89)  &  562.22(28) &  1749.48(86)  &  15.3708(76) \\
\hline
188.628 & 190.882(96)  &  594.31(55) &  1848.07(92)  &  16.2420(111) \\
\hline
191.583 & 194.271(118) &  604.12(31) &  1880.19(94)  &  16.5187(85) \\
\hline
195.519 & 197.320(123) &  614.11(31) &  1911.45(97)  &  16.7910(88) \\
\hline
199.516 & 199.497(103) &  620.53(33) &  1931.28(99)  &  16.9670(89) \\
\hline
201.624 & 200.200(104) &  622.65(33) &  1937.94(100) &  17.0254(89) \\
\hline
210.000 & 200.910(107) &  624.95(33) &  1945.00(103) &  17.0876(91) \\
\hline
\end{tabular} }
\caption{{\sc RacoonWW} predictions for W-pair production cross sections for
LEP2 CM energies including ${\cal O}(\alpha^2)$ ISR
(results for the Winter conferences 2000)}
\label{tab:lep2cs2}
\vspace*{2em}
\centerline{
\begin{tabular}{|l||l|l|l||l|}
\hline
$\sqrt{s}/\GeV$ & $\sigma_{\mathrm{leptonic}}/\fba$ & 
$\sigma_{\mathrm{semileptonic}}/\fba$ & 
$\sigma_{\mathrm{hadronic}}/\fba$ & $\sigma^{\PW\PW}/\pba$ \\
\hline\hline
168.000 & 115.639(57)  &  359.86(18) &  1120.03(55)  &  
\phantom{0}9.8392(49) \\
\hline
172.086 & 142.043(70)  &  442.36(36) &  1375.72(67)  &  12.0896(76)\\
\hline
176.000 & 160.192(78)  &  498.39(24) &  1551.17(75)  &  13.6271(66)\\
\hline
180.000 & 173.511(85)  &  539.96(27) &  1679.35(82)  &  14.7585(72)\\
\hline
182.655 & 180.669(89)  &  562.13(28) &  1749.21(86)  &  15.3684(76)\\
\hline
185.000 & 185.474(92)  &  576.88(29) &  1794.93(88)  &  15.7716(78)\\
\hline
188.628 & 190.959(95)  &  594.55(55) &  1848.84(91)  &  16.2486(111)\\
\hline
191.583 & 194.271(118) &  604.13(31) &  1880.21(94)  &  16.5188(85)\\
\hline
195.519 & 197.435(123) &  614.47(31) &  1912.58(97)  &  16.8009(87)\\
\hline
199.516 & 199.640(103) &  620.98(33) &  1932.66(99)  &  16.9791(88)\\
\hline
201.624 & 200.271(104) &  622.88(33) &  1938.65(100) &  17.0316(89)\\
\hline
205.000 & 200.742(105) &  624.77(33) &  1943.80(101) &  17.0792(89)\\
\hline    
208.000 & 200.973(106) &  625.33(33) &  1945.37(102) &  17.0942(90)\\
\hline    
210.000 & 200.888(107) &  624.88(33) &  1944.79(103) &  17.0858(91)\\
\hline
215.000 & 200.334(107) &  623.28(33) &  1938.88(103) &  17.0378(91)\\
\hline
\end{tabular} }
\caption{{\sc RacoonWW} predictions for W-pair production cross sections for
LEP2 CM energies including ${\cal O}(\alpha^3)$ ISR
(results for the Summer conferences 2000)}
\label{tab:lep2cs3}
\end{table} 
Note that for CC03 and negligible fermion masses the results are
independent of the final state within these channels. In the
calculation of these numbers no cuts have been applied.
The only difference between the results shown in the two tables lies
in the ISR beyond $\Oa$:
while \refta{tab:lep2cs2} includes only the ${\cal O}(\alpha^2)$
contributions of \refeq{SFexp},
\refta{tab:lep2cs3} includes the ${\cal O}(\alpha^3)$
contributions in addition.  There are no significant differences
between both tables, \ie the effect of the ${\cal O}(\alpha^3)$
contributions is smaller
than the integration error, which is at the level of 0.1\%.
The given errors are purely statistical. The errors for the total cross
sections were obtained by adding the (statistically correlated) errors
of the various channels linearly.

Figure~\ref{fig:wwcs2000} shows a comparison of the results of {\sc
  RacoonWW} as presented in \refta{tab:lep2cs2} (see also
\citere{lep2mcws}) and of other Monte Carlo predictions with recent
LEP2 data, as given by the LEP Electroweak Working Group
\cite{LEPEWWG} for the Winter 2000 conferences.
\begin{figure}
\setlength{\unitlength}{1cm}
\centerline{
\begin{picture}(10.5,10.5)
\put(-.5,-.8){\includegraphics{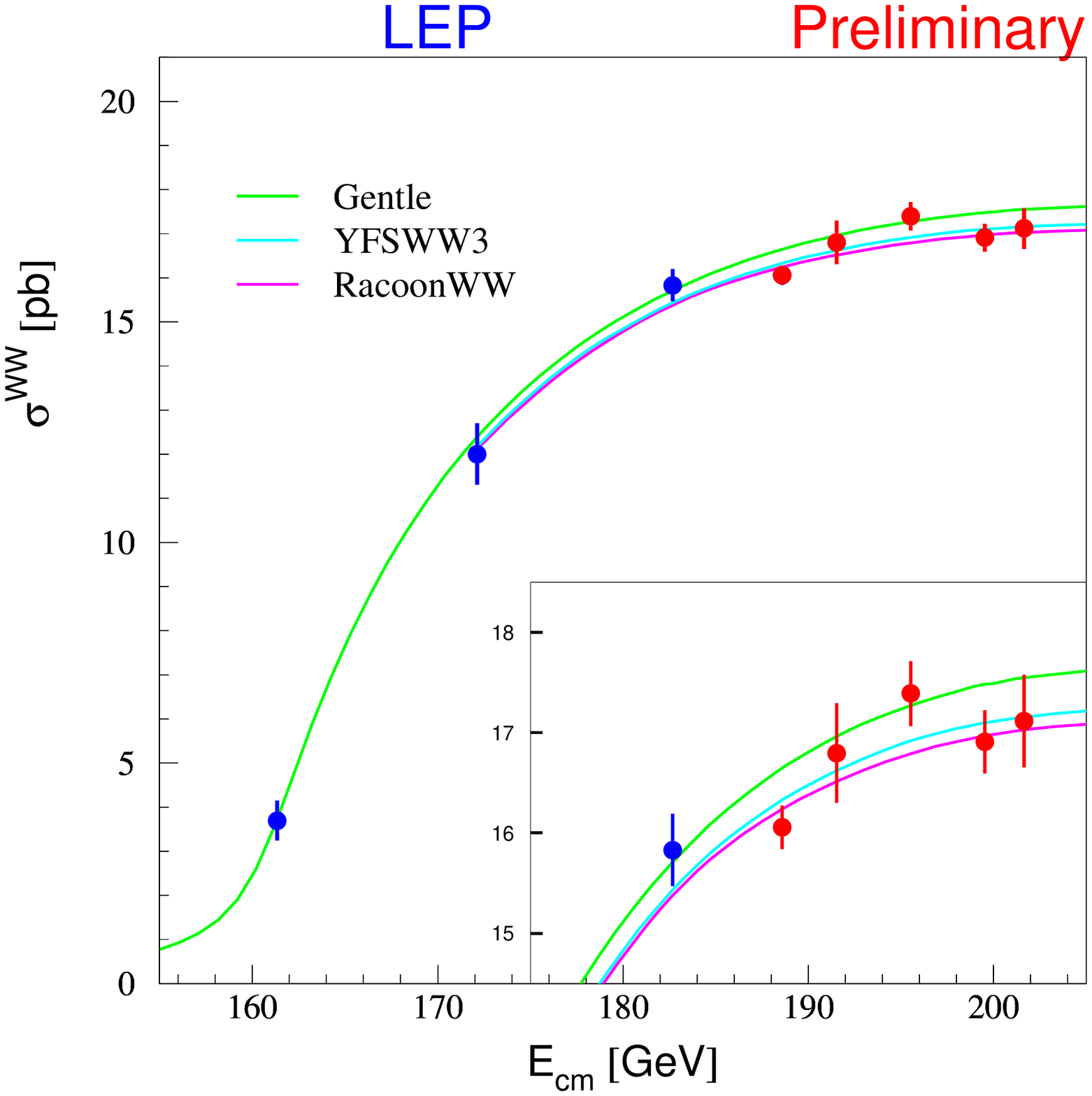}}
\end{picture}
}
\caption{Total WW production cross section at LEP2, as given by the
LEPEWWG \cite{LEPEWWG}}
\label{fig:wwcs2000}
\end{figure}

The data are in good agreement with the predictions of {\sc RacoonWW}
and {\sc YFSWW3} \cite{ja97,ja99}.  At the time of the conferences in
Winter 2000 the predictions of {\sc YFSWW3} where about 0.5--0.7\%
larger than those of {\sc RacoonWW}, which is somewhat larger than the
intrinsic DPA ambiguity. Meanwhile, however, the main source of this
discrepancy was found, and the improved {\sc YFSWW3} results differ
from the ones of {\sc RacoonWW} only by about 0.3\% at LEP2 energies.
More details on the conceptual differences of the two generators, as
well as a detailed comparison of numerical results, can be found in
\citere{lep2mcws}. A brief summary is given in \refse{se:yfsww}.
Figure~\ref{fig:wwcs2000} also includes the prediction provided by
{\sc GENTLE} \cite{gentle}, which is 2--2.5\% larger than those from
{\sc RacoonWW} and {\sc YFSWW3}.
This difference is due to the neglect of non-leading, non-universal
${\cal O}(\alpha)$ corrections in {\sc GENTLE}. The 1--2\% reduction
in the W-pair production cross section by these effects could already
be seen from the results of \citeres{bo92,lep2repWcs,di97}, where such
corrections to on-shell W-pair production were considered.  In
summary, the comparison between predictions of the electroweak
Standard Model and LEP2 data reveals empirical evidence of non-leading
electroweak radiative corrections beyond the level of universal
effects.

\subsection{Distributions for $\Pep\Pem\to\Pu\bar\Pd\mu^-\bar\nu_\mu$
at LEP2} 
\label{se:distri}

Now we turn to results obtained with the cut and recombination procedure
described at the beginning of \refse{se:numres}.
We consider the cases of a tight recombination cut $M_\recomb=
5\GeV$ (``\bare'') and of a loose recombination cut $M_\recomb= 25\GeV$
(``\calo''). 

In Table~\ref{ta:totcs_sl_sub_cuts} we provide the total
cross sections for leptonic, semileptonic and hadronic processes at
the CM energy $\sqrt{s}=200\GeV$ with the {\bare} cuts applied.  The
corrections are $-9.1\%$, $-5.8\%$,
and $-2.3\%$ for the leptonic, semileptonic and hadronic channel,
respectively.  The results obtained when using the phase-space-slicing
method agree well with those of the subtraction method, i.e.\  within
the statistical errors of about $0.05\%$.
\begin{table}
\bce
\begin{tabular}{|c|c|c|c|}
\hline
\multicolumn{2}{|c|}{\bf with bare cuts}&
\multicolumn{2}{|c|}{\bf$\sigma_{\mathrm{tot}}[\mathrm{fb}]$}\\
\hline
final state & program & Born & best \nl
\hline\hline
& { slicing}    & 211.166(36)  & 191.847(91) \nl
$\nu_\mu\mu^+\tau^-\bar\nu_\tau$
& { subtraction}& 211.034(39)  & 191.686(46) \nl
\cline{2-4}
& (sub--sli)/sli & $-0.06(3)$\% & $-0.08(5)$\% \nl
\hline\hline
& {slicing}     & 627.38(11)   & 591.11(25) \nl
$\Pu\bar\Pd\mu^-\bar\nu_\mu$
& {subtraction} & 627.22(12)   & 590.94(14) \nl
\cline{2-4}
& (sub--sli)/sli & $-0.03(3)$\% & $-0.03(5)$\% \nl
\hline\hline
& {slicing}     & 1864.79(32)  & 1821.72(66) \nl
$\Pu\bar\Pd\Ps\bar\Pc$
& {subtraction} & 1864.28(35)  & 1821.16(43) \nl
\cline{2-4}
& (sub--sli)/sli & $-0.03(3)$\% & $-0.03(4)$\% \nl
\hline
\end{tabular}
\ece
\caption[]{Total CC03 cross sections when using the subtraction method
  and the phase-space-slicing method at $\sqrt{s}=200\protect\GeV$ with {\bare}
  cuts. The numbers in parentheses are statistical errors
  corresponding to the last digits.}
\label{ta:totcs_sl_sub_cuts}
\end{table}
For {\calo} cuts we find the same good agreement between both
branches.  The differences between the cross sections for {\bare} and
{\calo} cuts are below $0.05\%$; of course, the lowest-order results
are not affected by the recombination procedure.

In the following we show a variety of distributions for the
semi-leptonic process $\Pep\Pem\to\Pu\bar\Pd\mu^-\bar\nu_\mu$ at
$\sqrt{s}=200\GeV$ at lowest order and when taking into account
radiative corrections (``best''). We also display the relative
corrections $\rd\sigma / \rd \sigma_{\Born}-1$ (in per cent) to
illustrate the effect of the radiative corrections.  For the relative
corrections we always provide the results of both branches of {\sc
  RacoonWW}, one employing the subtraction and one the
phase-space-slicing method.  Since we use the
subtraction-method-inspired definition of the finite virtual
corrections in both cases,
and thus the same DPA definition, the
results should agree within integration errors.

The invariant-mass distributions for the $\PWp$ and $\PWm$ bosons are
shown in \reffi{fi:mwpwm-sub} for the {\calo} recombination cut.  The
relative corrections to the $\PWp$ and $\PWm$ invariant-mass
distributions are shown in \reffis{fi:mwpsl-sub} and
\ref{fi:mwmsl-sub}, respectively, for both recombination cuts.
\begin{figure}
{\centerline{
\setlength{\unitlength}{1cm}
\begin{picture}(14,7.8)
\put(-4.2,-13.5){\includegraphics{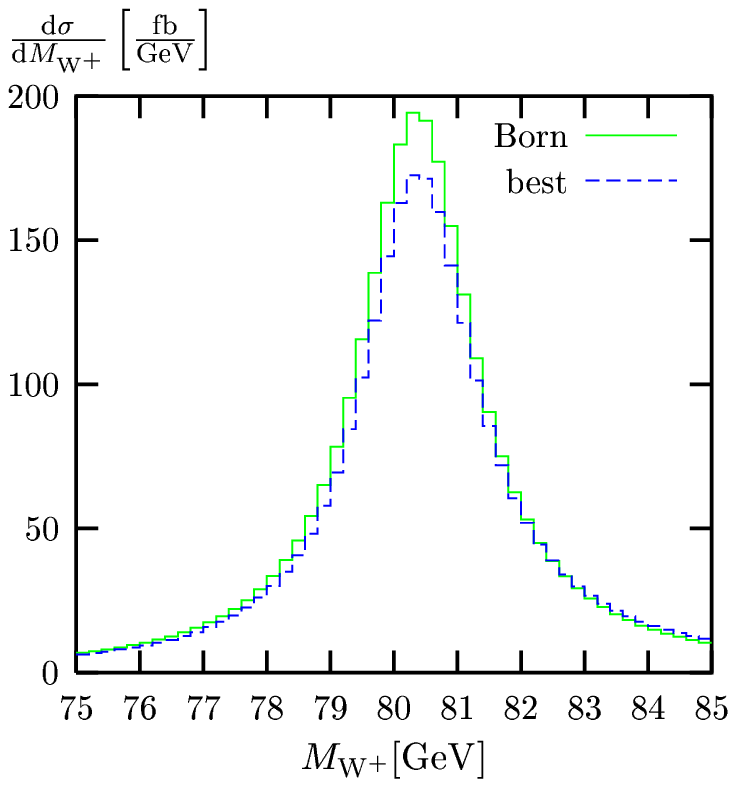}}
\put( 2.8,-13.5){\includegraphics{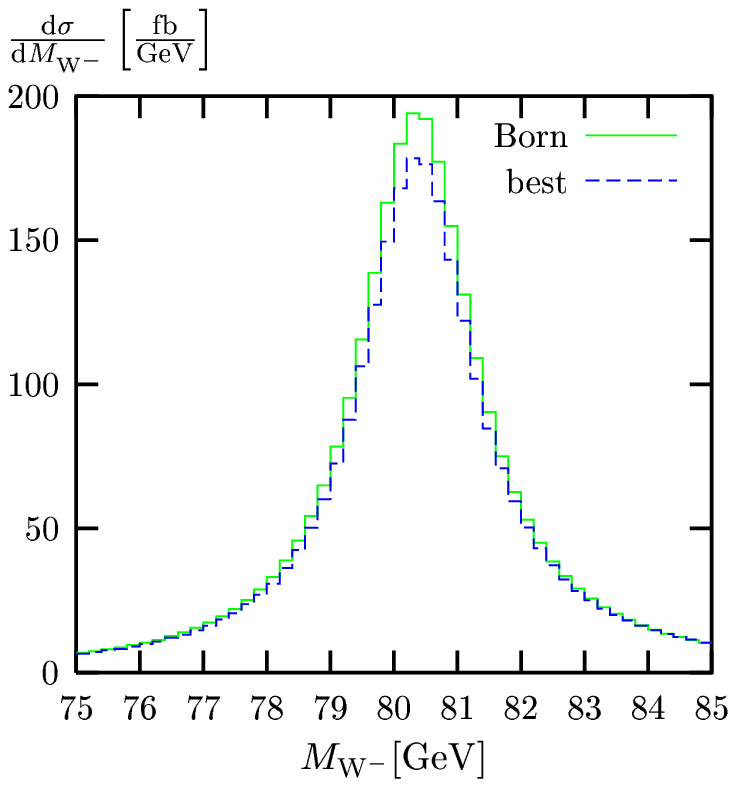}}
\end{picture} }}
\caption[]{Distributions in the $\protect\PWp$ (l.h.s.) and 
  $\protect\PWm$ (r.h.s.) invariant masses for a {\calo} setup for
  $\Pep\Pem\to \Pu \Pdbar \mu^- \bar\nu_{\mu}$ at
  $\protect\sqrt{s}=200\,\protect\GeV$}
\label{fi:mwpwm-sub}
\efi
\begin{figure}
{\centerline{
\setlength{\unitlength}{1cm}
\begin{picture}(14,7.8)
\put(-4.2,-13.5){\includegraphics{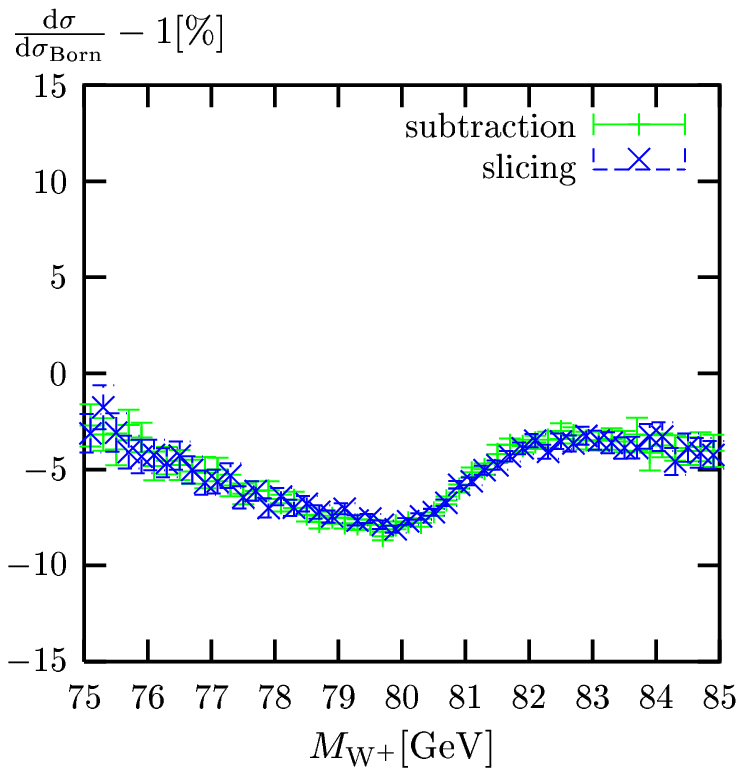}}
\put( 2.8,-13.5){\includegraphics{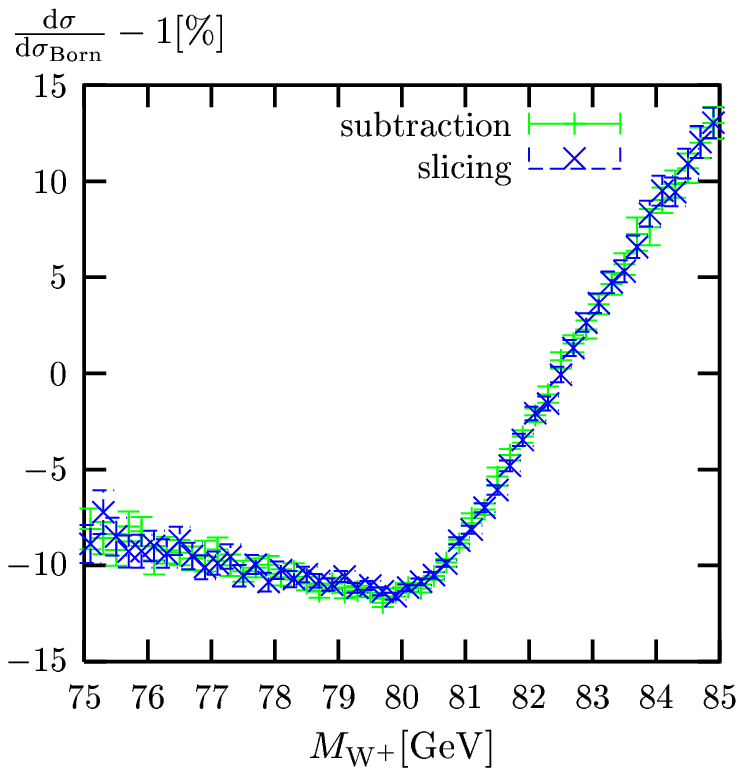}}
\put( 1.4,5.0){\footnotesize $M_\recomb=5\GeV$}
\put( 8.4,5.0){\footnotesize $M_\recomb=25\GeV$}
\end{picture} }}
\caption[]{Relative corrections to the 
  $\protect\PWp$ invariant-mass distributions for the {\bare} (l.h.s.)
  and {\calo} (r.h.s.) setup for $\Pep\Pem\to \Pu \Pdbar \mu^-
  \bar\nu_{\mu}$ at $\protect\sqrt{s}=200\,\protect\GeV$}
\label{fi:mwpsl-sub}
\vspace*{2em}
{\centerline{
\setlength{\unitlength}{1cm}
\begin{picture}(14,7.8)
\put(-4.2,-13.5){\includegraphics{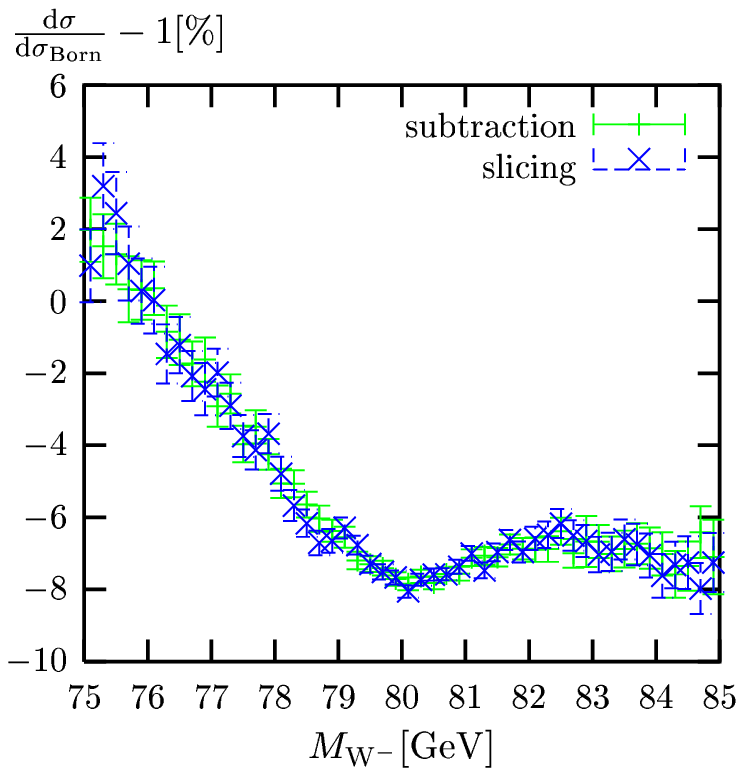}}
\put( 2.8,-13.5){\includegraphics{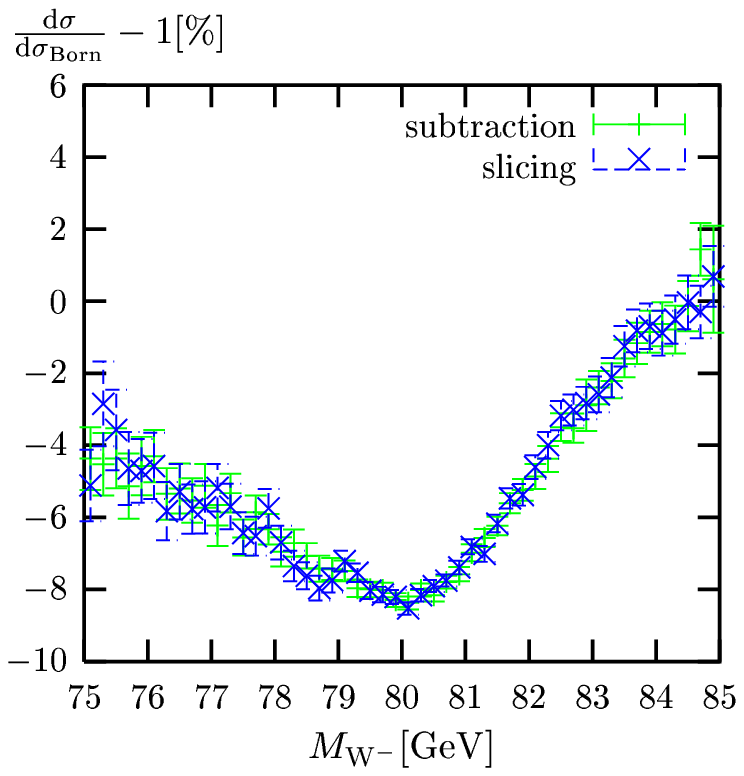}}
\put( 1.4,5.0){\footnotesize $M_\recomb=5\GeV$}
\put( 8.4,5.0){\footnotesize $M_\recomb=25\GeV$}
\end{picture} }}
\caption[]{Relative corrections to the 
$\protect\PWm$ invariant-mass distributions for the {\bare}
(l.h.s.) and {\calo} (r.h.s.) setup at 
for $\Pep\Pem\to \Pu \Pdbar \mu^- \bar\nu_{\mu}$ at
$\protect\sqrt{s}=200\,\protect\GeV$}
\label{fi:mwmsl-sub}
\efi
The invariant masses are obtained from the four-momenta of the decay
fermions of the W bosons after eventual recombination with the photon
four-momentum. They are particularly sensitive to the treatment of the
photons, and thus a strong dependence of the relative corrections on
the recombination cut can be observed. A detailed discussion of the
distortion of the invariant-mass distributions of the W bosons and
their origin has been given in \citere{de99a}.  The distributions
obtained from the phase-space-slicing and subtraction methods are
compatible with each other. The integration errors are larger further
away from the W resonance because of the decreasing statistics.

In \reffis{fi:costh_Wp-sl-sub}--\ref{fi:azimuthal_W+_angle} on the
l.h.s.\
we always show the distributions in lowest order and
including the radiative corrections for the recombination cut
$M_\recomb=25\GeV$, and on the r.h.s.\
the corresponding
relative corrections for both the subtraction and phase-space-slicing
methods.

We define all angles in the laboratory system, which is the CM system
of the initial state.  The distributions in the cosine of the $\PWp$
production angle $\theta_{\PWp}$ and of the decay angle $\theta_{\PWp
  \bar\Pd}$, together with the corresponding relative corrections, are
shown in \reffis{fi:costh_Wp-sl-sub} and \ref{fi:costh_Wpd-sl-sub},
respectively.
\begin{figure}
{\centerline{
\setlength{\unitlength}{1cm}
\begin{picture}(14,7.8)
\put(-4.2,-13.5){\includegraphics{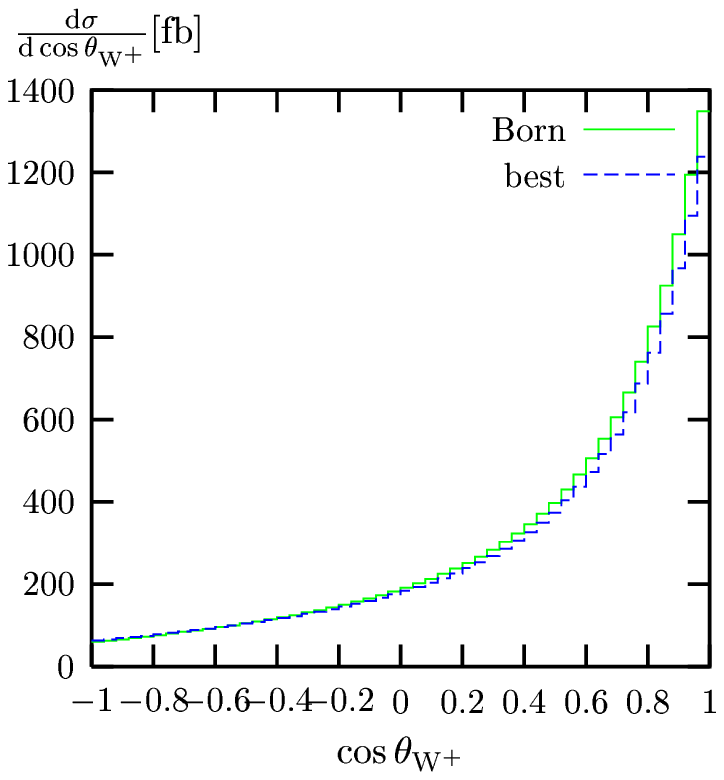}}
\put( 2.8,-13.5){\includegraphics{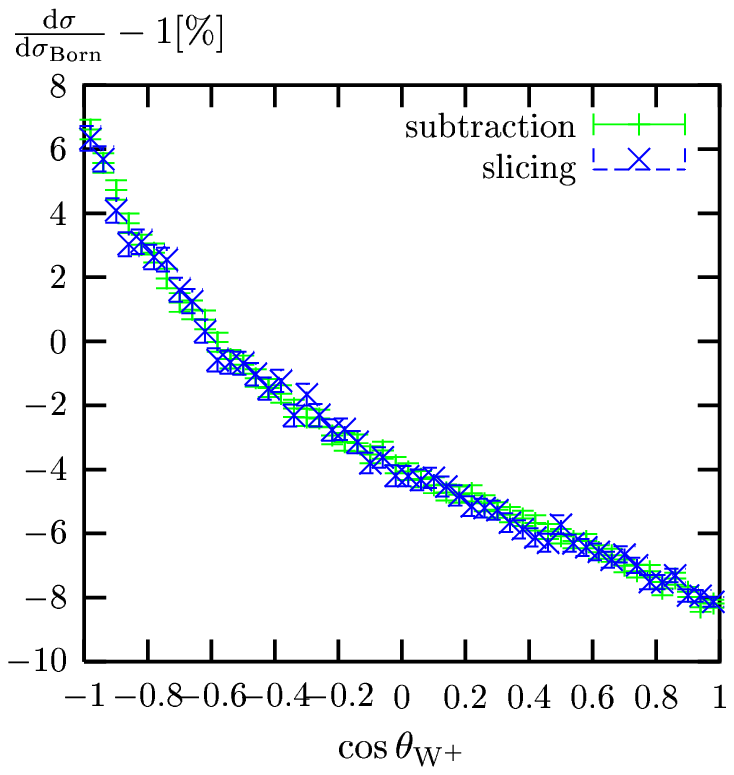}}
\end{picture} }}
\caption[]{Distributions in the cosine of the $\protect\PWp$ production angle
  with respect to the $\protect\Pep$ beam (l.h.s.) and the relative corrections
  (r.h.s.) for $\Pep\Pem\to \Pu \Pdbar \mu^- \bar\nu_{\mu}$
  at $\protect\sqrt{s}=200\,\protect\GeV$}
\label{fi:costh_Wp-sl-sub}
\vspace*{2em}
{\centerline{
\setlength{\unitlength}{1cm}
\begin{picture}(14,7.8)
\put(-4.2,-13.5){\includegraphics{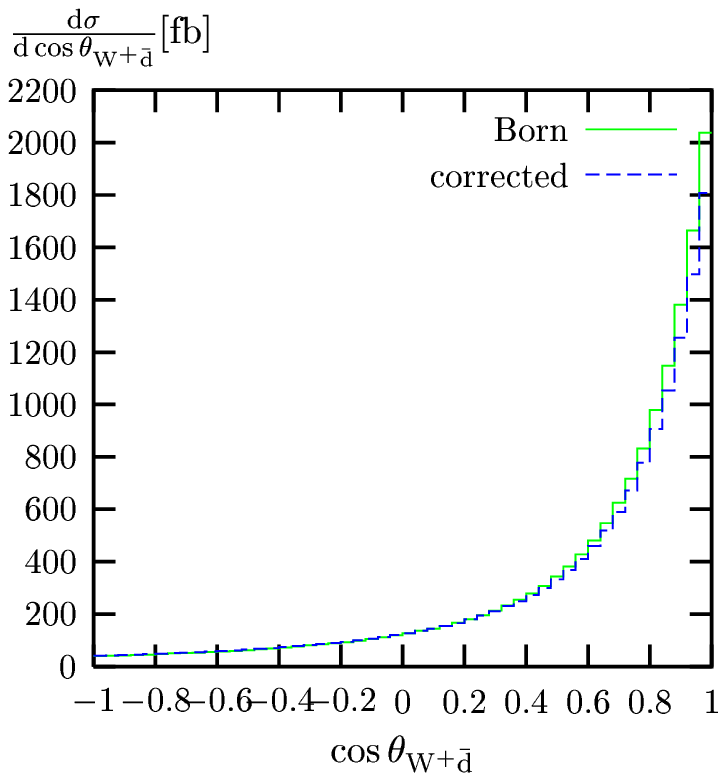}}
\put( 2.8,-13.5){\includegraphics{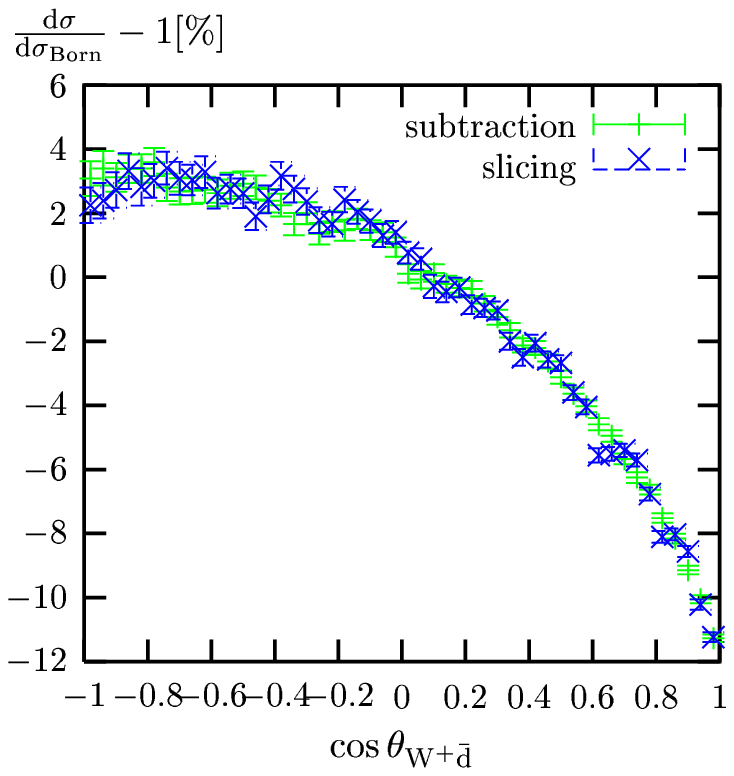}}
\end{picture} }}
\caption[]{Distributions in the cosine of the $\protect\Pdbar$ decay angle
  with respect to the $\protect\PWp$ direction (l.h.s.) and
  the relative corrections (r.h.s.) 
  for $\Pep\Pem\to \Pu \Pdbar \mu^- \bar\nu_{\mu}$
  at $\protect\sqrt{s}=200\protect\GeV$}
\label{fi:costh_Wpd-sl-sub}
\efi
In both cases the corrections depend strongly on the angle.
The distributions obtained with the subtraction 
and the phase-space-slicing branch of {\sc RacoonWW} agree 
with each other within 
integration errors, \ie in general to better than $0.5\%$.  The
increase of the integration errors for large decay angles is due to
the smallness of the corresponding cross section.

Finally, we consider distributions in azimuthal angles.
In \reffi{fi:azimuthal_W+_angle} we show the
distributions over the azimuthal decay angle $\phi_{\PWp}$ of the $\PWp$
boson,
\ie the angle between the decay plane of the $\PWp$ and the plane of 
\PW-pair production,
\beq
\cos\phi_{\PWp} = \frac{(\bk_+\times \bp_+)(\bk_+\times \bk_1)}
{|\bk_+\times \bp_+||\bk_+\times \bk_1|},\qquad
\sgn(\sin\phi_{\PWp}) = \sgn\left\{\bk_+\cdot
\left[(\bk_+\times \bp_+)\times(\bk_+\times \bk_1)\right]\right\}.
\eeq
In this distribution the corrections 
vary only weakly
with the angle.

The distributions over the angle $\phi$ between the two planes spanned
by the momenta of the two fermion pairs in which the \PW~bosons decay, \ie
(note that $\bk_+=-\bk_-$ for non-photonic events)
\beq
\cos\phi = \frac{(\bk_+\times \bk_1)(-\bk_-\times \bk_3)}
{|\bk_+\times \bk_1||{-\bk_-}\times \bk_3|},\qquad
\sgn(\sin\phi) = \sgn\left\{\bk_+\cdot
\left[(\bk_+\times \bk_1)\times(-\bk_-\times \bk_3)\right]\right\},
\eeq
are presented in \reffi{fi:azimuthal_angle}.
\begin{figure}
{\centerline{
\setlength{\unitlength}{1cm}
\begin{picture}(14,7.8)
\put(-4.2,-13.5){\includegraphics{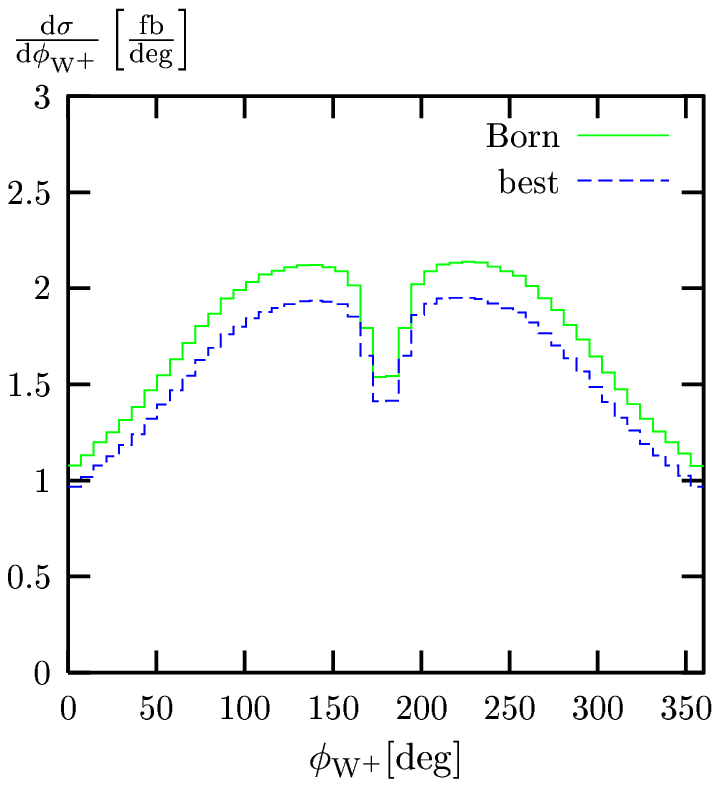}}
\put( 2.8,-13.5){\includegraphics{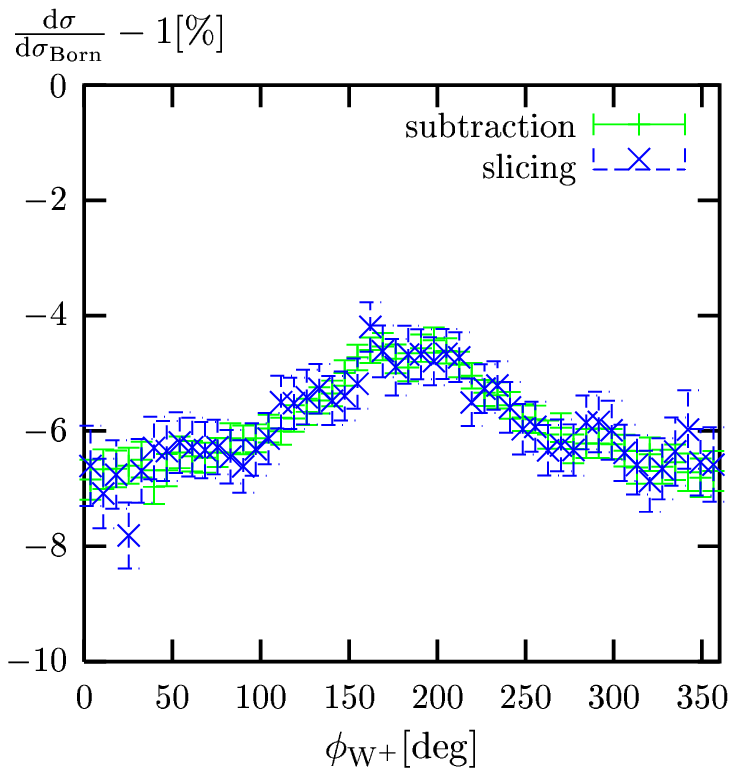}}
\end{picture} }}
\caption{Distributions in the azimuthal $\PWp$ angle
  (l.h.s.) and the relative corrections (r.h.s.) for $\Pep\Pem\to \Pu
  \Pdbar \mu^- \bar\nu_{\mu}$ at $\protect\sqrt{s}=200\GeV$}
\label{fi:azimuthal_W+_angle}
\vspace*{2em}
{\centerline{
\setlength{\unitlength}{1cm}
\begin{picture}(14,7.8)
\put(-4.2,-13.5){\includegraphics{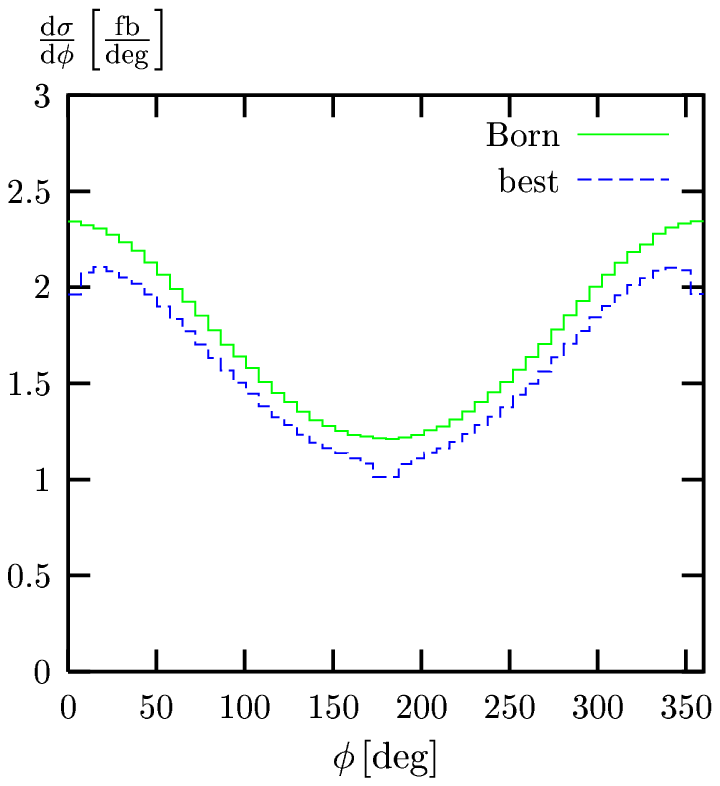}}
\put( 2.8,-13.5){\includegraphics{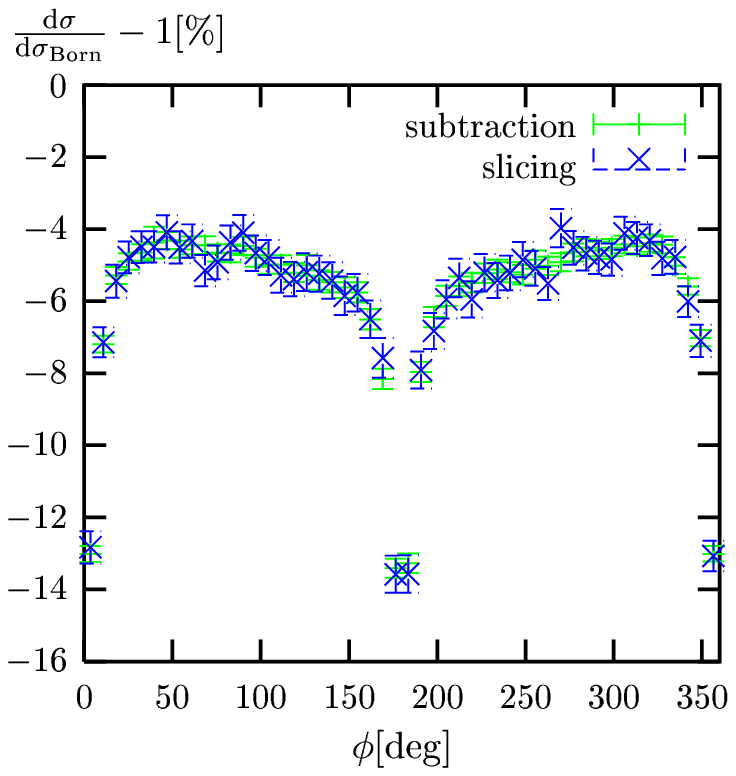}}
\end{picture} }}
\caption{Distributions in the azimuthal $\phi$ angle 
  (l.h.s.) and the relative corrections (r.h.s.) for $\Pep\Pem\to \Pu
  \Pdbar \mu^- \bar\nu_{\mu}$ at $\protect\sqrt{s}=200\GeV$}
\label{fi:azimuthal_angle}
\end{figure}
The corrections are about $-5\%$ except for angles $\phi$ about $0^\circ$
or $180^\circ$, i.e.\ if the two decay planes coincide.
This can be understood as follows:
for non-photonic events the angle $\phi$ is related to the azimuthal decay
angles $\phi_{\PWpm}$ by $\phi_{\PWp}\pm\phi_{\PWm}$ plus some constant
(depending on the precise definition of $\phi_{\PWm}$).
Since the virtual corrections depend on $\phi_{\PWpm}$ only via
$\sin\phi_{\PWpm}$ and $\cos\phi_{\PWpm}$, which are rather smooth
functions, the peaks for $\phi=0^\circ$ and $\phi =180^\circ$
cannot originate from the virtual corrections.
If, however, a hard photon is emitted, the
finite angle between $\bk_+$ and $-\bk_-$ leads to a finite angle
between the two decay planes except for the rare situation where
$\bk_1$ and $\bk_3$ are in the plane spanned by $\bk_+$ and $-\bk_-$. 
Thus, for hard photonic events
$\phi=0^\circ$ and $\phi =180^\circ$ are suppressed, and the large
negative corrections at these angles are remnants of the virtual
counterparts of the collinear singularities of the ISR, which are only
partially compensated by real photon emission. This
interpretation is supported by the numerical observation that the peaks
are mainly contained in the convolution over $x$ in 
$\rd\sigma_{\virt+\real,\sing}^{\eeffff}$.
If we had used $\bk_+$ instead
of $-\bk_-$ in the definition of $\phi$, these large corrections would
be absent.

Theoretically the shown azimuthal-angle distributions
are of particular interest,
since they are sensitive to the imaginary parts of the loop corrections.
As can be deduced from App.~A of \citere{Be98}, the contributions of
the imaginary parts of the one-loop corrections always involve a
factor $\sin\phi_{\PWp}$ or
 a factor $\sin\phi_{\PWm}$ together
with symmetric functions in these angles. Consequently, they average
to zero if the azimuthal angles of the decay fermions are integrated
over. This is obviously the case for all quantities discussed so far
except for the $\phi$ and $\phi_{\PWp}$ distributions. While we do not
find significant effects of imaginary loop parts in the $\phi$
distribution, we find a small impact on the $\phi_{\PWp}$
distribution.  In \reffi{fi:azimuthal_W+_angle_imag} we compare the
relative corrections to the $\phi_{\PWp}$ distribution with (``imag'')
and without (``def'') imaginary parts.
\begin{figure}
{\centerline{
\setlength{\unitlength}{1cm}
\begin{picture}(7,7.8)
\put(-4.2,-13.5){\includegraphics{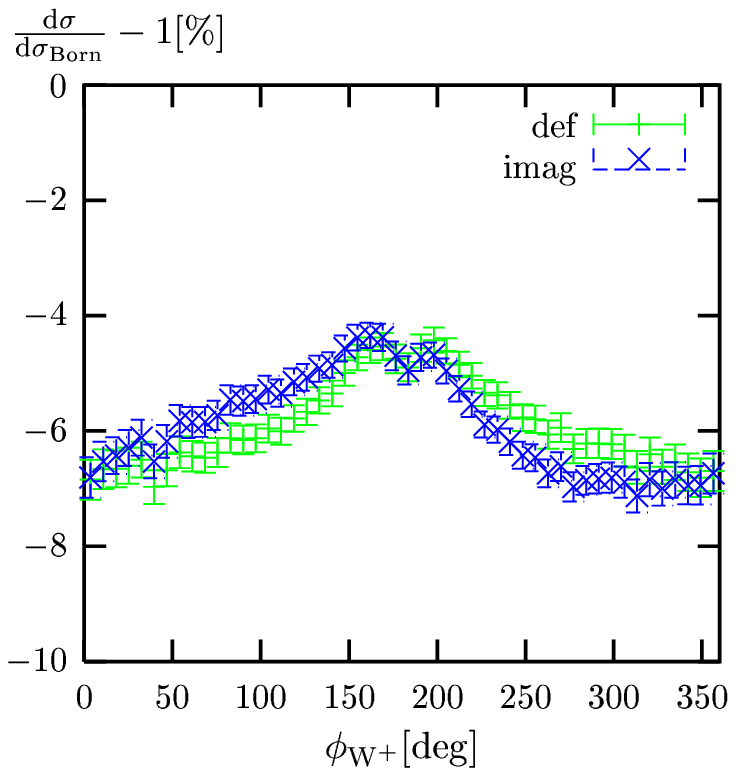}}
\end{picture} }}
\caption{Relative corrections to the distribution in the azimuthal
  angle $\phi_\PWp$ 
  with (``imag'') and without (``def'') imaginary parts of the loop
  integrals
  for $\Pep\Pem\to \Pu \Pdbar \mu^- \bar\nu_{\mu}$
  at $\protect\sqrt{s}=200\GeV$}
\label{fi:azimuthal_W+_angle_imag}
\end{figure}

\subsection{Discussion of intrinsic ambiguities and of other options}
\label{se:amb}

In order to investigate the accuracy of the DPA quantitatively, we
have modified the implementation of the DPA and compared the obtained
results. The differences should be of the order of
$\alpha\Gamma_\PW/\MW$.  Recall that in {\sc RacoonWW} only the finite
(\ie non-mass-singular)
virtual corrections are treated in DPA, while real photon emission is
based on the full $\eeffffg$ matrix element with the exact
five-particle phase space.  Thus, only finite
virtual corrections are affected by the following modifications.
Specifically, we consider four types of options:
\begin{itemize}
\item Different on-shell projections: \\
As explained in \refse{se:onshellprojection},
one has to specify a projection of the physical momenta to a set of 
momenta for on-shell W-pair production and decay%
\footnote{This option only illustrates the effect of different
  on-shell projections in the four-particle phase space; if real
  photonic corrections were treated in DPA the impact of different
  projections could be larger.}, in order to define a DPA.  This can
be done in an obvious way by fixing the direction of one of the
\PW~bosons and of one of the final-state fermions originating from
either \PW~boson in the CM frame of the incoming $\Pep\Pem$ pair. The
default in {\sc RacoonWW}, which is explicitly specified in
\refapp{app:onshell}, is to fix the directions of the momenta of the
fermions (not of the anti-fermions) resulting from the $\PWp$ and
$\PWm$ decays (``def''). A different projection is obtained by fixing
the direction of the anti-fermion from the $\PWp$ decay (``proj'')
instead of the fermion direction.
\item Different definitions of finite virtual corrections:\\
  In \refse{se:sing} the matching of soft and collinear
  singularities between virtual and real corrections has been
  explained in detail.  Recall that the redistribution of these
  singular parts fixes only the universal, singular parts, while the
  redistribution of non-singular parts is a mere convention. Owing to
  the asymmetric treatment of the corrections
  (finite virtual in DPA, real and mass-singular virtual from full
  matrix elements), different redistributions of non-singular
  contributions change the result by terms of the order
  $(\al/\pi)(\GW/\MW)$, which is beyond the accuracy of the DPA.
  As discussed in \refse{se:sing}, {\sc RacoonWW} allows to choose
  between two possible definitions of the finite part of the virtual
  photonic corrections to which the DPA is applied.  As default,
  the subtraction-method-inspired approach \refeq{eq:virt_sing_sub} is
  chosen (``def'') and compared to the YFS-inspired approach (``eik'')
  \refeq{eq:virt_sing_yfs}.  The two treatments differ in the finite
  parts of the contribution $\rd\sigma_{\virt,\sing}^{\eeffff}$ (see
  \refse{se:YFS}), which is subtracted from the virtual corrections in
  DPA and added in its exact form to the singular part of the real
  photon corrections.
  As can be seen from the difference $\Delta$ of
  \refeq{eq:diff_sl_sub},
  the ambiguity originates from terms of the
  form $(\alpha/\pi)\times\pi^2\times {\cal O}(1)$ which are either
  multiplied with the DPA or with the full off-shell Born cross
  sections.
\item On-shell versus off-shell Coulomb singularity: \\
  The Coulomb singularity is (up to higher orders) fully contained in
  the virtual ${\cal O}(\alpha)$ correction.  Performing the on-shell
  projection of the DPA to the full virtual correction leads to the
  on-shell Coulomb singularity, which is a simple factor of
  $\alpha\pi/(2\beta)$.  However, since the Coulomb singularity is an
  important correction in the LEP2 energy range and is also known
  beyond DPA \cite{coul}, {\sc RacoonWW} includes these extra off-shell
  parts of the
  Coulomb correction as default.  This replacement of the Coulomb
  singularity is performed by adding and subtracting the corresponding
  contributions in the virtual non-factorizable corrections, as
  described in \citere{nfc1a}.  Switching the extra off-shell parts of
  the Coulomb correction off (``Coul''), yields an effect of the order
  of the uncertainty of the DPA.  Note that for CM energies close to
  the W-pair-production threshold, the 
  on-shell Coulomb singularity is
  not adequate, and the difference between on-shell and off-shell
  Coulomb singularity cannot be viewed as a measure of the theoretical
  uncertainty.

\item Imaginary parts of virtual corrections:\\
  As default the imaginary parts of the loop integrals are neglected
  in {\sc RacoonWW}. However, {\sc RacoonWW} contains an option
  (``imag'') that takes into account the imaginary parts.
  Since all contributions of these imaginary parts are proportional to
  the sine of the azimuthal decay angle of one 
  of the W~bosons and
  otherwise involve only symmetric functions of these angles, the
  imaginary parts drop out in distributions where these decay angles
  are integrated over. The numerical check of this cancellations can
  serve as a consistency check of {\sc RacoonWW}.
  Moreover, this option might be useful for observables where the
  imaginary parts do not drop out. 
\end{itemize}

In the following table and figures the total cross section and various
distributions to $\Pep\Pem\to\Pu\bar\Pd\mu^-\bar\nu_\mu(\gamma)$ are
compared for the different versions of the DPA defined above in the
calo setup, \ie with the loose recombination cut 
$M_\recomb=25\GeV$. 
We consider the CM energies $172\GeV$,  $200\GeV$, and  $500\GeV$.
The used lowest-order cross section is based on the full $4f$ matrix
element.  ``Naive'' QCD correction factors and ISR corrections beyond
${\cal O}(\alpha)$ are not included in the results of this section.

The results for the total cross section are shown in
\refta{tab:sigma_DPA_unc}.
\begin{table}\centerline{        
\begin{tabular}{|c||c|c|c|c|}
\hline  
                     & def        & proj       & eik        & Coul \\
\hline \hline    
& \multicolumn{4}{c|}{$\sqrt{s}=172\GeV$} \\ \hline
$\sigma/\mathrm{pb}$ & 400.39(22) & 400.27(22) & 400.03(22) & 403.54(22) 
\\ \hline                                                        
$\delta/\%$          & 0          & $-0.03$    & $-0.09$    & 0.79
\\ \hline
& \multicolumn{4}{c|}{$\sqrt{s}=200\GeV$} \\ \hline
$\sigma/\mathrm{pb}$ & 570.10(37) & 569.93(37) & 570.04(37) & 570.85(37) 
\\ \hline                                                      
$\delta/\%$          & 0          & $-0.03$    & $-0.01$    & 0.13  
\\ \hline 
& \multicolumn{4}{c|}{$\sqrt{s}=500\GeV$} \\ \hline 
$\sigma/\mathrm{pb}$ & 190.30(20) & 190.28(20) & 190.45(20) & 190.31(20) 
\\ \hline
$\delta/\%$          & 0          & $-0.01$ & \phantom{-}0.08 & 0.01 
\\ \hline                                           
\end{tabular}
}  
\caption{{\sc RacoonWW} predictions for the total cross section to
$\Pep\Pem\to\Pu\bar\Pd\mu^-\bar\nu_\mu(\gamma)$ for the calo setup 
at different CM energies
in various versions of the DPA and the relative differences 
$\delta=\sigma/\sigma_{\mathrm{def}}-1$}
\label{tab:sigma_DPA_unc}
\end{table}                                                  
We find relative differences at the level of $0.1\%$ As expected, the
prediction that is based on the on-shell Coulomb correction is
somewhat higher than the exact off-shell treatment, since off-shell
effects screen the positive Coulomb singularity.  As mentioned above,
for the low CM energy of $172\GeV$ the difference between on-shell and
off-shell Coulomb singularity, which is quite large (0.79\%), cannot
be viewed as a measure of the theoretical uncertainty.  While the
difference between the two on-shell projections is below 0.03\%, the
effect of including different finite terms in the DPA (``eik'') is at
the level of 0.1\% in the considered energy range.
When switching on the imaginary parts in the corrections we find
the same results as without imaginary parts (``def'').

\begin{figure}
{\centerline{
\setlength{\unitlength}{1cm}
\begin{picture}(14,7.8)
\put(-4.2,-13.5){\includegraphics{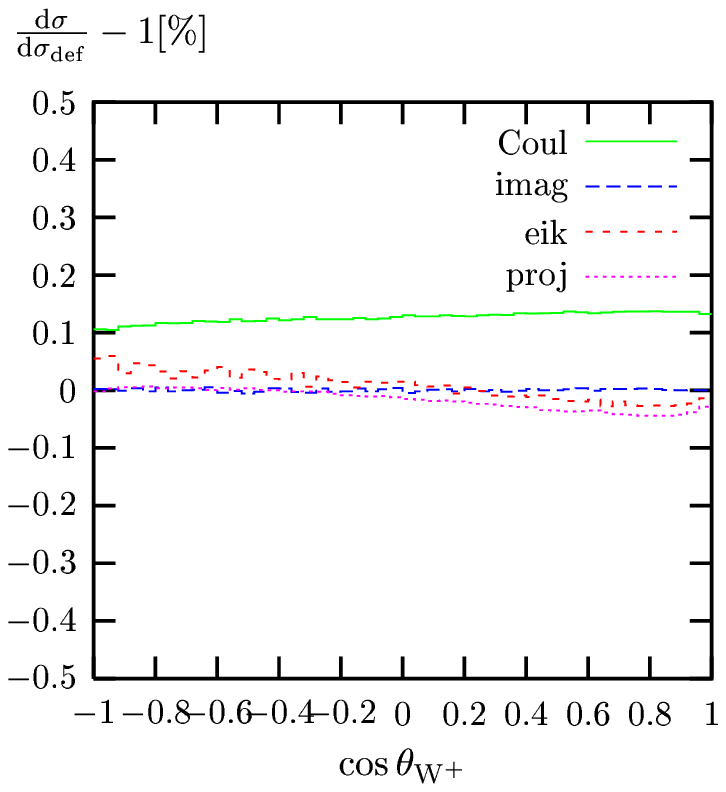}}
\put( 2.8,-13.5){\includegraphics{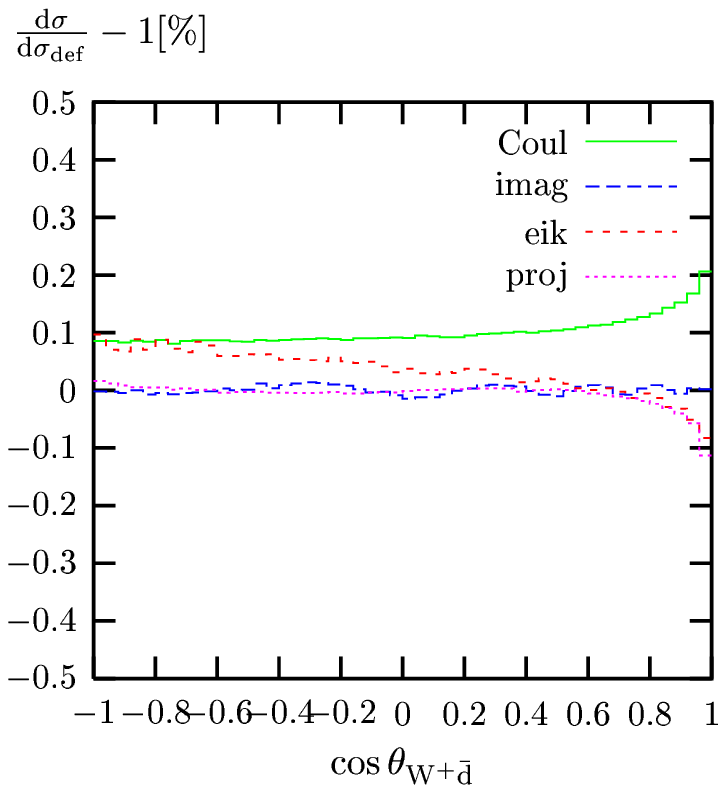}}
\end{picture} }}
\caption{Differences between different versions of the DPA
  for distributions in the cosines of the production $\theta_{\PWp}$
  (l.h.s.) and decay $\theta_{\PWp \Pdbar}$ (r.h.s.)  angles for
  $\Pep\Pem\to\Pu\bar\Pd\mu^-\bar\nu_\mu(\gamma)$ at
  $\sqrt{s}=200\GeV$, as described in the text}
\label{fi:tu_thwp_thwpmu}
\vspace*{2em}
{\centerline{
\setlength{\unitlength}{1cm}
\begin{picture}(14,7.8)
\put(-4.2,-13.5){\includegraphics{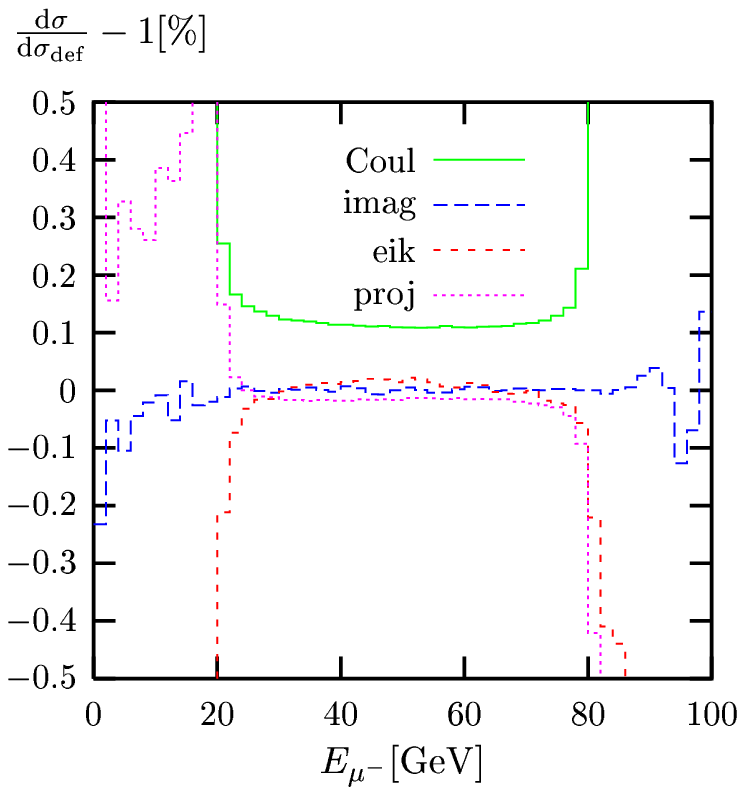}}
\put( 2.8,-13.5){\includegraphics{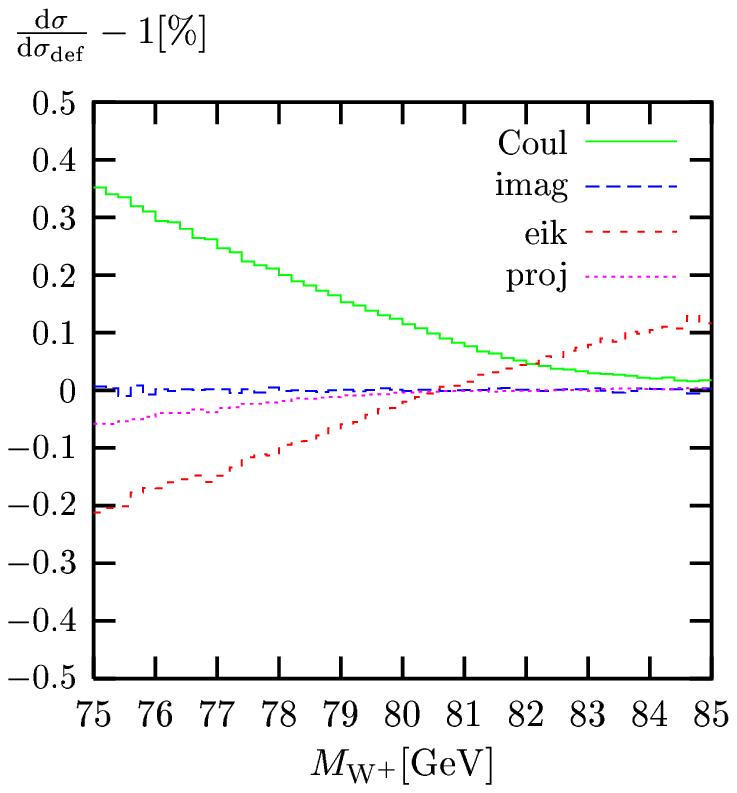}}
\end{picture} }}
\caption{Differences between different versions of the DPA
  for distributions in the $\mu$ energy (l.h.s.) and in the
  $\Pu\bar\Pd$ invariant mass (r.h.s.) for
  $\Pep\Pem\to\Pu\bar\Pd\mu^-\bar\nu_\mu(\gamma)$ at
  $\sqrt{s}=200\GeV$, as described in the text}
\label{fi:tu_Emu_mm}
\efi
\begin{figure}
{\centerline{
\setlength{\unitlength}{1cm}
\begin{picture}(14,7.8)
\put(-4.2,-13.5){\includegraphics{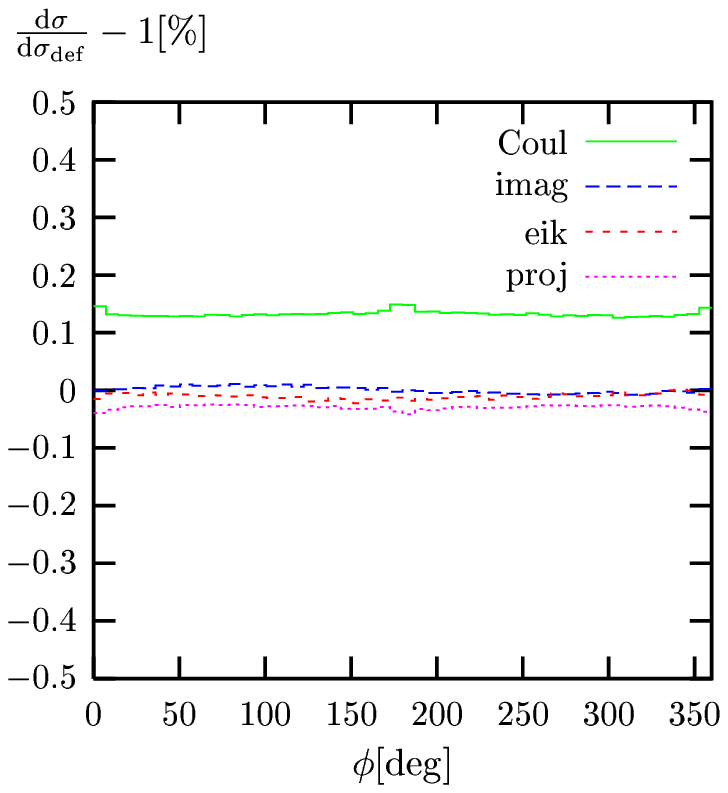}}
\put( 2.8,-13.5){\includegraphics{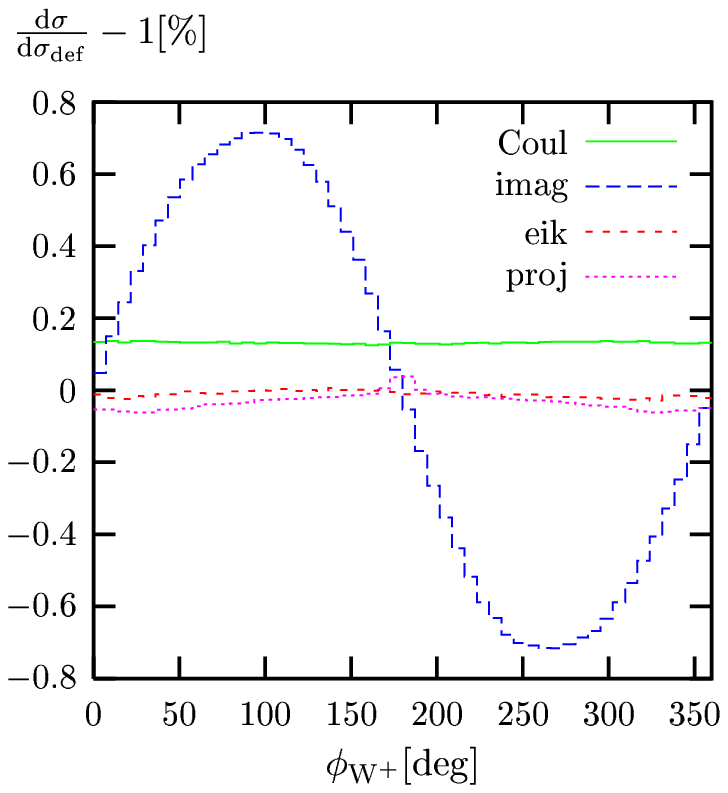}}
\end{picture} }}
\caption{Differences between different versions of the DPA
  for distributions in the azimuthal angles $\phi$ (l.h.s.) and
  $\phi_{\PWp}$ (r.h.s.) for
  $\Pep\Pem\to\Pu\bar\Pd\mu^-\bar\nu_\mu(\gamma)$ at
  $\sqrt{s}=200\GeV$, as described in the text}
\label{fi:tu_phi}
\efi
In \reffis{fi:tu_thwp_thwpmu}, \ref{fi:tu_Emu_mm}, and \ref{fi:tu_phi}
we show the differences of the ``proj'', ``eik'', ``Coul'', and
``imag'' modifications to the default version of the DPA for some
distributions at $\sqrt{s}=200\GeV$. For the distribution in the
cosines of the W-production angle $\theta_{\PW^+}$ and in the W-decay
angle $\theta_{\PW^+\bar\Pd}$ (\reffi{fi:tu_thwp_thwpmu}) the relative
differences are of the order of $0.1\%$ for all angles, which is of
the expected order for the intrinsic DPA uncertainty. For the
$\mu$-energy distribution, shown 
on the l.h.s.\ of
\reffi{fi:tu_Emu_mm}, the differences are typically of the same order,
as long as $E_\mu$ is in the range for W-pair production, which is
$20.2\GeV<E_\mu<79.8\GeV$ at $\sqrt{s}=200\GeV$.  Outside this region,
the four-fermion process is not dominated by the W-pair diagrams, and
the DPA is not reliable anymore, which is also indicated by large
intrinsic ambiguities. The r.h.s.\ of \reffi{fi:tu_Emu_mm} shows the
DPA ambiguities for the $\Pu\bar\Pd$ invariant-mass distribution.
Within a window of $2\Gamma_\PW$ around the W~resonance the relative
differences between the considered modifications are also at the level
of $0.1$--$0.3\%$. The differences grow with the distance from the
resonance point.  As expected, we find that the contribution of the
imaginary parts to the production-angle
and decay-angle distributions and to the invariant-mass distributions
are compatible with zero.

Finally, we show in \reffi{fi:tu_phi} the difference between different
versions of the DPA for the distributions in the azimuthal angle
between the decay planes of the two $\PW$ bosons and in the azimuthal
decay angle of the $\PWp$ boson.  The ambiguities resulting from
``proj'', ``eik'', and ``Coul'' are of the same size as for the
production- and decay-angle distributions considered above.
While for the $\phi$ distribution no significant impact of the
imaginary parts is visible, the effect influences the $\phi_{\PWp}$
distribution by a relative contribution that is roughly given by
$0.7\%\times\sin\phi_{\PWp}$.

The energy dependence of the ambiguities in the distributions
reflects the one of the total cross section, in particular if the
deviations are flat as for the production-angle
distribution.
\begin{figure}
{\centerline{
\setlength{\unitlength}{1cm}
\begin{picture}(14,7.8)
\put(-4.2,-13.5){\includegraphics{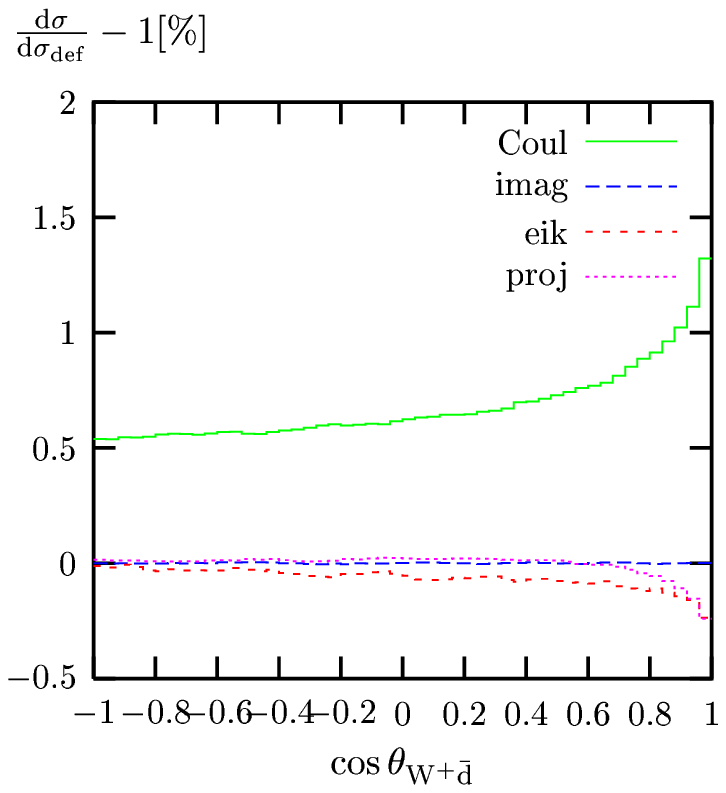}}
\put( 2.8,-13.5){\includegraphics{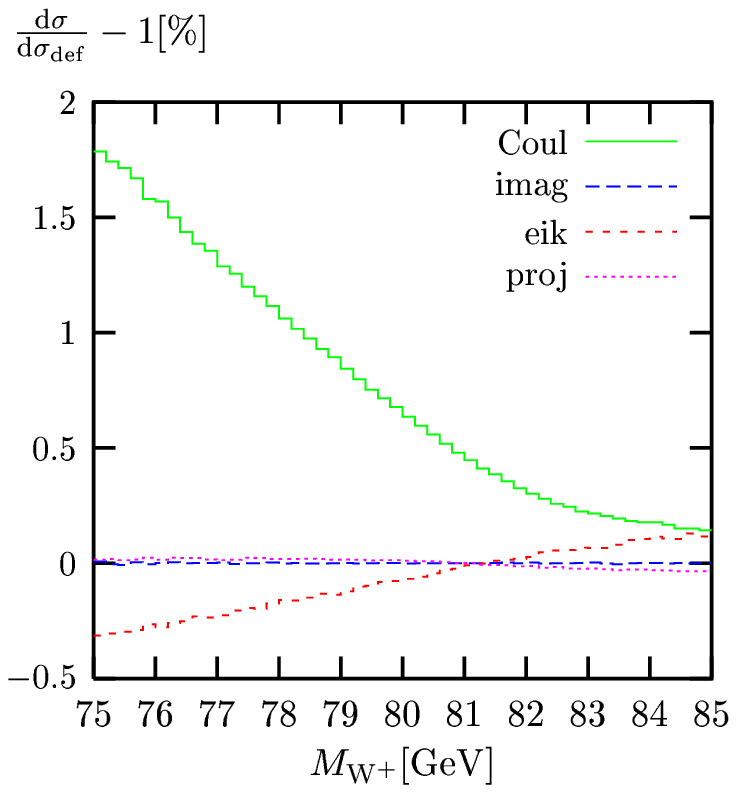}}
\end{picture} }}
\caption{Differences between different versions of the DPA 
  for distributions in the cosine of the decay angle $\theta_{\PWp
    \Pdbar}$ (l.h.s.)~and in the $\Pu\bar\Pd$ invariant mass
  (r.h.s.)~for $\Pep\Pem\to\Pu\bar\Pd\mu^-\bar\nu_\mu(\gamma)$ at
  $\sqrt{s}=172\GeV$, as described in the text}
\label{fi:tu_thwp_mp_172}
\vspace*{2em}
{\centerline{
\setlength{\unitlength}{1cm}
\begin{picture}(14,7.8)
\put(-4.2,-13.5){\includegraphics{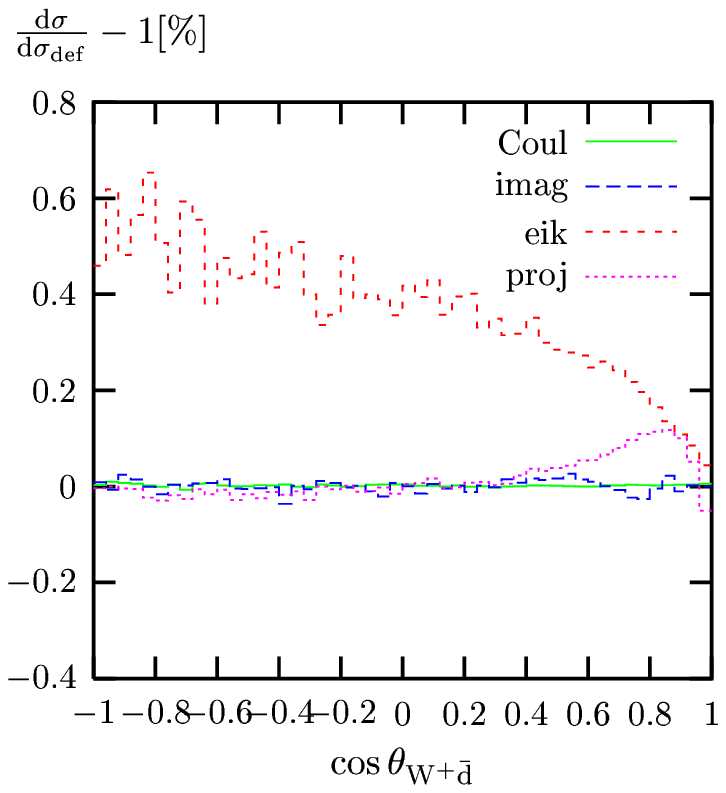}}
\put( 2.8,-13.5){\includegraphics{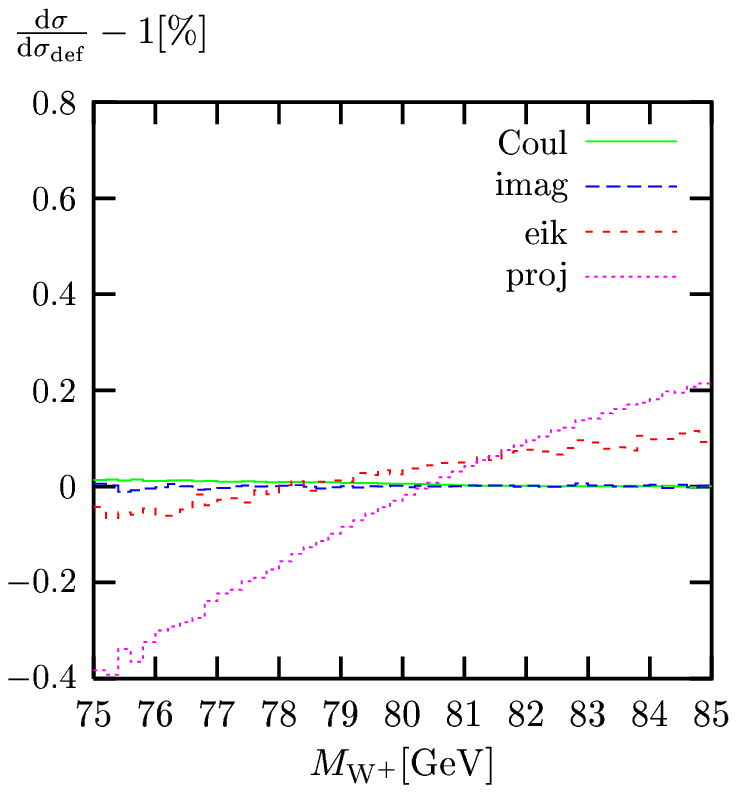}}
\end{picture} }}
\caption{Differences between different versions of the DPA
  for distributions in the cosine of the decay angle $\theta_{\PWp
    \Pdbar}$ (l.h.s.) and in the $\Pu\bar\Pd$ invariant mass (r.h.s.)
  for $\Pep\Pem\to\Pu\bar\Pd\mu^-\bar\nu_\mu(\gamma)$ at
  $\sqrt{s}=500\GeV$, as described in the text}
\label{fi:tu_thwp_mp_500}
\efi
The differences for 
the distributions in the W-decay angle $\theta_{\PW^+\bar\Pd}$ and in
the $\Pu\bar\Pd$ invariant mass, which are not flat, are illustrated
in \reffis{fi:tu_thwp_mp_172} and \ref{fi:tu_thwp_mp_500} at
$\sqrt{s}=172\GeV$ and $500\GeV$, respectively.  As already stated
above, for the low CM energy of $\sqrt{s}=172\GeV$ the difference
between on-shell and off-shell Coulomb singularity cannot be viewed as
an ambiguity.  On the other hand, the effect of the Coulomb
singularity is completely negligible for $\sqrt{s}=500\GeV$.  Note
also that the largest ambiguities, which can become of the order of
0.5\%, appear when the cross section becomes small. 
The large differences between the two different on-shell projections 
in the $\theta_{\PWp \Pdbar}$ distribution at $500\GeV$ are 
due to the increasing size and the larger angular variations of the 
relative corrections at higher energies. This uncertainty could be
reduced by treating the dominant corrections at high energies, which are
due to Sudakov-type corrections of the form
$(\al/\pi)\ln^2(s/\MW^2)$ \cite{be93}, not in DPA but in connection
with the full $\eeffff$ cross sections.

The discussed results illustrate that the intrinsic ambiguities of the
DPA, as applied in {\sc RacoonWW}, are at the level of a few per mil,
whenever resonant W-pair production dominates the considered observable.

\subsection{Comparison with existing results}
\label{se:comp}

In \citere{lep2mcws} we performed a detailed, tuned comparison of {\sc
  RacoonWW} with a semi-analytical calculation of the complete ${\cal
  O}(\alpha)$ corrections to $\Pep\Pem\to\PW\PW\to 4\,$leptons in DPA
by Beenakker, Berends and Chapovsky \cite{Be98} (called BBC in the
following) and with the Monte Carlo generator {\sc YFSWW3}
\cite{ja97,ja99}.  Here we restrict ourselves to a brief summary of
the outcome and refer to \citere{lep2mcws} for more details.
Moreover, we present a brief comparison of the real-photonic
corrections obtained by Jegerlehner and Ko\l odziej \cite{je99} (JK)
based on an exact matrix-element calculation.

\subsubsection{Comparison with Berends, Beenakker and Chapovsky}

The numerical comparison with BBC has been done for the purely
leptonic channel $\Pep\Pem\to\nu_\mu\mu^+\tau^-\bar\nu_\tau$.
The conceptual differences between {\sc RacoonWW} and the BBC approach
are as follows: BBC treat both the virtual and real corrections in
DPA, while the complete real bremsstrahlung corrections and the
universal leading-logarithmic part of the $\O(\al)$ ISR correction are
based on the exact CC11 matrix elements for $\eeffffg$ and $\eeffff$
in {\sc RacoonWW}.
The virtual corrections are treated in DPA in both programs, but {\sc
  RacoonWW} employs the full off-shell phase space with an appropriate
on-shell projection, whereas BBC have chosen the on-shell phase space
with decoupled invariant masses of the W~bosons. These differences are
(at least formally) beyond the accuracy of the DPA. Thus, if the
complete photonic phase space is integrated over and virtual
corrections are added, both approaches yield results in DPA accuracy.

For the total cross section, the differences between the two
approaches should be of the naively expected accuracy of the DPA of
0.5\% or better, whenever the DPA is applicable. 
We found that both calculations agree very well above $185\GeV$.
Below this energy the differences in the implementation of the DPA
become visible. The main effect originates probably from the 
different
treatment
of the $\O(\al)$ ISR and the phase space. 
While the uncertainty arising from the DPA is enhanced by leading ISR
logarithms in the BBC approach, this enhancement is absent in {\sc
  RacoonWW}.

For distributions unavoidable differences arise from the definition of
the phase-space variables in the presence of photon recombination.
When defining the momenta of the $\PW$~bosons for angular
distributions, BBC choose to assign the photon to one of the
production/decay subprocesses.  
The angles are then determined from the resulting $\PW$ momenta
and the original fermion momenta. In {\sc RacoonWW}, all angles are
defined from the fermion momenta after eventual photon recombination.
The two different angle definitions lead to a redistribution of events
in the angular distributions, which arises, in particular, from hard
photon emission.
For the considered W-production-angle and W-decay-angle distributions 
the differences between BBC and {\sc RacoonWW} turned out to be of the
order of 1\% at a typical LEP2 energy, with a tendency to increase
with increasing energy.

Invariant-mass distributions of the W bosons depend crucially on the
treatment of real photons, which is done in a fundamentally different
way in {\sc RacoonWW} compared to the treatment of BBC. Therefore, it
does not make sense to compare these distributions between the two
programs. Specifically, BBC define the W~invariant masses from the
fermion momenta only, i.e.\ without photon recombination at all. The
resulting shifts in the maxima of the distributions are negative, up
to $-77\MeV$ for $\Pe^\pm$ in the final state for $\sqrt{s}=184\GeV$,
and dominated by mass singularities \cite{Be98}.  On the other hand,
studies with {\sc RacoonWW} show that these shifts are in general
positive and of the order of several $10\MeV$ \cite{de99a,de99b} if
photon recombination is taken into account.

\begin{sloppypar}
Finally, we compared the photon-energy spectrum
$E_\gamma(\rd\sigma/\rd E_\gamma)$.
Since the {\sc RacoonWW} prediction for the $E_\gamma$ spectrum is not
based on a DPA, but on the full matrix element for $4f+\gamma$
production, this comparison can serve as a consistency check of the
DPA for the real-photonic corrections as used by BBC.  We find an
agreement between the two approaches within $\sim 10\%$, which is of
the order of the naive expectation for the DPA error of ${\cal
  O}(\Gamma_W/\Delta E)$.  Although these results illustrate the
reliability of the naive error estimate of the DPA used by BBC for the
real corrections at least for LEP2 energies, this observation cannot
serve as a proof that a DPA for real-photonic corrections will work
for any observable in any kinematical situation, as e.g.\ for higher
energies.
\end{sloppypar}

\subsubsection{Comparison with {\sc YFSWW3}}
\label{se:yfsww}

In contrast to the comparison with BBC, where some improvements of the
DPA were switched off in {\sc RacoonWW}, in the comparison with {\sc
  YFSWW3} only the input parameters, cuts, and recombination
procedures are chosen to be the same, but no further modifications of
the codes are made.  Both event generators produce their ``best''
results, which can also be directly compared to the LEP2 data.

We compared the total CC03 cross sections for leptonic, semileptonic,
and hadronic W-pair production channels and various distributions for
$\Pep\Pem\to \Pu \bar\Pd \mu^- \bar\nu_\mu$ at $\sqrt{s}=200\GeV$ for
the same setup as used in this paper.

The ``best'' results for the total cross sections obtained with 
{\sc RacoonWW} and {\sc {YFSWW3}} when no cuts, bare or calo cuts have
been applied are consistent with each other. 
The relative
differences amount to only $0.3\%$, which is within the expected DPA
accuracy of $0.5\%$.

The comparison of the distributions has been done for both, calo and
bare, recombination schemes.  We found that the W-production-angle and
W-invariant-mass distributions are statistically compatible with each
other, \ie they agree within 1\%.  We also compared several photon
observables, such as the distributions of the photon energy, of the 
photon production angle, and of the angle between the photon and the
nearest charged fermion and found relative differences at the 10\%
level. This is not surprising since visible photons are treated quite
differently in {\sc RacoonWW} and {\sc YFSWW3}: while the predictions
for photon observables by {\sc RacoonWW} are based on the
exact
lowest-order matrix element for $\eeffffg$, in {\sc YFSWW3}
multi-photon radiation to W-pair production (within the YFS scheme) is
combined with ${\cal O}(\alpha^2)$ LL photon radiation in the W decays
(done by PHOTOS).

\subsubsection{Comparison with Jegerlehner and Ko\l odziej}

In \citere{je99} JK have evaluated some cross sections and
distributions for semi-leptonic channels $\eeffffg$ of the CC11 class.
The calculation is based on the
exact matrix element with finite fermion masses, and finite
gauge-boson widths are introduced in the so-called fixed-width
scheme. The presented bremsstrahlung corrections to the total cross
sections of $\eeffff$, which have been regularized with a small photon
mass $\la$, can be compared with predictions made with {\sc RacoonWW}.
To this end, the virtual corrections are omitted in {\sc RacoonWW} and
consequently no DPA is needed.
The real correction provided by {\sc RacoonWW} corresponds to the full
$\eeffffg$ cross section in the successive asymptotic limits $\la\to
0$ and $m_f\to 0$. The relative difference to the results of JK should
thus be of the order of $m_f/\MW$, where $m_f$ is the largest fermion
mass of the process.  Table~\ref{tab:jk} shows a comparison of results
for the process $\Pep\Pem\to\Pu\bar\Pd\mu^-\bar\nu_\mu\gamma$, where
the input parameters of \citere{je99} are used consistently, and the
photon mass is set to $\la=10^{-6}\GeV$.
\begin{table}
\centerline{
\begin{tabular}{|c||l|l||l|}
\hline
$\sqrt{s}/\GeV$ & 189 & 500 & $\Delta E_\gamma/\GeV$ \\
\hline\hline
JK \cite{je99} & 1285(1)   & 571(1)   & 0.001 \\
               & 1285.0(8) & 570.0(6) & 0.1   \\
               & 1286.3(8) & 570.1(5) & 1.0   \\
\hline\hline
{\sc RacoonWW} -- slicing  
               & 1286(3)   & 575(3)   & 0.001 \\
               & 1286.2(4) & 570.4(3) & 0.1 \\
\hline
{\sc RacoonWW} -- subtraction & 
                 1284.9(5) & 570.4(3) & ---
\\
\hline
\end{tabular} }
\caption{Comparison of the total cross section $\sigma_\real/\fba$ for
$\Pep\Pem\to\Pu\bar\Pd\mu^-\bar\nu_\mu\gamma$ with $\la=10^{-6}\GeV$,
as obtained by JK \cite{je99} and {\sc RacoonWW} }
\label{tab:jk}
\end{table} 
The parameter $\Delta E_\gamma$ represents the cutoff that separates
the domains of hard and soft photon emission in the phase-space
slicing approach. The matching of hard and soft photons is correct up
to terms of ${\cal O}(\Delta E_\gamma)$, i.e.\ the exact results are
formally reproduced for $\Delta E_\gamma\to 0$.  Since the $4f\gamma$
matrix elements in {\sc RacoonWW} are based on massless fermions,
there is also a small acollinearity separation cut $\delcoll$ on all
photon emission angles in the slicing approach; in the numerics we
have set $\delcoll=0.001$.  The inclusion of the collinearity regions,
which involves finite fermion masses, is described in
\refse{se:singsli}.  The table shows acceptable agreement of the
slicing and subtraction branches of {\sc RacoonWW} with the results of
JK.

\section{Conclusions}
\label{se:concl}

A proper treatment of electroweak radiative corrections to the W-pair
production process $\eeWWffff$ is mandatory in order to account for
the experimental accuracy at LEP2 and future linear colliders.  A
practical calculation of the
${\cal O}(\alpha)$ corrections to the full $\eeffff$ process is beyond
present possibilities.  However, as long as W-pair channels dominate,
the combination of full lowest-order predictions for the four-fermion
processes $\eeffff$ with those corrections that are enhanced by two
resonant W~bosons approximates the full ${\cal O}(\alpha)$-corrected
cross sections within a relative accuracy of $0.5\%$, at least for
not too high energies.  A strategy for such a double-pole
approximation (DPA) for the ${\cal O}(\alpha)$ corrections and its
implementation in the Monte Carlo event generator {\sc RacoonWW} are
described in this paper.

In {\sc RacoonWW} only the non-leading 
virtual ${\cal O}(\alpha)$ corrections are treated in DPA, while real
photonic corrections are based on full matrix elements for the
radiative processes $\eeffffg$. In this way, potential problems in the
definition of a DPA for real photon emission are avoided, and possible
DPAs for the real corrections can be controlled.
We have shown explicitly how existing results for the corrections to
on-shell W-pair production and W~decay 
can be employed in the DPA; they define the class of so-called
factorizable corrections.  For the remaining, non-factorizable
contributions, which are due to photon exchange between the various
production and decay subprocesses, we have completed the results
already given in the literature.  Since virtual and real corrections
are not treated on equal footing, particular care is needed in the
treatment of soft and collinear singularities.  The proper
cancellation of these singularities has been discussed in detail; in
practice two different methods (a subtraction method and phase-space
slicing) have been applied.  Finally, the leading contributions from
initial-state radiation beyond ${\cal O}(\alpha)$ have been included
in {\sc RacoonWW}, rendering this program a state-of-the-art
calculation.

For LEP2 energies at detailed discussion of numerical results is
presented, including total cross sections, angular and invariant-mass
distributions.
The consistency of {\sc RacoonWW} is shown by comparing the results
obtained with the subtraction method with the ones of the slicing
approach.  Moreover, the reliability of the applied DPA is
demonstrated by various modifications of the DPA within its intrinsic
freedom.  The obtained intrinsic ambiguity supports the expectation
that the theoretical uncertainty of the used DPA is of the order of
$0.5\%$ for energies
between $170\GeV$ and $500\GeV$.  Above 500--$1000\GeV$,
leading-logarithmic electroweak corrections of higher orders have to
be taken into account.

As a first phenomenological application of {\sc RacoonWW} the
predictions for the total W-pair production at LEP2 have been compared
with LEP2 data and older ``improved Born predictions'' of GENTLE,
which were previously used in the data analyses.
The {\sc RacoonWW} results agree with the results of the new generator
{\sc YFSWW3} but lie 2--2.5\% below the GENTLE predictions.  The fact
that the experimental LEP2 data
favour predictions of {\sc RacoonWW} and {\sc YFSWW3} can be viewed as
empirical evidence for non-leading electroweak radiative corrections
beyond the level of universal effects.

\section*{Acknowledgement}

We thank all the authors of \citeres{Be98,ja99,ku99,je99}
for helpful information and discussions about their results.

\appendix
\section*{Appendix}

\section{Explicit form of an on-shell projection for the four-fermion final
state}
\label{app:onshell}

When performing the projection to on-shell \PW-boson momenta, care
should be taken that the projected momenta lie in the physical region.
This can be ensured by fixing angles while performing the limit
$k_\pm^2\to\MW^2$. An obvious way is to fix the direction of one of
the \PW~bosons and of one of the final-state fermions originating from
either \PW~boson in the CM frame of the incoming $\Pep\Pem$ pair.  We
fix the direction of the \PWp~boson, of the fermion $f_1$, and of the
fermion $f_3$.  Thus, we find for the on-shell-projected momenta
\beqar
\kon_{+0} &=& \frac{1}{2}\sqrt{s}, \qquad
\bkon_+   = \frac{\bk_+}{|\bk_+|}\be\frac{\sqrt{s}}{2},\qquad
\kon_-^\mu = p_+^\mu + p_-^\mu - \kon_{+}^\mu, \nl
\kon_{1}^\mu &=& k_{1}^\mu\frac{\MW^2}{2 \kon_+ k_1},\qquad
\kon_2^\mu = \kon_+^\mu - \kon_1^\mu, \nl
\kon_{3}^\mu &=& k_{3}^\mu\frac{\MW^2}{2 \kon_- k_3},\qquad
\kon_4^\mu = \kon_-^\mu - \kon_3^\mu
\eeqar
with $\beta=\sqrt{1-4\MW^2/s}$.

\section{Explicit expressions for standard matrix elements}
\label{app:smes}

The SMEs $\M^\sigma_n$, which have been defined in \refse{se:fRC},
have been calculated in the Weyl--van der Waerden spinor formalism
using the conventions of \citere{wvdw}. Since we consistently neglect
fermion masses, the final-state fermions are polarized according to
$\sigma_1 = -\sigma_2 = \sigma_3 = -\sigma_4 = -$, and the index
$\sigma$ determines the polarization of the incoming $\Pe^\pm$
according to $\sigma = \sigma_- = -\sigma_+$. The explicit expressions
for the two sets $\M^\pm_n$ read
\beqar
\M^+_1 &=& -4A\CSPppkb\SPpmkc\Big(\CSPppkd\SPppka + \CSPkbkd\SPkakb\Big), 
\nn\\
\M^+_2 &=& 2A \CSPkbkd\SPkakc
\Big(\CSPppka\SPpmka + \CSPppkb\SPpmkb\Big),
\nn\\
\M^+_3 &=& 2A
\CSPppkb\SPpmka\Big(\CSPkakd\SPkakc + \CSPkbkd\SPkbkc\Big),
\nn\\
\M^+_4 &=& -2A
\CSPppkd\SPpmkc\Big(\CSPkbkc\SPkakc + \CSPkbkd\SPkakd\Big),
\nn\\
\M^+_5 &=& 2A \CSPppkb\CSPpmkd\SPpmka\SPpmkc,
\nn\\
\M^+_6 &=& -2A \CSPppkb\CSPppkd\SPppka\SPpmkc,
\nn\\
\M^+_7 &=& A
\Big(\CSPkbkc\SPkakc + \CSPkbkd\SPkakd\Big)
\Big(\CSPkakd\SPkakc + \CSPkbkd\SPkbkc\Big)
\nn\\ && {} \times
\Big(\CSPppka\SPpmka + \CSPppkb\SPpmkb\Big),
\nn\\
\M^+_8 &=& A \CSPppkb\CSPpmkd\SPppka\SPpmkc
\Big(\CSPppka\SPpmka + \CSPppkb\SPpmkb\Big),
\nn\\
\M^+_9 &=& A
\CSPpmkd\SPpmkc\Big(\CSPkbkc\SPkakc + \CSPkbkd\SPkakd\Big)
\nn\\ && {} \times
\Big(\CSPppka\SPpmka + \CSPppkb\SPpmkb\Big),
\nn\\
\M^+_{10} &=& A
\CSPppkb\SPppka\Big(\CSPkakd\SPkakc + \CSPkbkd\SPkbkc\Big)
\nn\\ && {} \times
\Big(\CSPppka\SPpmka + \CSPppkb\SPpmkb\Big),
\nn\\[1em]
\M^-_1 &=& -4A\CSPpmkd\SPppka\Big(\CSPppkb\SPppkc - \CSPkakb\SPkakc\Big),
\nn\\
\M^-_2 &=& 2A \CSPkbkd\SPkakc
\Big(\CSPpmka\SPppka + \CSPpmkb\SPppkb\Big),
\nn\\
\M^-_3 &=& 2A
\CSPpmkb\SPppka\Big(\CSPkakd\SPkakc + \CSPkbkd\SPkbkc\Big),
\nn\\
\M^-_4 &=& -2A
\CSPpmkd\SPppkc\Big(\CSPkbkc\SPkakc + \CSPkbkd\SPkakd\Big),
\nn\\
\M^-_5 &=& 2A \CSPpmkb\CSPpmkd\SPppka\SPpmkc,
\nn\\
\M^-_6 &=& -2A\CSPppkb\CSPpmkd\SPppka\SPppkc,
\nn\\
\M^-_7 &=& A
\Big(\CSPkbkc\SPkakc + \CSPkbkd\SPkakd\Big)
\Big(\CSPkakd\SPkakc + \CSPkbkd\SPkbkc\Big)
\nn\\ && {} \times
\Big(\CSPpmka\SPppka + \CSPpmkb\SPppkb\Big),
\nn\\
\M^-_8 &=& A \CSPppkb\CSPpmkd\SPppka\SPpmkc
\Big(\CSPpmka\SPppka + \CSPpmkb\SPppkb\Big),
\nn\\
\M^-_9 &=& A
\CSPpmkd\SPpmkc\Big(\CSPkbkc\SPkakc + \CSPkbkd\SPkakd\Big)
\nn\\ && {} \times
\Big(\CSPpmka\SPppka + \CSPpmkb\SPppkb\Big),
\nn\\
\M^-_{10} &=& A
\CSPppkb\SPppka\Big(\CSPkakd\SPkakc + \CSPkbkd\SPkbkc\Big)
\nn\\ && {} \times
\Big(\CSPpmka\SPppka + \CSPpmkb\SPppkb\Big),
\eeqar
where $\langle pk\rangle$ denotes the spinor product for two light-like
momenta $p^\mu$ and $k^\mu$, and $A$ is the global factor
\beq
A = \frac{e^2}{2\sw^2} \,
\frac{1}{k_+^2-\MW^2+\ri\MW\GW} \,
\frac{1}{k_-^2-\MW^2+\ri\MW\GW}. 
\eeq

\section{Scalar integrals for the virtual non-factorizable corrections}
\label{app:scalints}

In \refse{se:nfRC} we have given the virtual non-factorizable
corrections in terms of scalar one-loop integrals.
In this appendix we list the explicit expressions for those scalar
integrals that are not already given in \citere{nfc1a}:
\beqar
\lefteqn{
D_0(p_\pm,k_\pm,k_i,\lambda,\Me,M,m_i) \sim
\frac{1}{t_{\pm i}K_\pm} \biggl\{
2\ln\biggl(\frac{\Me m_i}{-t_{\pm i}}\biggr)
\ln\biggl(\frac{\lambda\MW}{-K_\pm}\biggr)
-\ln^2\biggl(\frac{\Me\MW}{\MW^2-t}\biggr)
} \hspace*{23em} 
\nn\\
\lefteqn{
-\ln^2\biggl(\frac{m_i}{\MW}\biggr)
-\frac{\pi^2}{3}-\Li\biggl(1-\frac{t-\MW^2}{t_{\pm i}}\biggr)
\biggr\},
} \hspace*{6em} 
\nn\\[.5em] 
\lefteqn{
D_0(p_\pm,k_\mp,k_i,\lambda,\Me,M,m_i) \sim
D_0(p_\pm,k_\pm,k_i,\lambda,\Me,M,m_i) 
\Big|_{t\to u, t_{\pm i}\to u_{\pm i}, K_\pm\to K_\mp}, 
} \hspace*{23em} 
\nn\\[.5em]
C_0(p_\pm,k_\pm,0,\Me,M)
-\Bigl[C_0(p_\pm,k_\pm,\lambda,\Me,\MW)\Bigr]_{k_\pm^2=\MW^2} &&
\nn\\
\lefteqn{ \sim
\frac{1}{t-\MW^2} \biggl\{ 
\ln\biggl(\frac{\Me\MW}{\MW^2-t}\biggr)
\biggl[ \ln\biggl(\frac{-K_\pm}{\MW^2-t}\biggr)
+\ln\biggl(\frac{-K_\pm}{\lambda^2}\biggr)
+\ln\biggl(\frac{\Me}{\MW}\biggr) \biggr] + \frac{\pi^2}{6} \biggr\},
} \hspace*{20em}
\nn\\[.5em] 
C_0(p_\pm,k_\mp,0,\Me,M)
-\Bigl[C_0(p_\pm,k_\mp,\lambda,\Me,\MW)\Bigr]_{k_\mp^2=\MW^2} &&
\nn\\
\lefteqn{ \sim
\biggl\{ C_0(p_\pm,k_\pm,0,\Me,M)
-\Bigl[C_0(p_\pm,k_\pm,\lambda,\Me,\MW)\Bigr]_{k_\pm^2=\MW^2} \biggr\}
_{t\to u, K_\pm\to K_\mp},
} \hspace*{20em}
\nn\\[.5em] 
C_0(k_\pm,k_i,0,M,m_i)
-\Bigl[C_0(k_\pm,k_i,\lambda,\MW,m_i)\Bigr]_{k_\pm^2=\MW^2} &&
\nn\\
\lefteqn{ \sim
-\frac{1}{\MW^2} \biggl\{ 
\ln\biggl(\frac{m_i^2}{\MW^2}\biggr)
\ln\biggl(\frac{-K_\pm}{\lambda\MW}\biggr)
+\ln^2\biggl(\frac{m_i}{\MW}\biggr) + \frac{\pi^2}{6} \biggr\},
} \hspace*{20em}
\nn\\[.5em] 
\frac{B_0(k_\pm^2,0,M)-B_0(M^2,0,M)}{k_\pm^2-M^2}
-B'_0(\MW^2,\lambda,\MW) &\sim&
\frac{1}{\MW^2}\biggl\{ \ln\biggl(\frac{\lambda\MW}{-K_\pm}\biggr)+1
\biggr\}.
\hspace{2em}
\label{eq:B0}
\eeqar
Recall that all expression are given for
$k_\pm^2\to\MW^2$ and
 $\GW\to 0$.


\begin{thebibliography}{99}
\frenchspacing
\newcommand{\vj}[4]{{\sl #1~}{\bf~#2 }\ifnum#3<100 (19#3) \else (#3) \fi #4}
 \newcommand{\ej}[3]{{\bf #1~}\ifnum#2<100 (19#2) \else (#2) \fi #3}
 \newcommand{\vjs}[2]{{\sl #1~}{\bf #2}}
 \renewcommand{\vj}[4]{{\sl #1 }{\bf #2 }\ifnum#3<100 (19#3) \else (#3) \fi #4}
 \renewcommand{\ej}[3]{{\bf #1 }\ifnum#2<100 (19#2) \else (#2) \fi #3}
 \renewcommand{\vjs}[2]{{\sl #1~}{\bf #2}}

 \newcommand{\am}[3]{\vj{Ann.~Math.}{#1}{#2}{#3}}
 \newcommand{\ap}[3]{\vj{Ann.~Phys.}{#1}{#2}{#3}}
 \newcommand{\app}[3]{\vj{Acta~Phys.~Pol.}{#1}{#2}{#3}}
 \newcommand{\cmp}[3]{\vj{Commun. Math. Phys.}{#1}{#2}{#3}}
 \newcommand{\cnpp}[3]{\vj{Comments Nucl. Part. Phys.}{#1}{#2}{#3}}
 \newcommand{\cpc}[3]{\vj{Comput. Phys. Commun.}{#1}{#2}{#3}}
 \newcommand{\epj}[3]{\vj{Eur. Phys. J.}{#1}{#2}{#3}}
 \newcommand{\epjdir}[3]{{\sl EPJdirect} {\bf #1} (#2) #3}
 \newcommand{\epl}[3]{\vj{Europhys. Lett.}{#1}{#2}{#3}}
 \newcommand{\fp}[3]{\vj{Fortschr. Phys.}{#1}{#2}{#3}}
 \newcommand{\hpa}[3]{\vj{Helv. Phys.~Acta}{#1}{#2}{#3}}
 \newcommand{\ijmp}[3]{\vj{Int. J. Mod. Phys.}{#1}{#2}{#3}}
 \newcommand{\jetp}[3]{\vj{JETP}{#1}{#2}{#3}}
 \newcommand{\jetpl}[3]{\vj{JETP Lett.}{#1}{#2}{#3}}
 \newcommand{\jmp}[3]{\vj{J. Math. Phys.}{#1}{#2}{#3}}
 \newcommand{\jp}[3]{\vj{J. Phys.}{#1}{#2}{#3}}
 \newcommand{\lnc}[3]{\vj{Lett. Nuovo Cimento}{#1}{#2}{#3}}
 \newcommand{\mpl}[3]{\vj{Mod. Phys. Lett.}{#1}{#2}{#3}}
 \newcommand{\nc}[3]{\vj{Nuovo Cimento}{#1}{#2}{#3}}
 \newcommand{\nim}[3]{\vj{Nucl. Instr. Meth.}{#1}{#2}{#3}}
 \newcommand{\np}[3]{\vj{Nucl. Phys.}{#1}{#2}{#3}}
 \newcommand{\npbps}[3]{\vj{Nucl. Phys. B (Proc. Suppl.)}{#1}{#2}{#3}}
 \newcommand{\pl}[3]{\vj{Phys. Lett.}{#1}{#2}{#3}}
 \newcommand{\prp}[3]{\vj{Phys. Rep.}{#1}{#2}{#3}}
 \newcommand{\pr}[3]{\vj{Phys. Rev.}{#1}{#2}{#3}}
 \newcommand{\prl}[3]{\vj{Phys. Rev. Lett.}{#1}{#2}{#3}}
 \newcommand{\ptp}[3]{\vj{Prog. Theor. Phys.}{#1}{#2}{#3}}
 \newcommand{\rpp}[3]{\vj{Rep. Prog. Phys.}{#1}{#2}{#3}}
 \newcommand{\rmp}[3]{\vj{Rev. Mod. Phys.}{#1}{#2}{#3}}
 \newcommand{\rnc}[3]{\vj{Revista del Nuovo Cim.}{#1}{#2}{#3}}
 \newcommand{\sjnp}[3]{\vj{Sov. J. Nucl. Phys.}{#1}{#2}{#3}}
 \newcommand{\sptp}[3]{\vj{Suppl. Prog. Theor. Phys.}{#1}{#2}{#3}}
 \newcommand{\zp}[3]{\vj{Z. Phys.}{#1}{#2}{#3}}

\bibitem{wwrev}
W. Beenakker and A. Denner, \ijmp{A9}{94}{4837}.

\bibitem{lep2rep}
G.~Altarelli, T.~Sj\"o\-strand and F.~Zwirner (eds.),
{\sl Physics at LEP2} (Report CERN 96-01, Geneva, 1996).

\bibitem{lep2mcws}
M.~Gr\"unewald et al., 
``Four Fermion Production in Electron Positron Collisions'',
report of the four-fermion working group of the {\it LEP2 Monte Carlo
workshop}, CERN, 1999/2000, hep-ph/0005309.

\bibitem{bu99}
C.~Burgard, talk given at the {\it International Workshop on Linear Colliders},
Sitges, Barcelona, Spain, 1999.

\bibitem{lep2repWevgen}
D.~Bardin et al., in \citere{lep2rep}, Vol.~2, p.~3, hep-ph/9709270.

\bibitem{bo92}
M.~B\"ohm, A.~Denner and S.~Dittmaier, \np{B376}{92}{29};
E: \ej{B391}{1993}{483}.

\bibitem{lep2repWcs}
W.~Beenakker et al., in \citere{lep2rep}, Vol.~1, p.~79, hep-ph/9602351.

\bibitem{di97}
S.~Dittmaier, \app{B28}{97}{619}; \\
A.~Denner and S.~Dittmaier,
{\it Proceedings of the Joint ECFA/DESY
Study: Physics and Detectors for a Linear Collider},
Frascati, London, Munich, Hamburg, 1996,
ed.\ R.~Settles, DESY 97-123E (Hamburg, 1997), p.~131,
hep-ph/9706388.

\bibitem{Vi98} 
A. Vicini, \app{B29}{98}{2847}.

\bibitem{rcwprod1}
M.~B\"ohm et al., \np{B304}{88}{463}.

\bibitem{rcwprod2}
J. Fleischer, F. Jegerlehner and M. Zra\l ek, \zp{C42}{89}{409}.

\bibitem{rcwdecay2}
D.Yu. Bardin, S. Riemann and T. Riemann,  \zp{C32}{86}{121};\\
F. Jegerlehner, \zp{C32}{86}{425}.

\bibitem{rcwdecay1}
A.~Denner and T. Sack, \zp{C46}{90}{653}.

\bibitem{me96}
K. Melnikov and O.I. Yakovlev, \np{B471}{96}{90}.

\bibitem{nfc2}
W. Beenakker,  A.P. Chapovsky and F.A. Berends,
\pl{B411}{97}{203} and \np{B508}{97}{17}.

\bibitem{nfc1a}
A. Denner, S. Dittmaier and M. Roth, \np{B519}{98}{39}.

\bibitem{ro99}
M. Roth, dissertation ETH Z\"urich No.~13363, 1999.

\bibitem{Ae94}
A. Aeppli, G.J. van Oldenborgh and D. Wyler, \np{B428}{94}{126}.

\bibitem{Be98}
W. Beenakker,  A.P. Chapovsky and F.A. Berends,
\np{B548}{99}{3}.

\bibitem{ja97}
S.~Jadach et al., \pl{B417}{98}{326}.

\bibitem{ja99}
S.~Jadach et al., \pr{D61}{2000}{113010}.

\bibitem{Ch99} 
A.P. Chapovsky and V.A. Khoze, \epj{C9}{99}{449}.

\bibitem{ku99} 
Y. Kurihara, M. Kuroda and D. Schildknecht,
\np{B565}{2000}{49}.

\bibitem{be93} 
W.~Beenakker et al., \pl{B317}{1993}{622} and
\np{B410}{1993}{245}; \\
M.~Kuroda and D.~Schildknecht, \np{B531}{1998}{24}.

\bibitem{de99a}
A. Denner, S. Dittmaier, M. Roth and D. Wackeroth,
\pl{B475}{2000}{127}.

\bibitem{de99b}
A. Denner, S. Dittmaier, M. Roth and D. Wackeroth,
\epjdir{Vol.~2 C4}{2000}{1} (hep-ph/9912447).

\bibitem{di99} 
S.~Dittmaier, \np{B565}{2000}{69}.

\bibitem{ee4fa}
A. Denner, S. Dittmaier, M. Roth and D. Wackeroth, 
\np{B560}{1999}{33}.

\bibitem{polescheme}
R.G. Stuart, \pl{B262}{91}{113};\\ 
H. Veltman, \zp{C62}{94}{35};\\ 
A. Aeppli, F. Cuypers and G.J. van Oldenborgh, \pl{B314}{93}{413}.

\bibitem{bhf2}
W. Beenakker et al., \np{B500}{97}{255}.

\bibitem{Yennie:1961ad}
D.R. Yennie, S.C. Frautschi and H. Suura, \ap{13}{1961}{379}.

\bibitem{FA} J.~K\"ublbeck, M.~B\"ohm and A.~Denner, \cpc{60} {90}{165}; \\
H.~Eck and J.~K\"ublbeck, {\it Guide to FeynArts 1.0\/},
University of W\"urzburg, 1992.

\bibitem{math} S.~Wolfram, {\it Mathematica --- A System for Doing
Mathematics by Computer} (Addison-Wesley, Redwood City, CA, 1988).

\bibitem{de93}
A.~Denner, \fp{41}{93}{307}.

\bibitem{de95}
A.~Denner, S.~Dittmaier and G.~Weiglein, \np{B440}{95}{95}.

\bibitem{pa79}  
G.\ Passarino and M.\ Veltman, \np{B160}{79}{151}.

\bibitem{nfc1b}
A. Denner, S. Dittmaier and M. Roth, \pl{B429}{98}{145}.

\bibitem{ca96}
S. Catani and M.H. Seymour, \pl{B378}{96}{287} and
\np{B485}{97}{291},
E: \ej{B510}{98}{503}.

\bibitem{be82}
F.A.~Berends et al., \np{B206}{82}{61}; \\
R.~Kleiss, \zp{C33}{87}{433}.

\bibitem{bo93}
M.~B\"ohm and S.~Dittmaier,
\np{B409}{93}{3} and {\bf B412} (1994) 39.

\bibitem{ba99}
U.~Baur, S.~Keller and D.~Wackeroth, \pr{D59}{99}{013002}.

\bibitem{'tHooft:1979xw}
G.~'t Hooft and M.~Veltman,
Nucl.\ Phys.\  {\bf B153}, 365 (1979).

\bibitem{Ki62} T. Kinoshita, \jmp{3}{62}{650};\\
T.D.Lee and M. Nauenberg, \pr{133}{64}{1549}.

\bibitem{Wackeroth:1997hz}
D.~Wackeroth and W.~Hollik,
\pr{D55}{97}{6788}.

\bibitem{sf}
E.A.~Kuraev and V.S.~Fadin, 
\vj{Yad.\ Fiz.}{41}{1985}{753} [\sjnp{41}{1985}{466}];\\
G.~Altarelli and G.~Martinelli, in {\it ``Physics at LEP''}, 
eds. J.~Ellis and R.~Peccei, CERN 86-02 (CERN, Geneva, 1986), Vol.~1, p.~47;\\
O.~Nicrosini and L.~Trentadue, \pl{B196}{1987}{551}; \zp{C39}{1988}{479};\\
F.A.~Berends, G.~Burgers and W.L.~van Neerven, \np{B297}{1988}{429},
E:\ej{\bf B304}{88}{921}.

\bibitem{LEPEWWG} 
Homepage of the LEP Electroweak Working Group, \\
http://lepewwg.web.cern.ch/LEPEWWG/. 

\bibitem{gentle}
D.~Bardin, M.~Bilenky, A.~Olchevski and T.~Riemann,
\pl{B308}{93}{403};\\
D.~Bardin et al., \cpc{104}{97}{161}.

\bibitem{je99}
F. Jegerlehner and K. Ko\l odziej, 
\epj{C12}{2000}{77}.

\bibitem{coul}
V.S.~Fadin, V.A.~Khoze and A.D.~Martin, \pl{B311}{93}{311};\\
D.~Bardin, W.~Beenakker and A.~Denner, \pl{B317}{93}{213};\\
V.S.~Fadin et al., \pr{D52}{95}{1377}.

\bibitem{wvdw} 
S.~Dittmaier, \pr{D59}{99}{016007}.

\end{thebibliography}
\end{document}